\documentclass[ALICE,manyauthors]{cernphprep}
\usepackage[comma,square,numbers,sort&compress]{natbib}
\usepackage{hyperref}
\usepackage{lineno}
\usepackage{xspace}
\usepackage{xcolor}
\usepackage{booktabs}
\usepackage{multicol}
\usepackage{multirow}
\usepackage[T1]{fontenc}
\usepackage{orcidlink}

\begin{document}
%

\newcommand{\pp}           {pp\xspace}
\newcommand{\ppbar}        {\mbox{$\mathrm {p\overline{p}}$}\xspace}
\newcommand{\XeXe}         {\mbox{Xe--Xe}\xspace}
\newcommand{\PbPb}         {\mbox{Pb--Pb}\xspace}
\newcommand{\pA}           {\mbox{pA}\xspace}
\newcommand{\pPb}          {\mbox{p--Pb}\xspace}
\newcommand{\AuAu}         {\mbox{Au--Au}\xspace}
\newcommand{\dAu}          {\mbox{d--Au}\xspace}
\newcommand{\Pbp}          {\mbox{Pb--p}\xspace}

\newcommand{\s}            {\ensuremath{\sqrt{s}}\xspace}
\newcommand{\snn}          {\ensuremath{\sqrt{s_{\mathrm{NN}}}}\xspace}
\newcommand{\pt}           {\ensuremath{p_{\rm T}}\xspace}
\newcommand{\meanpt}       {$\langle p_{\mathrm{T}}\rangle$\xspace}
\newcommand{\ycms}         {\ensuremath{y_{\rm CMS}}\xspace}
\newcommand{\ylab}         {\ensuremath{y_{\rm lab}}\xspace}
\newcommand{\etalab}       {\ensuremath{\eta_{\rm lab}}\xspace}
\newcommand{\etarange}[1]  {\mbox{$\left | \eta \right |~<~#1$}}
\newcommand{\yrange}[1]    {\mbox{$\left | y \right |~<~#1$}}
\newcommand{\dndy}         {\ensuremath{\mathrm{d}N_\mathrm{ch}/\mathrm{d}y}\xspace}
\newcommand{\dndeta}       {\ensuremath{\mathrm{d}N_\mathrm{ch}/\mathrm{d}\eta}\xspace}
\newcommand{\dndetalab}       {\ensuremath{\mathrm{d}N_\mathrm{ch}/\mathrm{d}\eta_{\rm lab}}\xspace}
\newcommand{\dndetaphoton} {\ensuremath{\mathrm{d}N_\mathrm{\gamma}/\mathrm{d}\eta}\xspace}
\newcommand{\dndetaphotonlab} {\ensuremath{\mathrm{d}N_\mathrm{\gamma}/\mathrm{d}\eta_{\rm lab}}\xspace}
\newcommand{\avdndeta}     {\ensuremath{\langle\dndeta\rangle}\xspace}
\newcommand{\dNdy}         {\ensuremath{\mathrm{d}N_\mathrm{ch}/\mathrm{d}y}\xspace}
\newcommand{\Npart}        {\ensuremath{N_\mathrm{part}}\xspace}
\newcommand{\Ncoll}        {\ensuremath{N_\mathrm{coll}}\xspace}
\newcommand{\dEdx}         {\ensuremath{\textrm{d}E/\textrm{d}x}\xspace}
\newcommand{\RpPb}         {\ensuremath{R_{\rm pPb}}\xspace}

\newcommand{\nineH}        {$\sqrt{s}~=~0.9$~Te\kern-.1emV\xspace}
\newcommand{\seven}        {$\sqrt{s}~=~7$~Te\kern-.1emV\xspace}
\newcommand{\twoH}         {$\sqrt{s}~=~0.2$~Te\kern-.1emV\xspace}
\newcommand{\twosevensix}  {$\sqrt{s}~=~2.76$~Te\kern-.1emV\xspace}
\newcommand{\five}         {$\sqrt{s}~=~5.02$~Te\kern-.1emV\xspace}
\newcommand{\twosevensixnn}{$\sqrt{s_{\mathrm{NN}}}~=~2.76$~Te\kern-.1emV\xspace}
\newcommand{\fivenn}       {$\sqrt{s_{\mathrm{NN}}}~=~5.02$~Te\kern-.1emV\xspace}
\newcommand{\LT}           {L{\'e}vy-Tsallis\xspace}
\newcommand{\GeVc}         {Ge\kern-.1emV/$c$\xspace}
\newcommand{\MeVc}         {Me\kern-.1emV/$c$\xspace}
\newcommand{\TeV}          {Te\kern-.1emV\xspace}
\newcommand{\GeV}          {Ge\kern-.1emV\xspace}
\newcommand{\MeV}          {Me\kern-.1emV\xspace}
\newcommand{\GeVmass}      {Ge\kern-.2emV/$c^2$\xspace}
\newcommand{\MeVmass}      {Me\kern-.2emV/$c^2$\xspace}
\newcommand{\lumi}         {\ensuremath{\mathcal{L}}\xspace}

\newcommand{\ITS}          {\rm{ITS}\xspace}
\newcommand{\TOF}          {\rm{TOF}\xspace}
\newcommand{\ZDC}          {\rm{ZDC}\xspace}
\newcommand{\ZDCs}         {\rm{ZDCs}\xspace}
\newcommand{\ZNA}          {\rm{ZNA}\xspace}
\newcommand{\ZNC}          {\rm{ZNC}\xspace}
\newcommand{\SPD}          {\rm{SPD}\xspace}
\newcommand{\SDD}          {\rm{SDD}\xspace}
\newcommand{\SSD}          {\rm{SSD}\xspace}
\newcommand{\TPC}          {\rm{TPC}\xspace}
\newcommand{\TRD}          {\rm{TRD}\xspace}
\newcommand{\VZERO}        {\rm{V0}\xspace}
\newcommand{\VZEROA}       {\rm{V0A}\xspace}
\newcommand{\VZEROC}       {\rm{V0C}\xspace}
\newcommand{\Vdecay} 	   {\ensuremath{V^{0}}\xspace}
\newcommand{\PMD}          {\rm{PMD}\xspace}
\newcommand{\ZPA}          {\rm{ZPA}\xspace}
\newcommand{\ZPC}          {\rm{ZPC}\xspace}

\newcommand{\ee}           {\ensuremath{e^{+}e^{-}}} 
\newcommand{\pip}          {\ensuremath{\pi^{+}}\xspace}
\newcommand{\pim}          {\ensuremath{\pi^{-}}\xspace}
\newcommand{\pizero}        {\ensuremath{\pi^{0}}\xspace}
\newcommand{\kap}          {\ensuremath{\rm{K}^{+}}\xspace}
\newcommand{\kam}          {\ensuremath{\rm{K}^{-}}\xspace}
\newcommand{\pbar}         {\ensuremath{\rm\overline{p}}\xspace}
\newcommand{\kzero}        {\ensuremath{{\rm K}^{0}_{\rm{S}}}\xspace}
\newcommand{\lmb}          {\ensuremath{\Lambda}\xspace}
\newcommand{\almb}         {\ensuremath{\overline{\Lambda}}\xspace}
\newcommand{\Om}           {\ensuremath{\Omega^-}\xspace}
\newcommand{\Mo}           {\ensuremath{\overline{\Omega}^+}\xspace}
\newcommand{\X}            {\ensuremath{\Xi^-}\xspace}
\newcommand{\Ix}           {\ensuremath{\overline{\Xi}^+}\xspace}
\newcommand{\Xis}          {\ensuremath{\Xi^{\pm}}\xspace}
\newcommand{\Oms}          {\ensuremath{\Omega^{\pm}}\xspace}
\newcommand{\degree}       {\ensuremath{^{\rm o}}\xspace}

\newcommand{\pyperu}       {\rm{PYTHIA~6~Perugia~2011}\xspace}
\newcommand{\pymonash}     {\rm{PYTHIA~8~Monash~2013}\xspace}
\newcommand{\phojet}       {\rm{PHOJET}\xspace}
\newcommand{\eposlhc}      {\rm{EPOS~LHC}\xspace}
\newcommand{\dpmjet}       {\rm{DPMJET}\xspace}
\newcommand{\hijing}       {\rm{HIJING}\xspace}
\newcommand{\ampt}         {\rm{AMPT}\xspace}

\begin{titlepage}
  \PHyear{2023}       
  \PHnumber{026}      
  \PHdate{27 February}  
  
  \title{Inclusive photon production at forward rapidities in \pp and \pPb collisions at
    $\sqrt{\textbf{\textit{s}}_{\rm \textbf{NN}}}~=~\textbf{5.02}$~TeV}
  \ShortTitle{Inclusive photon production in \pp and \pPb}   
  
  \Collaboration{ALICE Collaboration\thanks{See Appendix~\ref{app:collab} for the list of collaboration members}}
  \ShortAuthor{ALICE Collaboration} 
  
  \begin{abstract}
    A study of multiplicity and pseudorapidity distributions of inclusive photons measured in
    \pp and \pPb collisions at a center-of-mass energy per nucleon--nucleon collision of \fivenn using the
    ALICE detector in the forward pseudorapidity region 2.3~$<~\etalab~<$~3.9 is presented. Measurements
    in \pPb collisions are reported for two beam configurations in which the directions of the proton
    and lead ion beam were reversed. The pseudorapidity distributions in
    \pPb collisions are obtained for seven centrality classes which are defined based on different event
    activity estimators, i.e., the charged-particle multiplicity measured at midrapidity as well as the
    energy deposited in a calorimeter at beam rapidity.
    The inclusive photon multiplicity distributions for both \pp and \pPb collisions are described by
    double negative binomial distributions. The pseudorapidity distributions of inclusive photons are
    compared to those of charged particles at midrapidity in \pp collisions and for different centrality
    classes in \pPb collisions. The results are compared to predictions from various Monte Carlo event
    generators. None of the generators considered in this paper reproduces the inclusive photon multiplicity
    distributions in the reported multiplicity range. The pseudorapidity distributions are, however, better
    described by the same generators.
        
  \end{abstract}
\end{titlepage}

\setcounter{page}{2} 


\section{Introduction} 
The primary goal of high-energy heavy-ion physics is the study of a new state of nuclear matter,
the quark--gluon plasma (QGP), a thermalized system of partons (quarks and gluons)~\cite{QGP-1,QGP-2,ALICE:ReviewPaper}.
The study of proton--proton (\pp) and proton--nucleus (p--A) collisions provides the baseline for the
interpretation of the effects of the QGP formation that are observed in heavy-ion collisions.
In addition, the study of p--A collisions helps to understand the effects of cold nuclear matter
on the production of final-state particles. Among these effects, an important role is played by
transverse momentum ($k_{\rm T}$)  broadening of initial- and final-state partons~\cite{CNM_effect2},
and by the modification of the parton distributions functions in bound nucleons compared to those of
free nucleons, leading in particular to a reduction of parton densities (shadowing) at small parton
fractional momentum $x$~\cite{CNM_effect1}. Interestingly, recent experimental results in \pp and
p--A collisions particularly at high multiplicities have shown features such as collective flow and
strangeness enhancement which are usually attributed to the formation of a QGP in heavy-ion
collisions~\cite{CMS:2010ifv,CMS:2015fgy,ATLAS:2015hzw,ALICE:2012eyl,ATLAS:2012cix,CMS:2012qk,ATLAS:2014qaj,ALICE:2016fzo}.
The origin of these phenomena in \pp and p--A systems is not yet fully understood and under
active scrutiny by the community. It is therefore important to understand the global properties
of the system produced in \pp and p--A collisions. Multiplicity and pseudorapidity distributions
of produced particles are some examples of global observables, providing important information
about the particle production mechanisms in these collisions. The integrated yield of particle
production is mostly dominated by soft quantum chromodynamics (QCD) interactions, i.e.~small
momentum transfer ($Q^2$) processes, and is described by non-perturbative phenomenological models.
On the other hand, the hard particle production (large $Q^2$) can be described by the
well-established theory of perturbative quantum chromodynamics (pQCD). Measurements of multiplicity and
pseudorapidity distributions provide important constraints on these models. Moreover, the variation of
pseudorapidity ($\eta$) density of produced particles ($\mathrm{d}N/\mathrm{d}\eta$) with collision
centrality can be parametrized by using a two-component model to extract the relative contributions to
particle production from hard scatterings and soft processes~\cite{two_component_model1,two_component_model2}.

The measurement of the multiplicity of inclusive photons provides complementary information with
respect to those of charged particles as the inclusive photons are mostly produced in the decay of
neutral pions (\pizero)~\cite{ALICE:PMDpaper,STAR:PMD_prl,STAR:PMD_prc}. A comparative
study of charged particles and inclusive photons can shed light on the possible similarities
and/or differences in the underlying mechanisms of charged and neutral particle production. 

At the Large Hadron Collider (LHC) energies, the underlying mechanisms of particle production could be
different at central and forward rapidities. Charged-particle multiplicity measurements at the LHC were
previously performed by ALICE, ATLAS, CMS, and LHCb experiments in \pp, \pPb, and \PbPb collisions~\cite{ALICE:ppMidChMultpaper900,ALICE:ppMidChMultpaper900and2360,ALICE:ppMidChMultpaper7000,ALICE:ppMidChMultpaper900to8000,ALICE:ppFrdChMultpaper,ALICE:pPbMidChdNdEtapaper5020MB,ALICE:pPbMidChdNdEtapaper5020Cent,ALICE:pPbMidChdNdEtapaper8160,ALICE:ppMultDependent,ALICE:MBMidrapidity2760,ALICE:PbPb_CentMidrapidity2760,ALICE:PbPb_CentFrdrapidity2760,ALICE:PbPbsatellite,ALICE:PbPb_CentMidrapidity5020,ALICE:PbPb_CentFrdrapidity5020,ATLAS:ChPrMidpp900,ATLAS:ChPrMidpp900_2760_7000,ATLAS:ChPrMidpp8000,ATLAS:ChPrMidpPb5020,ATLAS:PbPb_CentMidrapidity2760,CMS:ChPrMidpp900_2360_7000,CMS:ChPrMidpp13,CMS:ChPrMidpPb5TeV,CMS:ChPrMidPbPb2760,LHCb:ChPrFrdpp7000}. 
Inclusive photon production at forward rapidity was studied in \pp collisions at \s~$=$~0.9, 2.76, and
7~\TeV by ALICE~\cite{ALICE:PMDpaper}. At midrapidity, the mean charged-particle multiplicity is found
to follow a power law dependence on \s~\cite{ALICE:ppMidChMultpaper900and2360,ALICE:ppMidChMultpaper7000,ALICE:ppMidChMultpaper900to8000}
whereas the mean multiplicity of inclusive photons at forward rapidity can be described by both a
logarithmic and a power law dependence with \s at LHC energies~\cite{ALICE:PMDpaper}.

In the present study, measurements of multiplicity and pseudorapidity distributions of inclusive photons
at forward rapidity are reported for \pp, \pPb, and \Pbp collisions at \fivenn with ALICE. The dependence
of the inclusive photon production on the centrality of the collision, which is related to the impact
parameter of the p--A collision, and its scaling behavior with the number of participating nucleons (\Npart)
at forward rapidity is also studied for the first time in \pPb collisions at \fivenn. This measurement
enables an extension of our understanding of particle production at forward rapidity.

This paper is organized as follows. The ALICE sub-detectors relevant for the measurement of inclusive
photons are described in Sec.~\ref{exsetup}. The data samples used in this analysis and the selection
of events are discussed in Sec.~\ref{datasample}. Section~\ref{centdef} discusses the details of the
used centrality estimators. A discussion on Monte Carlo (MC) event generators and the simulation
framework is given in Sec.~\ref{Simdetails}. The reconstruction of inclusive photons is presented
in Sec.~\ref{PhRecons}. The correction for instrumental effects using the unfolding method is
described in Sec.~\ref{correction}. Section~\ref{SysUncEst} discusses the estimation of systematic
uncertainties from various sources. Section~\ref{results} presents the results of inclusive photon
multiplicity and pseudorapidity distributions obtained in this analysis and the outcome of this
study is summarized in Sec.~\ref{summary}.

\section{Experimental setup}
\label{exsetup}
A comprehensive description of the ALICE detectors and their performances can be found in
Refs.~\cite{ALICE:Exp,ALICE:performance}. The Photon Multiplicity Detector (\PMD) is used for the
detection of inclusive photons at forward rapidity whereas for the purpose of event selection and
centrality determination, the Silicon Pixel Detector (\SPD), the \VZERO detector and the Zero Degree
Calorimeter (\ZDC) are used in this analysis. The ALICE reference frame is defined with the $z$ axis
directed along the beam line, the nominal interaction point (IP) at $z = 0$, and the positive $z$
direction pointing towards the \PMD.

The \PMD is a preshower detector to measure inclusive photon multiplicity and its spatial distribution
in the forward rapidity region. It is a gaseous detector placed at $z = 367$~cm covering the pseudorapidity
region of 2.3~$<~\etalab~<$~3.9 with full azimuth coverage. The \PMD consists of two fine granular planes:
the Charged Particle Veto plane and the preshower (PRE) plane. A lead converter with a thickness of
three radiation lengths (3$X_0$) is placed between these two planes. The PMD consists of 184,320
hexagonal cells of size 0.22~cm$^{2}$ and depth of 0.5~cm with a copper honey-comb shaped cathode
extended towards a 20~$\mu$m thick gold-plated tungsten wire at ground potential at the center of
each cell. The cells are arranged in 40 modules in two planes. Each cell is filled with a gas mixture
of Ar and CO$_2$ with a 70:30 ratio. Due to the presence of the lead converter,
the incident photons produce electromagnetic showers by pair production and bremsstrahlung radiation and hit
several cells in the PRE plane. On the other hand, charged hadrons generally fire only one or two cells
in the PRE plane and produce a signal typical of a minimum ionizing particle. After receiving signals in the
PRE plane a photon reconstruction algorithm is applied and a photon-rich sample of clusters is obtained by
applying suitable rejection criteria to remove clusters from hadrons and from secondary
particles as discussed in Sec.~\ref{PhRecons}.
The performance of the \PMD is described in Refs.~\cite{ALICE:PMDtdr1,ALICE:PMDtdr2,ALICE:PMDpaper}.

The \SPD makes up the two innermost cylindrical layers of the ALICE Inner Tracking System
(\ITS)~\cite{ALICE:ITStdr} surrounding the beam pipe. The two layers cover the pseudorapidity
ranges $|\etalab| < 2$ (inner layer) and $|\etalab| < 1.4$ (outer layer). For this analysis,
the SPD is mainly used in the determination of the position of the interaction vertex,
the selection of the minimum bias (MB) trigger (combined with \VZERO), and the centrality
estimation~\cite{ALICE:ppMultDependent,ALICE:pPbMidChdNdEtapaper5020Cent,ALICE:pPbMidChdNdEtapaper8160}.
The \VZERO detector~\cite{ALICE:FrdDettdr,ALICE:V0performance} is made of two scintillator arrays
placed on either side of the IP at $z = 330$~cm (\VZEROA, covering the pseudorapidity
interval 2.8~$<$~$\etalab$~$<$~5.1), and $z = -90$~cm (\VZEROC, covering $-3.7$~$<$~$\etalab$~$<$~$-1.7$).
The \VZERO is used for event selection and the determination
of collision centrality~\cite{ALICE:pPbMidChdNdEtapaper5020Cent,ALICE:pPbMidChdNdEtapaper8160,ALICE:PbPb_CentMidrapidity2760,ALICE:PbPb_CentMidrapidity5020}. The \ZDC~\cite{ALICE:ZDCtdr} consists of two sets of neutron
(\ZNA and \ZNC) and proton (\ZPA and \ZPC) calorimeters positioned at $\pm$ 112.5~m from the IP,
on both sides. The ZPA and ZPC calorimeters cover the pseudorapidity range 6.5~$<$~$|\etalab|$~$<$~7.4,
while the \ZNA and \ZNC calorimeters have a geometric coverage $|\etalab| > 8.8$. The energy detected
by the ZN calorimeter on the Pb-remnant side (\ZNA) is used to determine the centrality in \pPb
collisions~\cite{ALICE:pPbMidChdNdEtapaper5020Cent,ALICE:pPbMidChdNdEtapaper8160}.

\section{Data sample and event selection}
\label{datasample}
The \pPb and \pp data samples used in this analysis were collected in 2013 during LHC Run 1 and in 2015 during LHC Run 2,
respectively. The \pPb collisions were recorded for two beam configurations: in one (denoted as \pPb), the proton
beam at 4~\TeV energy was pointing towards the negative $z$ direction in the ALICE reference
system~\cite{ALICE:CordSys}, while the lead ions at 1.58~\TeV per nucleon energy moved in the opposite direction;
in the other configuration (denoted as \Pbp), the directions of both proton and lead ion beams were reversed.
The nucleon--nucleon center-of-mass energy for both configurations resulted in \fivenn. Due to the asymmetry of the
beam energies, the nucleon--nucleon center-of-mass system in \pPb collisions is shifted in rapidity by
$\Delta y$~$=$~0.465 with respect to symmetric pp collisions in the direction of the proton beam. In the following,
the variable $\etalab$ is used to indicate the pseudorapidity in the laboratory reference frame. For \pp collisions, the
center-of-mass frame is same as the laboratory frame.

In \pp collisions, a sample of inelastic (INEL) events was selected using a MB trigger condition,
which required the detection of at least one particle in either of the two \VZERO scintillator arrays or in the \SPD.
For \pPb and \Pbp data samples, non-single diffractive (NSD) events were selected using a MB trigger condition, which
required signals in both \VZEROA and \VZEROC (\VZERO-AND requirement). The background events such as beam--gas or
beam--halo interactions occurring outside the interaction region were rejected using the timing information from the
\VZERO detector. Pile-up events, in which two or more collisions are detected as single events, were minimized using
the procedure outlined in Ref.~\cite{ALICE:ppMidChMultpaper900to8000}. Events were further selected by restricting the
reconstructed primary vertex position along the beam axis within $\pm$ 10~cm from the nominal IP. After applying all
selection criteria, about 23\,M INEL \pp collisions, 90\,M NSD \pPb events, and 1.2\,M NSD \Pbp collisions are considered
for this analysis.

\section{Centrality determination in \pPb collisions}
\label{centdef}
In \pPb collisions, the centrality is determined using two centrality estimators. The first is based on the number of
clusters in the outer layer of the \SPD (CL1) at midrapidity and the second one makes use of the energy
deposited in the neutron calorimeter in the Pb-remnant side at large rapidity (\ZNA for the \pPb beam configuration).
Multiplicity-based centrality selection is performed by fitting the multiplicity distribution of the number of
\SPD clusters with a Glauber MC model~\cite{GlauberModel,dEnterria:2020dwq} combined with a simple model for
particle production, which assumes that the multiplicity at midrapidity has an average value proportional to
\Npart and a probability distribution described by a Negative Binomial Distribution (NBD). Details about this
method can be found in Ref.~\cite{ALICE:pPbMidChdNdEtapaper5020Cent}. However, in \pPb collisions, multiplicity
fluctuations have a large influence on the centrality determination compared to the same effect in heavy-ion
collisions as these fluctuations are sizable compared to the width of the \Npart distribution. Other effects,
originating, e.g., from the fragmentation of partons produced in hard scatterings, were found to bias the
estimation of the centrality based on measurements of particle multiplicity at
midrapidity~\cite{ALICE:pPbMidChdNdEtapaper5020Cent}.

In contrast, the centrality estimation using the energy measured at very large rapidity by the \ZNA neutron
calorimeter in the Pb-going direction introduces minimal bias. The \ZNA detects slow neutrons emitted from
nuclear deexcitation processes or knocked out by wounded nucleons~\cite{SlowNucleons,ALICE:PhysicsPerfReportII}.
The value of \Npart is calculated using two procedures. The first one is that the energy spectrum
obtained from the calorimeter is fitted by a Slow-Nucleon Emission (SNM) model (which is developed based on
the parametrization of results at lower energy p--A experiments~\cite{SlowNucleons,ALICE:pPbMidChdNdEtapaper5020Cent})
coupled to the Glauber MC model. However, the estimation of \Npart with this method is model-dependent.
The second procedure is based on a hybrid method as discussed in detail in Ref.~\cite{ALICE:pPbMidChdNdEtapaper5020Cent}.
In this method, the values of \Npart for each centrality class are obtained under the following assumptions:

\begin{Description}
\item $N_\mathrm{part}^{\mathrm{high}\text-\pt}$: The yield of high-\pt charged particles
  at midrapidity is proportional to the number of inelastic nucleon--nucleon collisions (\Ncoll); 
\item $N_\mathrm{part}^{\mathrm{Pb\text-side}}$: The charged-particle multiplicity in the
  Pb-going direction is proportional to the number of wounded target nucleons.
\end{Description}
The detailed calculations are discussed in Ref.~\cite{ALICE:pPbMidChdNdEtapaper5020Cent}. The centrality classes used
in this analysis are 0--5\% (most central (higher multiplicity) collisions), 5--10\%, 10--20\%, 20--40\%, 40--60\%,
60--80\% and 80--100\% (most peripheral (lower multiplicity) collisions) for both CL1 and \ZNA estimators.

\section{Event generators and simulation framework}
\label{Simdetails}
In the measurements of inclusive photon multiplicity, the corrections for instrumental 
effects and the evaluation of the systematic uncertainties are performed with the help of
MC event generators. For \pPb collisions, \hijing~(v1.36)~\cite{hijing} 
and \dpmjet~(v3.0-5)~\cite{dpmjet} are used whereas PYTHIA~8 (v8.243) with the Monash 2013 tune~\cite{pythia8_monash}
and \eposlhc~\cite{EPOS_LHC} are used for \pp collisions. The generated particles from these event
generators are transported through the experimental setup using the GEANT~3 software package~\cite{geant3},
which includes a detailed description of the apparatus geometry and of the detector response via the
AliRoot software framework~\cite{aliroot}.  The response of the PMD (in terms of the number of fired
cells and energy deposition) to the incident particles in simulation is found to reproduce fairly well
the results from the test beam data as discussed in Ref.~\cite{ALICE:PMDpaper}.

For \pp collisions 1\,M (0.5\,M) minimum bias INEL events were generated using the PYTHIA~8 (EPOS LHC)
event generator whereas for \pPb and \Pbp collisions 13.7\,M (16\,M) and 3.4\,M (1\,M) NSD events were
produced using \hijing (\dpmjet).

The PYTHIA 8~\cite{pythia8pt3} event generator is a standard tool for the generation of high energy
physics collisions dominantly based on 2~$\rightarrow$~2 hard scattering processes.
In \pp interactions, PYTHIA 8 introduces an impact-parameter dependent MultiParton Interaction (MPI)
activity~\cite{MPI_pythia} to model the soft underlying event (UE). PYTHIA 8 uses
the Lund string fragmentation model for the hadronization of partons. An MPI-based model
is implemented in PYTHIA 8 to introduce beam remnants and color reconnection wherein all the gluons of
lower-\pt interactions are merged with the color-flow dipoles of the highest-\pt one, in such a way that
the total string length is minimized~\cite{pythia8pt2,Color_recon}. In the Monash tune of PYTHIA~8,
the parameters relevant for initial-state radiation and MPI are tuned by using the MB,
Drell-Yan, and UE data from the Tevatron, SPS and LHC. The EPOS event generator is a parton-based MC model
with flux tube initial condition for hadron--hadron collisions. It uses Gribov--Regge theory for
describing soft interactions. The \eposlhc~\cite{EPOS_LHC} generator is tuned to LHC data to describe the results
from various collision systems at different center-of-mass energies, particularly the observed collective behavior
in \pp and \pPb collisions at LHC energies. 

The \hijing event generator is a pQCD inspired MC model aimed particularly at the study of jet and minijet
production and jet-medium interactions in heavy-ion collisions. It also allows the study of particle production 
in high energy hadron--hadron and hadron--nucleus collisions. In p--A and heavy-ion collisions, multiple
nucleon--nucleon interactions are simulated using binary approximation and the Glauber 
model. Nuclear shadowing effects are taken into account using parameterized parton 
distribution functions inside the nucleus.
The soft interactions in \hijing are described by Lund FRITIOF~\cite{lund_fritiof} and the Dual Parton Model
(DPM)~\cite{dpm}. For jet fragmentation and hadronization, \hijing uses the prescription described in the
Lund fragmentation model. The \dpmjet event generator is a multi-purpose MC model based on the DPM capable of
simulating hadron--hadron, hadron--nucleus, nucleus--nucleus, photon--hadron,
photon--photon and photon--nucleus interactions from a few~\GeV up to the highest
cosmic-ray energies. It uses the Glauber--Gribov multiple scattering formalisms
to calculate nuclear cross sections and utilizes the Reggeon theory and pQCD to describe soft interactions 
and hard interactions respectively.

\section{Photon reconstruction}
\label{PhRecons}
A procedure similar to that described in Ref.~\cite{ALICE:PMDpaper} is adopted to reconstruct photons incident
on the preshower plane of the PMD. An incident photon is expected to deposit a larger amount of energy
(measured in ADC units) and produce a signal in a larger number of cells compared to an incident charged hadron
on the PRE plane of the PMD. Using a nearest-neighbor clustering algorithm, contiguous cells having
non-zero energy deposition are grouped together forming clusters. Each cluster carries information, such as the
number of fired cells contained in the cluster, the position of the centroid of the cluster, and the total energy
deposition in the cluster. 

Suitable thresholds on the number of fired cells and the energy deposited in clusters are applied to discriminate 
between photon and hadron clusters. The number of clusters that satisfy the discrimination threshold conditions is 
termed as $N_{\gamma\text-\rm like}$ clusters. A similar condition in discrimination threshold as used in
Ref.~\cite{ALICE:PMDpaper}, namely number of cells greater than 2 and energy deposition greater than 9 times the
MPV (Most Probable Value of the charged pion ADC distribution from a Landau distribution function fitted to the data), is
applied to obtain the distributions of $N_{\gamma\text-\rm like}$ clusters. The $N_{\gamma\text-\rm like}$ clusters provide
a photon-rich sample, however, there are contaminations due to residual background from hadron clusters and from
secondary particles (i.e.~particles produced in interactions with the upstream material in front of the PMD) that reduce
the purity of the sample. The application of the rejection criteria also removes some of the actual photons from
the sample. The distributions of $N_{\gamma\text-\rm like}$ clusters are therefore required to be corrected for these
effects as discussed in the next section.

\section{Correction procedure}
\label{correction}
The measured distributions of $N_{\gamma\text-\rm like}$ clusters are distorted by several instrumental effects such as 
photon detection inefficiency, limited acceptance, contaminations from hadrons and secondary particles etc. 
A set of correction procedures is adopted to obtain the corrected $N_{\gamma}$ distributions.

These instrumental effects are corrected for via a response matrix $R_{mt}$ extracted from MC simulations using
PYTHIA~8 with the Monash 2013 tune to generate \pp collisions and the HIJING event generator for \pPb collisions.
These simulations include a detailed description of the experimental conditions and the detector settings and the
simulated events are reconstructed and analyzed with the same algorithms used for the experimental data. Inelastic
\pp collisions can be classified into two processes: diffractive and non-diffractive (ND). The ND interactions are
the dominant processes in \pp collisions. According to Regge theory~\cite{regge_theory},
diffractive scattering occurs via the exchange of Pomerons. In single diffractive (SD) processes, the exchanged
Pomeron interacts and one of the protons breaks up, producing particles of the diffractive system of mass $M_X$.
In double diffractive (DD) processes, both protons break up. It was found that the $M_X$ distribution for the SD
processes in experimental data~\cite{CDF_SD_1,CDF_SD_2} differs from the one predicted by PYTHIA~8
simulations~\cite{ALICE:SD_DD_Xsections}. To account for this effect in simulation, diffraction-tuned event
generators (which can reproduce diffraction cross sections and the shapes of the $M_X$ distribution obtained from data)
were used in previous charged-particle measurements by ALICE~\cite{ALICE:ppMidChMultpaper900to8000,ALICE:ppFrdChMultpaper}.
In this work, the $M_X$ distribution of SD events in PYTHIA~8 with the Monash 2013 tune is reweighted to match the
diffraction-tuned PYTHIA simulation used in previous 7~\TeV analysis~\cite{ALICE:ppMidChMultpaper900to8000}.
The response matrix in \pp collisions is constructed using this reweighted PYTHIA~8 simulation.

\begin{figure}[h!]
  \subfigure[]{
    \label{response_matrix_pp}
    \includegraphics[scale=0.39]{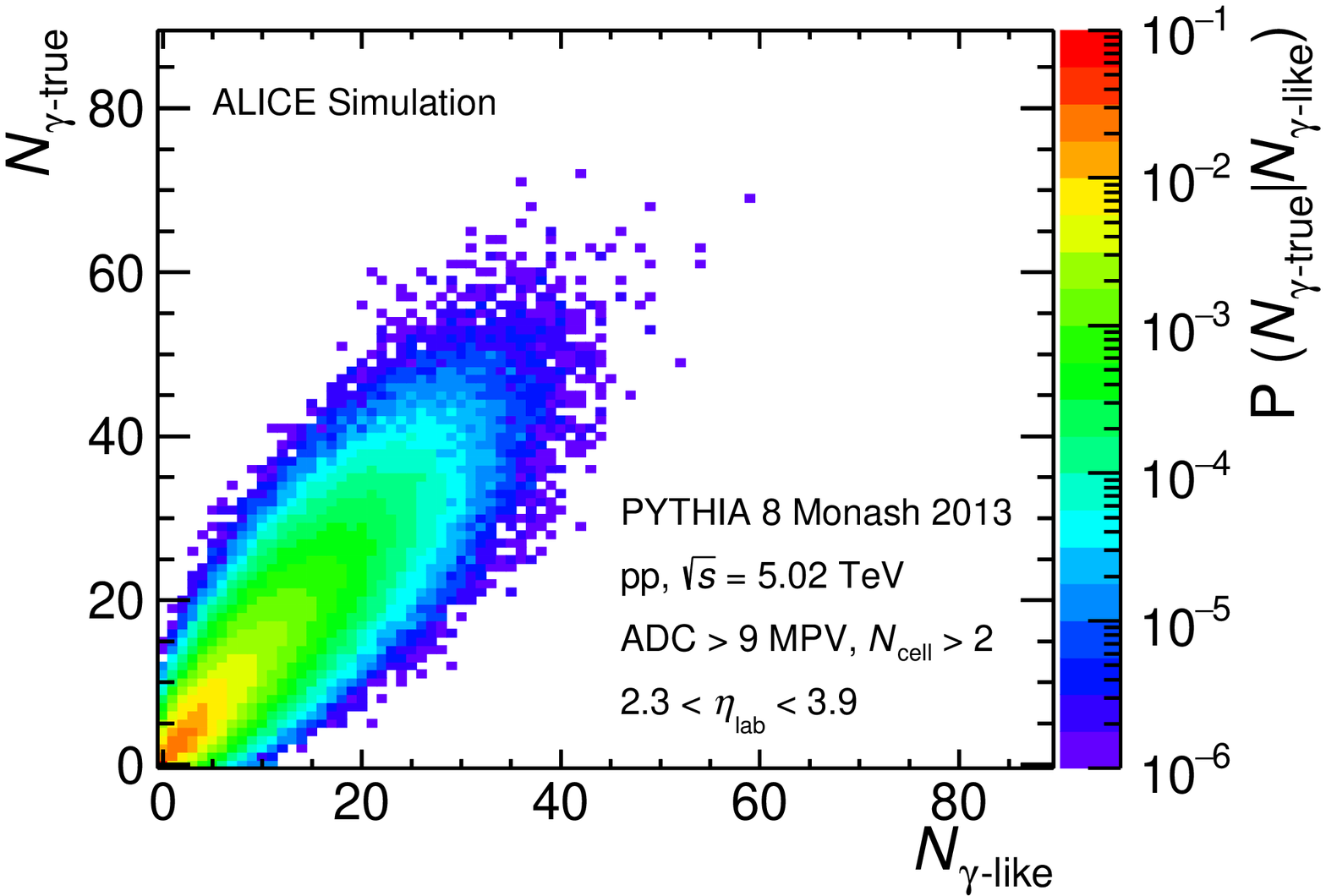}}
  \subfigure[]{
    \label{response_matrix_pPb}
    \includegraphics[scale=0.39]{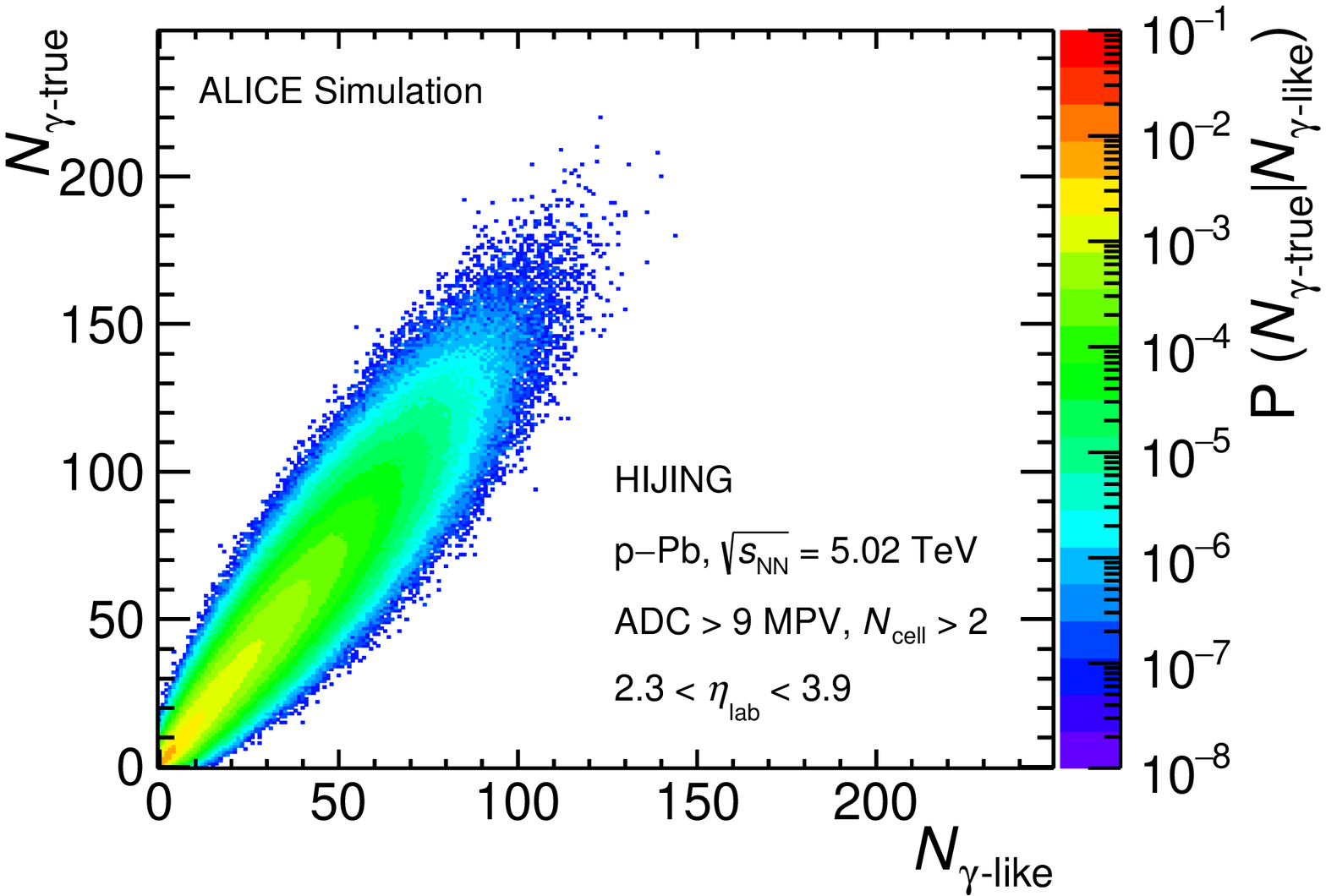}}
  \subfigure[]{
    \label{response_matrix_Pbp}
    \includegraphics[scale=0.39]{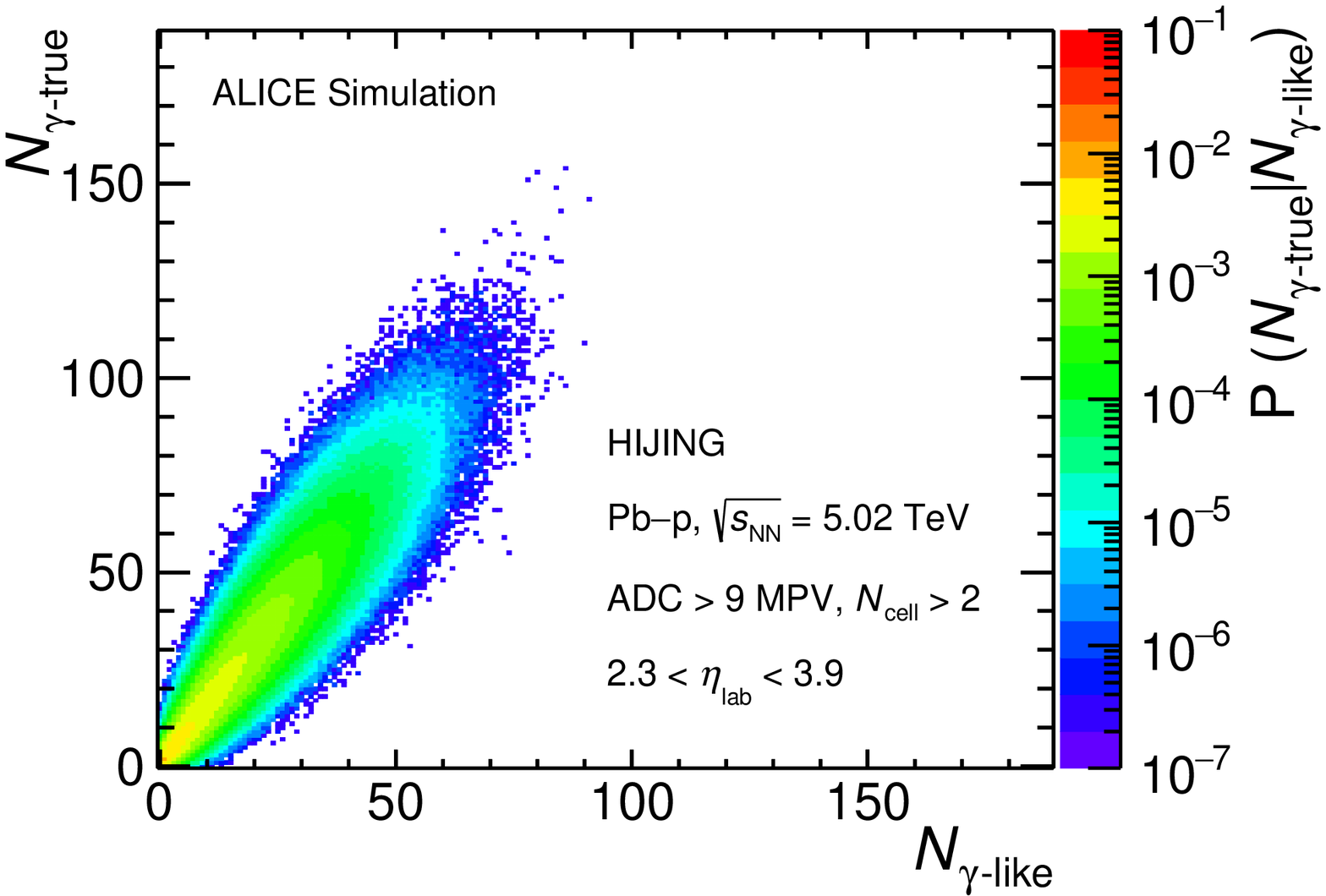}}
  \caption{Graphical representation of the detector response matrices for three different collision systems:
    \pp (a), \pPb (b) and \Pbp (c). The reconstructed and true photon multiplicities are denoted as
    $N_{\gamma\text-\rm like}$ and $N_{\gamma\text-\rm true}$, respectively, obtained from simulations using PYTHIA~8
    with the Monash 2013 tune for \pp collisions and \hijing for \pPb and \Pbp collisions.}
  \label{response_matrix}
\end{figure}

Figure~\ref{response_matrix} shows the graphical representation of the detector response matrices for minimum bias
\pp, \pPb, and \Pbp collisions in terms of the correlation between the true photon multiplicity ($N_{\gamma\text-\rm true}$)
from the event generator and the reconstructed photon multiplicity ($N_{\gamma\text-\rm like}$) in the simulated events.
The mean ($\langle N_{\gamma\text-\rm like}\rangle$) and width ($\sigma_{N_{\gamma\text-\rm like}}$) of the distribution of
$N_{\gamma\text-\rm like}$ are presented in Fig.~\ref{ResMatrix_resolution_pp} and Fig.~\ref{ResMatrix_resolution_pPb}
as a function of $N_{\gamma\text-\rm true}$ for \pp and \pPb collisions at \fivenn.
The resolution of photon multiplicity reconstruction in \pp (\pPb) collisions at $N_{\gamma\text-\rm true}$ = 44 (140)
is found to be $\sim$~12\% ($\sim$~6\%).
Simulation studies have shown that the PMD is sensitive to transverse momenta (\pt) as low as
$\sim$~50~MeV/$c$~\cite{ALICE:PMDpaper}. In the correction procedure, true photons of all \pt are considered,
which makes the present photon measurement inclusive. In simulation, the fraction of inclusive photons having
\pt below 50~MeV/$c$ is estimated to be about 16\% for both \pp and \pPb collisions.
The detection efficiency of inclusive photons as a function
of $\eta$ varies from 27\% to 52\%. The matrix element $R_{mt}$ represents the conditional probability that an
event with true multiplicity $t$ is measured as an event with multiplicity $m$. The distorted
measured distribution ($M$)
can be expressed as
\begin{equation}
  M = R_{mt}T.
  \label{MequalRT}
\end{equation}
One can therefore recover the true distribution ($T$) for given $M$ as
\begin{equation}
  T = R_{mt}^{-1}M.
  \label{TequalRM}
\end{equation}

However, the matrix $R_{mt}$ may be singular and one can calculate inverse matrix $R_{mt}^{-1}$ only
if $R_{mt}$ is not singular.
Furthermore, even if $R_{mt}$ can be inverted, the results obtained with Eq.~\ref{TequalRM} contain oscillations
mainly because of finite statistics in the response matrix. To overcome this problem, a regularized unfolding
method based on Bayes' theorem~\cite{BayesUnfolding} using the RooUnfold software package~\cite{roounfold} is
used to correct the measured $N_{\gamma\text-\rm like}$ distributions following the description outlined in
Refs.~\cite{ALICE:PMDpaper,ALICE:ppMidChMultpaper900and2360,ALICE:ppMidChMultpaper7000,ALICE:ppMidChMultpaper900to8000,ALICE:ppFrdChMultpaper}.

\begin{figure}[h!]
  \subfigure[]{
    \label{PerUnfoldEtaIntMBCut2_pp}
    \includegraphics[scale=0.39]{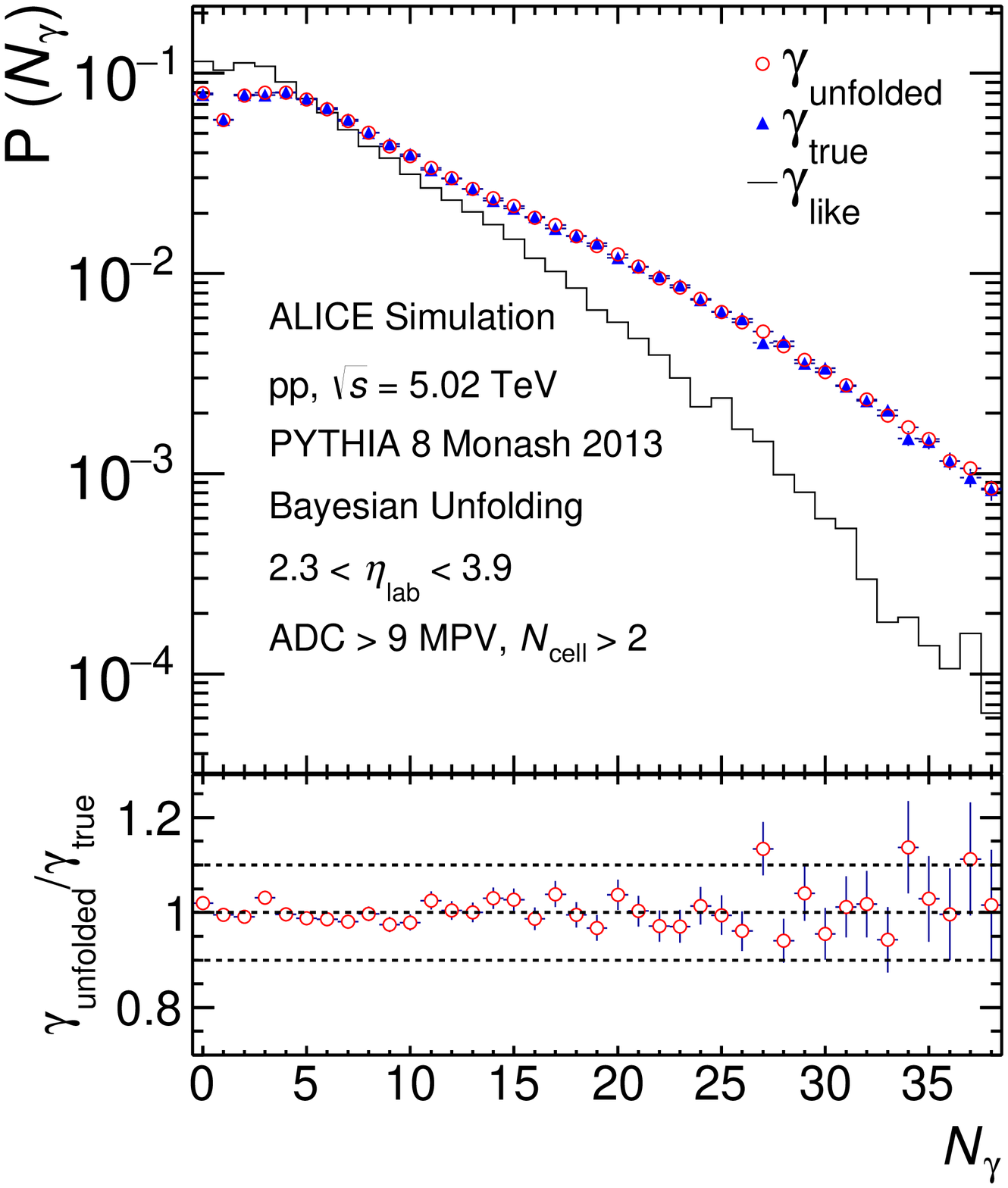}}
  \subfigure[]{
    \label{PerUnfoldEtaIntMBCut2_pPb}
    \includegraphics[scale=0.39]{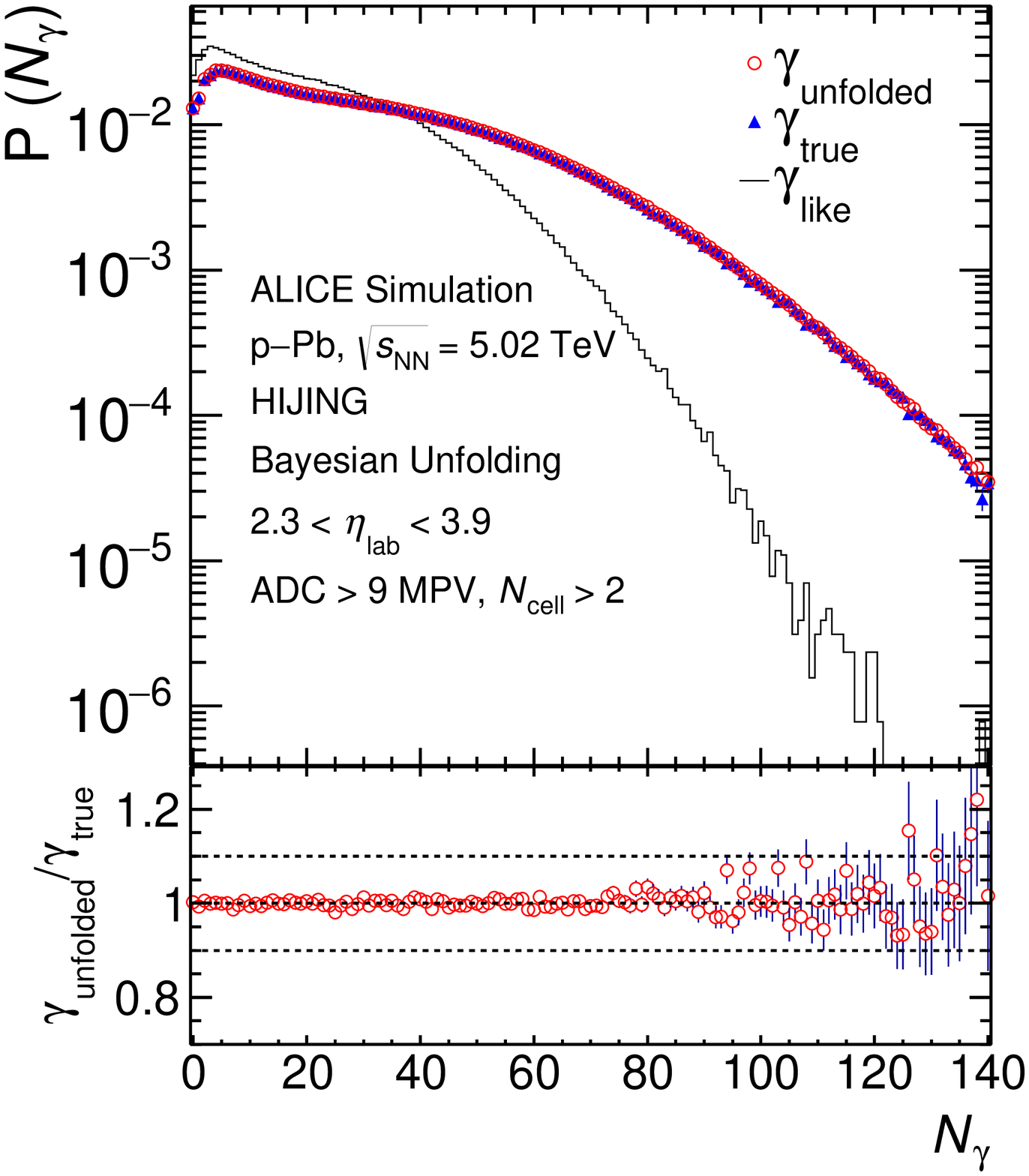}}
  \subfigure[]{
    \label{PerUnfoldEtaIntMBCut2_Pbp}
    \includegraphics[scale=0.39]{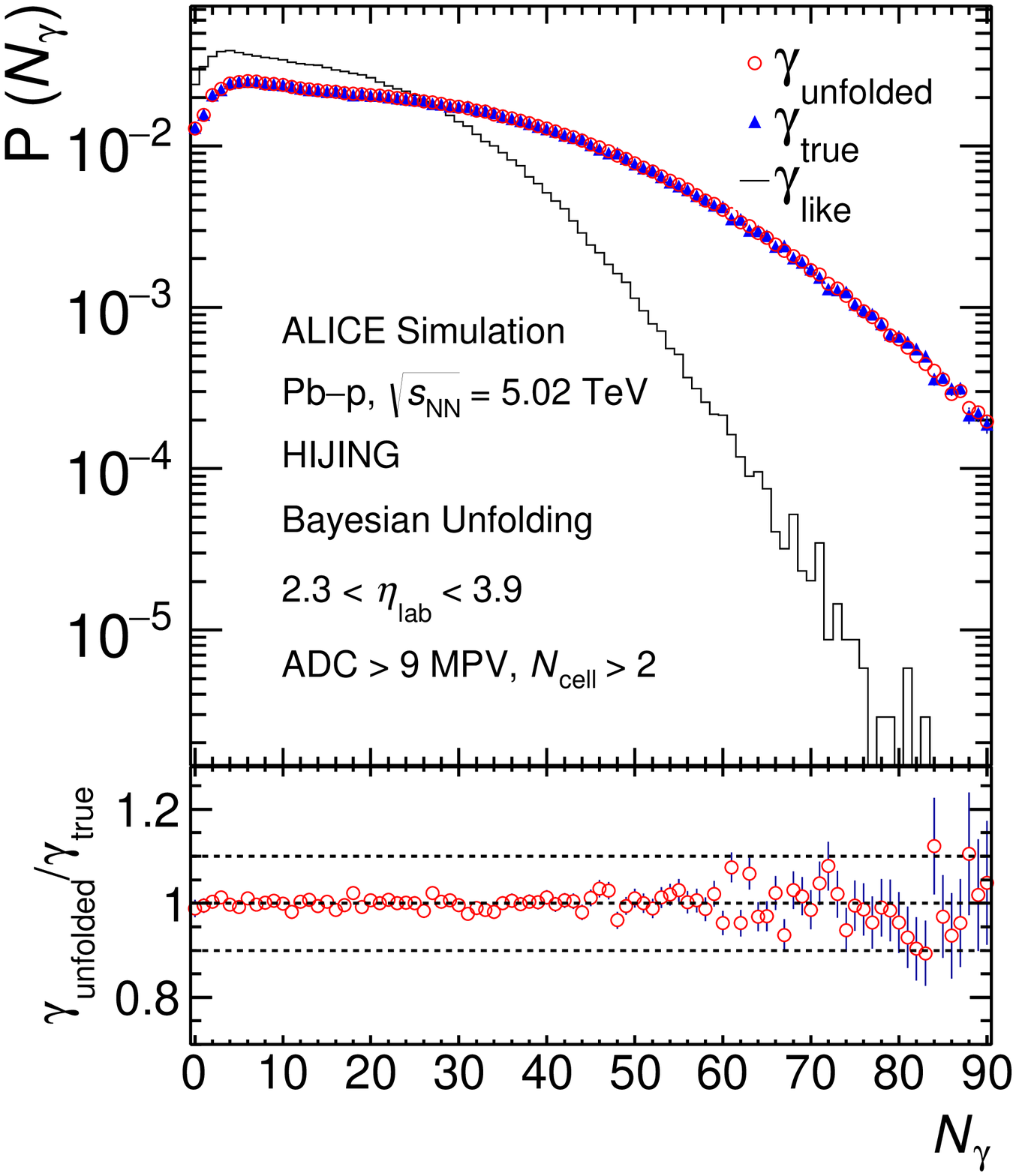}}
  \caption{MC closure test of the unfolding method for \pp (a), \pPb (b), and \Pbp (c) collisions
    at \fivenn. The reconstructed, unfolded and true photon multiplicity distributions are
    presented in the top panels. The ratios of unfolded to true multiplicity distributions
    are shown in the bottom panels.}
  \label{PerUnfoldEtaIntMBCut2}
\end{figure}

In order to optimize the parameters (number of iterations) of the Bayesian unfolding method, a MC closure test
was performed using simulations. The simulated data were divided into two statistically independent samples for
the closure test. The response matrix was built using one set whereas the other set was used to obtain the
reconstructed ($N_{\gamma\text-\rm like}$) distributions, correct them using Eq.~\ref{TequalRM}, and compare the
result to the true distributions. The sensitivity of the results was checked by varying the number of iterations
and a value corresponding to the best performance of the closure test was chosen as the optimized number of
iterations. Figure~\ref{PerUnfoldEtaIntMBCut2} shows the performance of the closure tests for
\pp (Fig.~\ref{PerUnfoldEtaIntMBCut2_pp}), \pPb (Fig.~\ref{PerUnfoldEtaIntMBCut2_pPb}), and \Pbp
(Fig.~\ref{PerUnfoldEtaIntMBCut2_Pbp}) collisions. The upper panels show the true,
reconstructed ($N_{\gamma\text-\rm like}$), and unfolded photon multiplicity distributions. In the bottom panels, the
ratios of the unfolded to the true photon multiplicity distributions are plotted for the optimized number of
iterations (1 for \pp, 4 for \pPb, and 2 for \Pbp collisions). The ratio plots indicate that there is good
agreement between the unfolded and true distributions except at high multiplicities where fluctuations arise due
to the limited size of the simulated sample. The measured photon distributions in experimental data
were unfolded using the optimized number of iterations.
An alternate method of unfolding, Singular Value Decomposition
(SVD)~\cite{SVDUnfolding}, was also used and consistent results were obtained. 
In this method, the response matrix can be factorized
in the form $R_{mt} = USV^{T}$, where $U$ and $V$ are orthogonal matrices and $S$ is a diagonal matrix with non-negative
diagonal elements. These diagonal elements are called singular values of the
matrix $R_{mt}$. Regularization is applied by using a smooth cut-off on small singular-value contributions suppressing
large fluctuations in the unfolded distribution.
The same procedure was applied to extract the pseudorapidity distributions, but the unfolding of photon multiplicity
distributions was performed separately in eight pseudorapidity intervals of a width of 0.2 units.

The unfolded results in MB collisions are then corrected for the trigger ($\epsilon_{\rm Trig}$)
and vertex reconstruction ($\epsilon_{\rm Vtx}$) efficiencies. A similar procedure as discussed
in~\cite{ALICE:PMDpaper} is adopted to estimate $\epsilon_{\rm Trig}$ and $\epsilon_{\rm Vtx}$
using MC simulations. These efficiencies can be defined as
\begin{equation}
  \epsilon_{\rm Trig} = \frac{N_{\rm Trig}}{N_{\rm All}}, \hspace{0.4cm}
  \epsilon_{\rm Vtx} = \frac{N_{\rm TrigVtx}}{N_{\rm Trig}}
\end{equation}
where $N_{\rm All}$ and $N_{\rm Trig}$ correspond to the number of all simulated events and of
triggered events, respectively, and $N_{\rm TrigVtx}$ corresponds to triggered events with
reconstructed vertex. The values of $\epsilon_{\rm Trig}$ and $\epsilon_{\rm Vtx}$ for \pp, \pPb,
and \Pbp collisions are reported in Table~\ref{TriggVtxEff_tab}.
The unfolded pseudorapidity distributions are corrected for these efficiencies using an overall
correction factor whereas multiplicity dependent correction factors are used to correct the
multiplicity distributions.

The \VZERO-AND trigger used to select \pPb events is not fully efficient for NSD events. The
correction for the \VZERO-AND trigger inefficiency and the imperfect description in the MC of
the vertex reconstruction efficiency would mainly concern the most peripheral event class
(80--100\%). In this work, centrality classes have been defined as percentiles of the visible
cross section and the centrality dependent pseudorapidity distributions are not corrected for
trigger inefficiencies as was also reported in Ref.~\cite{ALICE:pPbMidChdNdEtapaper5020Cent}.

\begin{table}[h!]
  \caption{Trigger and vertex reconstruction efficiencies for INEL events in \pp collisions and for NSD events in
    \pPb and \Pbp collisions. Uncertainties listed are total uncertainties. Statistical uncertainties are negligible.}
  \label{TriggVtxEff_tab}
  \begin{center}
    \begin{tabular}{l l l}
      \toprule
      System & Trigger efficiency (\%) & Vertex reconstruction efficiency (\%) \\
      \midrule
      \pp & 94.5 $\pm$~1.2 & 95.0 $\pm$~0.6 \\ [0.15cm]			
      \pPb & 98.5 $\pm$~0.9 & 99.8 $\pm$~0.1 \\[0.15cm]			
      \Pbp & 98.5 $\pm$~0.9 & 99.8 $\pm$~0.1 \\[0.15cm]
      \bottomrule
    \end{tabular}
  \end{center}	
\end{table}

\section{Estimation of Systematic Uncertainties}
\label{SysUncEst}
The contributions of the different sources of systematic uncertainties associated with the
measurements of multiplicity and pseudorapidity distributions of inclusive photons are
summarized in Table~\ref{SysUncTable_MB} for the measurements in MB events of \pp, \pPb,
and \Pbp collisions and in Tables~\ref{SysUncForCentbinsTableCL1}
and~\ref{SysUncForCentbinsTableZNA} for the measurements in centrality classes of \pPb collisions.
The results of \pPb collisions in centrality classes are not affected by the uncertainties on the
event selection and diffraction mass. The total systematic uncertainties are calculated by adding
the contributions from the individual sources in quadrature. The methods used for the estimation
of the systematic uncertainties are discussed in the following subsections. 

\subsection{Effect of upstream material}
One of the major contributions to the systematic uncertainties comes from the uncertainty in the
implementation of upstream material in front of the PMD (i.e., the material between the nominal
IP and the PMD) in the GEANT~3 description of the ALICE apparatus. At forward rapidity, the
material budget was studied in detail for the charged-particle multiplicity
measurements~\cite{ALICE:PbPb_CentFrdrapidity2760,ALICE:PbPbsatellite,ALICE:ppFrdChMultpaper}.
Based on this study, an upper limit of the material budget uncertainty of 10\% was considered
in this analysis. To study the effect of this on photon counting in \pPb and \Pbp collisions, two response
matrices using the \hijing event generator were considered, one with the default material description
and the other with a 10\% increase in the overall material of the ALICE apparatus. The measured
distribution in data is then unfolded using these two different matrices. The difference between
the two unfolded multiplicities determines the systematic uncertainty due to the material in front
of the PMD. A similar procedure is applied for each $\etalab$ interval to obtain the systematic
uncertainties for the pseudorapidity distributions of inclusive photons in MB events and in all
centrality classes. The systematic uncertainty due to material effects for \pp collisions is
taken from the previously published paper on inclusive photon measurement in \pp collisions
at \s = 0.9, 2.76, and 7~\TeV~\cite{ALICE:PMDpaper}.

\begin{table}[h!]
  \footnotesize
  \begin{center}
    \caption{Contributions to systematic uncertainties (in percent) in the measurements of pseudorapidity and
      multiplicity distributions of inclusive photons in \pp, \pPb, and \Pbp collisions. For multiplicity
      distributions, numbers are given at multiplicity values of 0, the mean $\langle m \rangle$ and when
      P($N_{\gamma}$) $=$ 10$^{\text-3}$.}
    \label{SysUncTable_MB}
    \renewcommand{\arraystretch}{1.3}
    \begin{tabular}{|c|c|c|c|c|c|c|c|c|c|c|c|c|}
      \hline
      \multirow{4}{*}{Sources} & \multicolumn{3}{c|}{\dndetaphotonlab analysis} & \multicolumn{9}{c|}{P($N_{\gamma}$) analysis} \\ 
      \cline{2-13}
      & \multirow{3}{*}{\pp} & \multirow{3}{*}{\pPb} & \multirow{3}{*}{\Pbp} & \multicolumn{3}{c|}{\pp} & \multicolumn{3}{c|}{\pPb} & \multicolumn{3}{c|}{\Pbp} \\
      \cline{5-13}
      & & & & $N_{\gamma}$ & $N_{\gamma}$ & P($N_{\gamma}$) & $N_{\gamma}$ & $N_{\gamma}$ & P($N_{\gamma}$) & $N_{\gamma}$ & $N_{\gamma}$ & P($N_{\gamma}$) \\
      & & & & 0 & $\langle m \rangle$ & 10$^{\text-3}$ & 0 & $\langle m \rangle$ & 10$^{\text-3}$ & 0 & $\langle m \rangle$ & 10$^{\text-3}$ \\
      \hline
      Upstream & 7.0 & 8.2--10.1 & 8.7--9.9 & 3.0 & 2.5 & 20.0 & 15.2 & 3.4 & 34.0 & 18.5 & 3.7 & 37.2 \\
      material &  &  &  &  &  &  &  &  &  &  &  &  \\[0.3cm]
      
      Hadron and & 1.9 & 1.9 & 2.0 & 1.4 & 1.4 & 7.5 & 6.7 & 1.2 & 1.2 & 0.7 & 1.1 & 7.2 \\
      secondary & & & & & & & & & & & & \\
      contamination & & & & & & & & & & & & \\[0.3cm]

      Event generator & 2.5 & 3.2 & 1.0 & 2.6 & 3.0 & 5.0 & 3.2 & 3.6 & 16.0 & 3.0 & 3.5 & 14.4 \\
      for response &  &  &  &  &  &  &  &  &  &  &  & \\
      matrix &  &  &  &  &  &  &  &  &  &  &  &  \\[0.3cm]
      
      Unfolding & 3.1 & 0.6 & 1.1 & 25.0 & 3.7 & 23.0 & 11.0 & 1.2 & 1.2 & 11.0 & 3.0 & 3.0 \\
      method &  &  &  &  &  &  &  &  &  &  &  &  \\[0.3cm]

      Event selection & 1.2 & 0.9 & 0.9 & 19.0 & 1.6 & 1.6 & 33.5 & 1.2 & 0.7 & 34.0 & 1.0 & 0.8 \\
      efficiency &  &  &  &  &  &  &  &  &  &  &  &  \\[0.3cm]
      
      Diffraction & 2.1 & n/a & n/a & 1.3 & 4.0 & 5.3 &  n/a & n/a & n/a & n/a & n/a & n/a \\
      Shape &  &  &  &  &  &  &  &  &  &  &  &  \\[0.3cm]
      
      Diffraction & 2.3 & n/a & n/a & 7.4 & 4.6 & 0.2 & n/a & n/a & n/a & n/a & n/a & n/a \\
      ratio &  &  &  &  &  &  &  &  &  &  &  &  \\
      
      \hline
      Total & 8.9 & 9.1--10.8 & 9.1--10.2 & 32.8 & 8.4 & 32.0 & 39.3 & 5.4 & 37.6 & 40.3 & 6.0 & 40.7 \\
      \hline
    \end{tabular}
  \end{center}
\end{table}
\begin{table}[h!]
  \small
  \begin{center}
    \caption{Contributions to systematic uncertainties (in percent) in the measurements of centrality
      dependent pseudorapidity distributions of inclusive photons using the CL1 estimator in \pPb collisions.}
    \label{SysUncForCentbinsTableCL1}
    \renewcommand{\arraystretch}{1.2}
    \begin{tabular}{c|c c c c c c c}
      \hline
      Sources & 0--5\%& 5--10\%& 10--20\% & 20--40\% &
      40--60\% & 60--80\%& 80--100\%\\
      \hline
      Upstream material & 8.2--10.1 & 8.2--10.1 & 8.2--10.1 & 8.2--10.1 & 8.2--10.1 & 8.2--10.1 & 8.2--10.1 \\[0.15cm]
      Hadron and secondary & 2.1 & 2.3 & 2.1 & 2.3 & 2.4 & 2.1 & 3.1 \\
      contamination & & & & & & & \\[0.15cm]
      Event generator & 3.0 & 3.0 & 3.1 & 3.0 & 2.9 & 3.1 & 3.0 \\
      for response matrix & & & & & & & \\[0.15cm]
      Unfolding method & 1.7 & 2.9 & 2.6 & 2.9 & 3.1 & 2.8 & 2.8 \\[0.15cm]
      \hline
      Total & 9.2--10.9 & 9.5--11.2 & 9.4--11.0 & 9.5--11.1 & 9.6--11.2 & 9.5--11.1 & 9.7--11.3 \\
      \hline
    \end{tabular}
  \end{center}
\end{table}
\begin{table}[h!]
  \small
  \begin{center}
    \caption{Contributions to systematic uncertainties (in percent) in the measurements of centrality
      dependent pseudorapidity distributions of inclusive photons using the \ZNA estimator in \pPb collisions.}
    \label{SysUncForCentbinsTableZNA}
    \renewcommand{\arraystretch}{1.2}
    \begin{tabular}{c|c c c c c c c}
      \hline
      Sources & 0--5\%& 5--10\%& 10--20\% & 20--40\% &
      40--60\% & 60--80\%& 80--100\%\\
      \hline
      Upstream material & 8.2--10.1 & 8.2--10.1 & 8.2--10.1 & 8.2--10.1 & 8.2--10.1 & 8.2--10.1 & 8.2--10.1 \\[0.15cm]
      Hadron and secondary & 2.1 &  1.3 & 1.6 & 1.6 & 2.1 & 1.7 & 2.6 \\
      contamination & & & & & & & \\[0.15cm]
      Event generator & 1.4 & 1.8 & 2.2 & 3.2 & 3.8 & 3.3 & 2.7 \\
      for response matrix & & & & & & & \\[0.15cm]
      Unfolding method & 2.8 & 1.9 & 2.0 & 1.7 & 2.7 & 2.8 & 3.0 \\[0.15cm]
      \hline
      Total & 9.1--10.8 & 8.7--10.5 & 9.0--10.7 & 9.2--10.9 & 9.7--11.3 & 9.5--11.1 & 9.6--11.2 \\
      \hline
    \end{tabular}
  \end{center}
\end{table}

\subsection{Contamination in the photon sample}
The photon--hadron discrimination conditions in terms of energy deposition and number of cells in
clusters were optimized to minimize contaminations from hadronic clusters and secondary particles.
The purity of the photon sample with the default threshold condition (energy deposition greater
than 9 times the MPV and number of cells greater than 2) is found to be 65\%. To estimate the
systematic uncertainty due to the correction for the contamination, the analysis was repeated with a
different threshold condition on energy deposition (greater than 6 times the MPV). The unfolded
distributions are obtained for both cases and the difference between these two distributions is quoted
as the systematic uncertainty. By applying a higher threshold condition on energy deposition (greater
than 12 times the MPV), the estimated systematic uncertainty with respect to the default threshold
condition is found to be of the same order as the quoted one.

\subsection{Unfolding procedure}
As discussed in Sec.~\ref{correction}, we rely on MC models to unfold the measured photon multiplicity
and pseudorapidity distributions. The sensitivity of the unfolded results to the choice of the MC models 
used in the unfolding procedure is estimated by comparing the results obtained using the response matrices
from two different event generators. For \pp collisions, two separate response matrices are built using
PYTHIA~8 with the Monash 2013 tune and \eposlhc, whereas for \pPb and \Pbp collisions those are constructed using
\hijing and \dpmjet. These matrices are used to unfold the measured distributions in data. The difference
in the unfolded distributions is considered as the systematic uncertainty. The effect of the choice of the
unfolding method was determined by using an alternative unfolding method, SVD, in addition to the default
Bayesian method. The difference in the unfolded results is considered as the systematic uncertainty.  

\subsection{Event selection efficiency}
The systematic uncertainty due to possible imperfections in the correction for the event selection efficiency
is determined by estimating the trigger and vertex reconstruction efficiencies using two different event
generators. For \pp collisions, the efficiency values are computed from events simulated with the PYTHIA~8
with Monash 2013 tune and \eposlhc generators, while \hijing and \dpmjet are used to obtain these efficiencies
in \pPb and \Pbp collisions. This uncertainty mostly influences the zero multiplicity bin in the multiplicity
distributions and it reduces significantly at large multiplicity because single and double diffraction
contributions become smaller when going to higher multiplicity.

\subsection{Diffraction mass distributions}
As described in Sec.~\ref{correction}, the central values of the final results in \pp collisions are determined
using simulations with the PYTHIA~8 event generator with the Monash 2013 tune reweighted to match the measured
$M_X$ distribution in SD events. The uncertainty associated with this procedure is estimated by taking the
difference between the results obtained with the reweighted mass distribution and the default one from PYTHIA~8
with Monash 2013 tune. The uncertainty related to the fractions of SD and DD events produced by PYTHIA~8 is also
evaluated by varying those fractions by $\pm$ 30\% of their nominal values.

\section{Results and discussions}
\label{results}
The multiplicity and pseudorapidity distributions of inclusive photons in the forward pseudorapidity
interval 2.3~$<~\etalab~<$~3.9 measured using the PMD for \pp collisions at \five and \pPb and \Pbp
collisions at $\sqrt{s_{\rm NN}} = 5.02$~TeV are presented and discussed in this section.
The distributions are obtained without any selection on the transverse momentum of the photons.
The obtained results are compared with predictions from MC models and available experimental
measurements of charged particles.

\subsection{Multiplicity distributions}
\label{results-mult}
\begin{figure}[h!]
  \subfigure[]{
    \label{MultDist_pp}
    \includegraphics[scale=0.4]{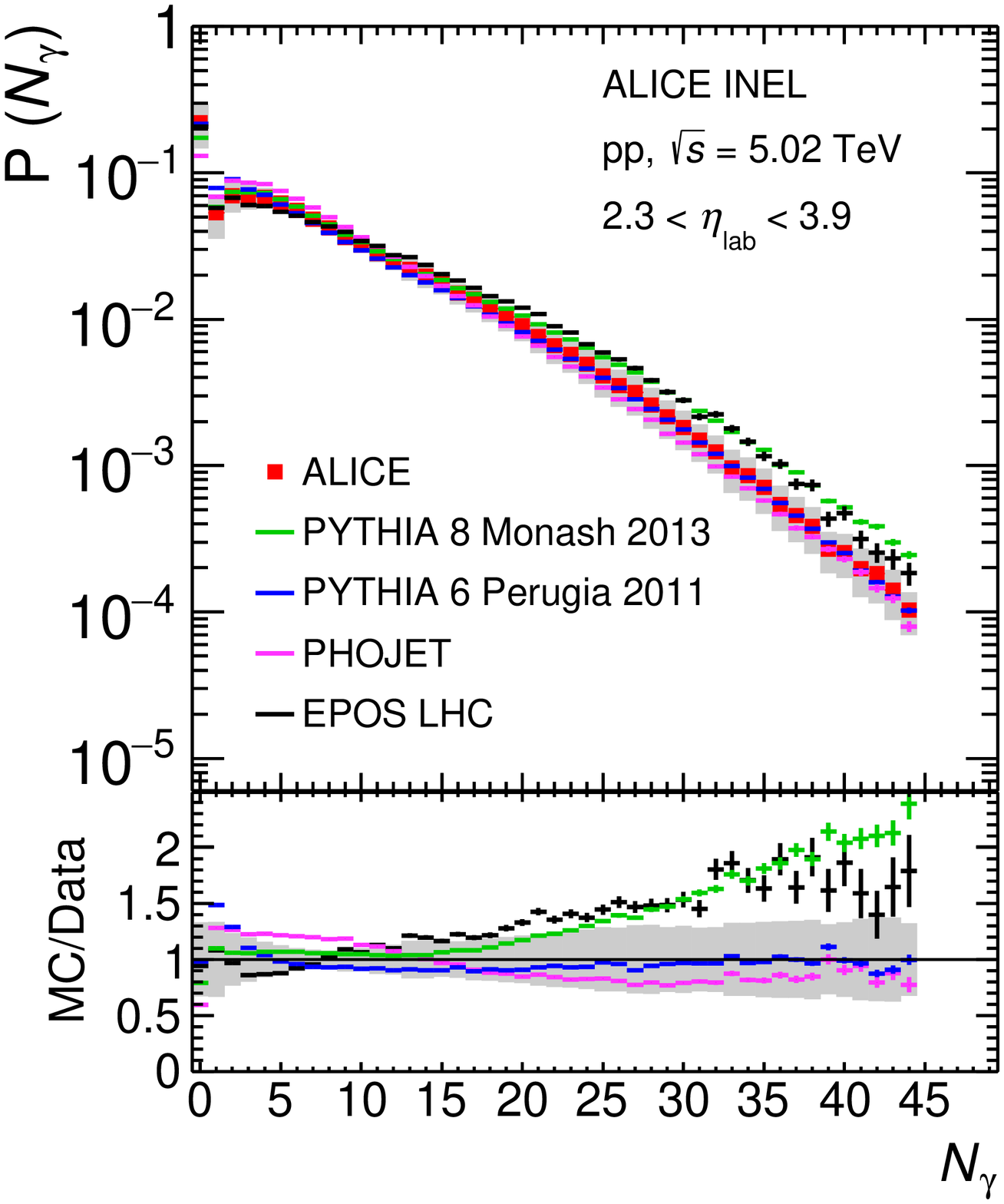}}
  \subfigure[]{
    \label{MultDist_pPb}
    \includegraphics[scale=0.4]{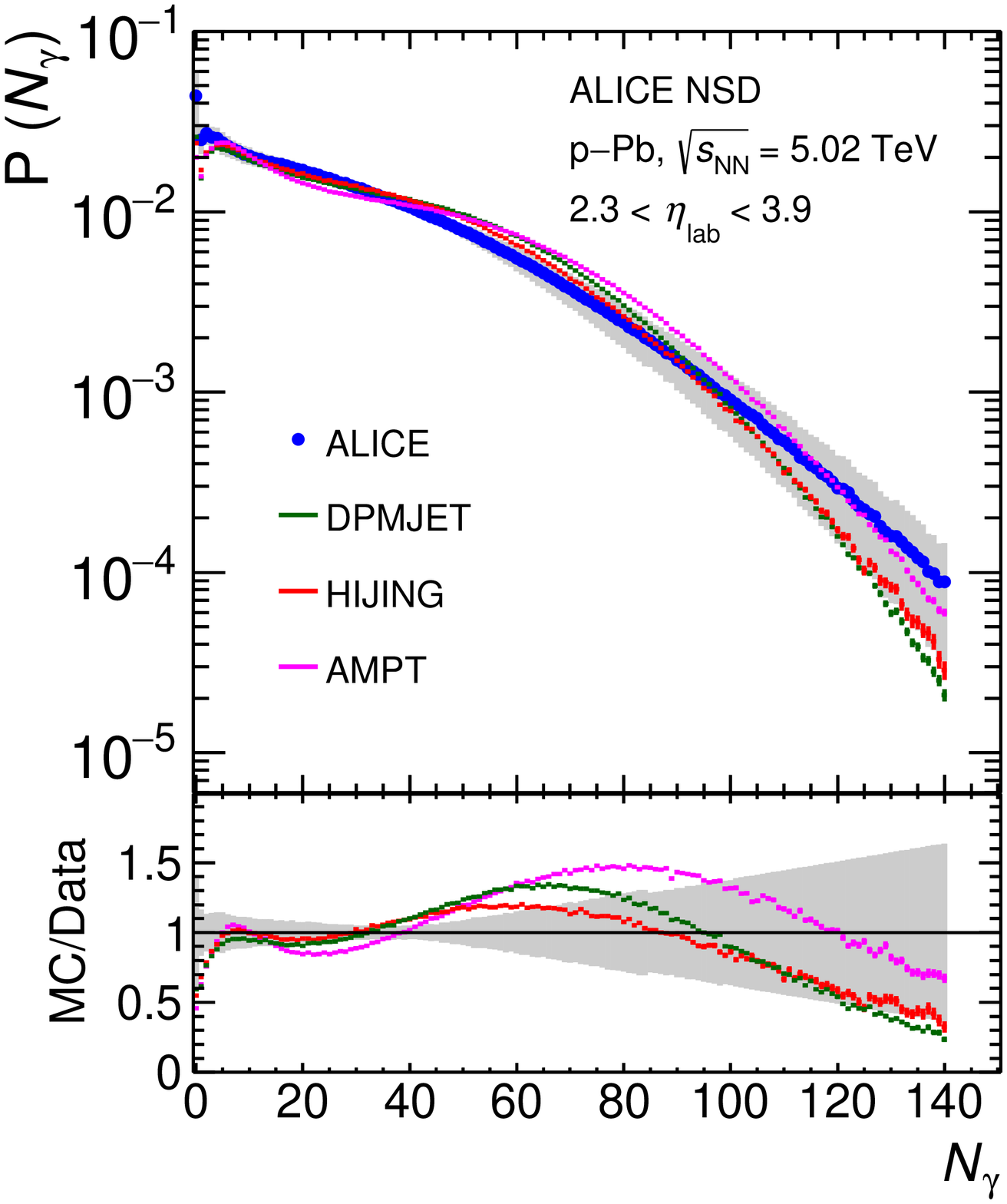}}
  \subfigure[]{
    \label{MultDist_Pbp}
    \includegraphics[scale=0.4]{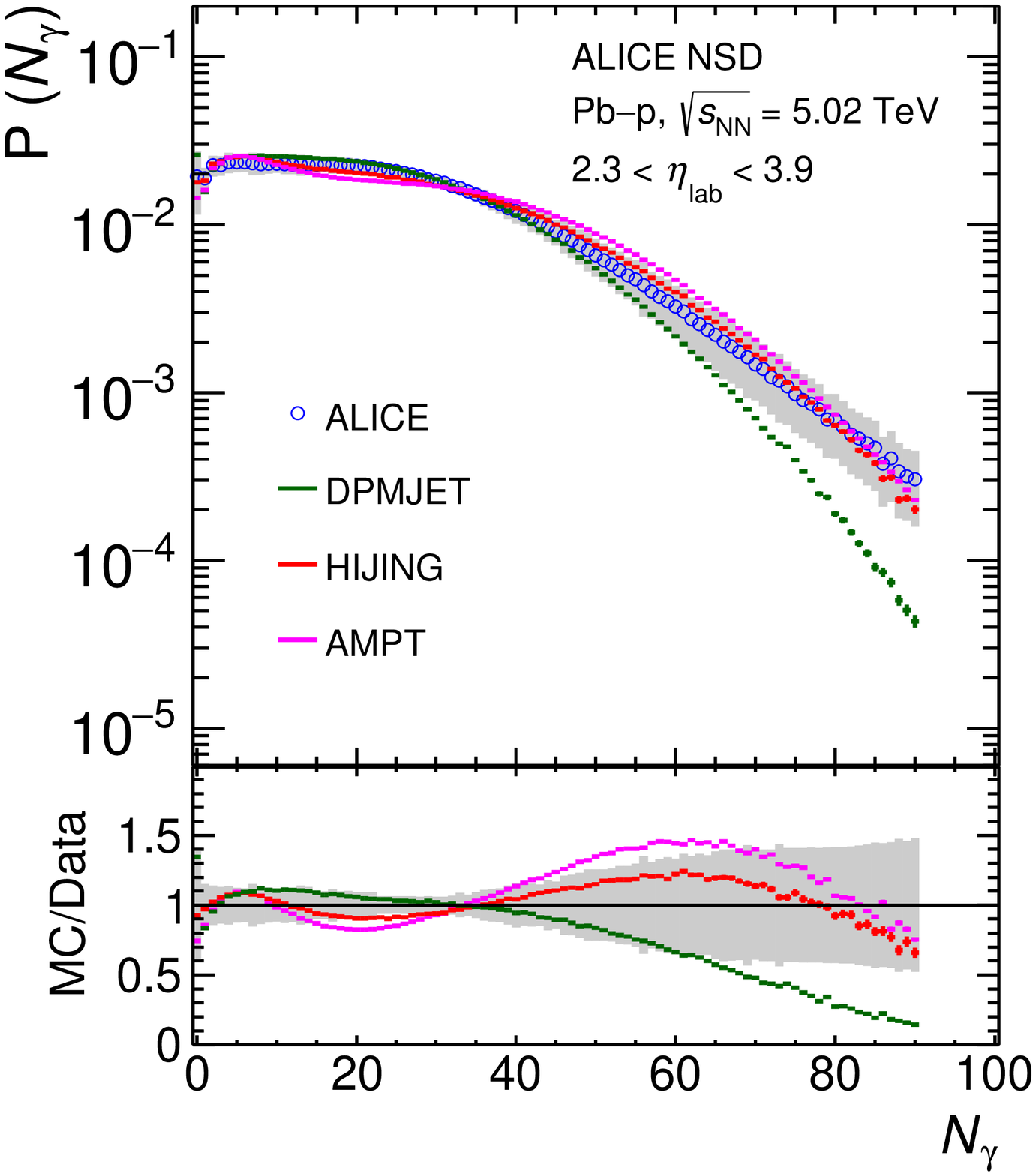}}
  \caption{Top panels: Inclusive photon multiplicity distributions measured within 2.3~$<$~$\etalab$~$<$~3.9 in
    \pp (a), \pPb (b), and \Pbp (c) collisions at \fivenn. Results from various MC predictions are superimposed.
    Bottom panels: The ratios between MC results and data are shown. Shaded boxes represent the systematic
    uncertainties.}
  \label{MultDist}
\end{figure}

The measured multiplicity distributions of inclusive photons are presented as probability distributions
(P($N_{\gamma}$)) as a function of $N_{\gamma}$ in Fig.~\ref{MultDist}. The systematic uncertainty is
represented by the gray bands and the statistical uncertainty is smaller than the marker size.
The measurements are obtained for \pp and for both configurations of \pPb collisions in the pseudorapidity
interval 2.3~$<~\etalab~<$~3.9. The average photon multiplicity for \pp collisions is 6.44 $\pm$ 0.36 (sys),
for \pPb collisions is 29.38 $\pm$ 1.88 (sys), and for \Pbp collisions is 24.59 $\pm$ 1.52 (sys).
The multiplicity range in \pPb collisions reaches up to 140 which is about 5 times larger than its mean
multiplicity. The average multiplicity in \pp collisions is about 4.5 times smaller than that in \pPb
collisions. The different multiplicity distributions and average values in \pPb and \Pbp collisions arise from
the different particle production in the p-going and Pb-going hemispheres as well as from the difference in the
rapidity coverage of the \PMD in the center-of-mass frame because of the $\Delta y$ shift of $\pm$~0.465 units
between \pPb and \Pbp collisions.

The results in \pp collisions (Fig.~\ref{MultDist_pp}) are compared to the predictions from
PYTHIA~8 with the Monash 2013 tune~\cite{pythia8_monash}, PYTHIA~6~(v6.425) with the Perugia 2011
tune~\cite{pythia6_perugia2011}, \phojet~(v1.12)~\cite{phojet}, and EPOS LHC~\cite{EPOS_LHC} event generators.
The ratios between MC predictions and data are shown in the bottom panels. 
It is observed that PYTHIA~8 with the Monash 2013 tune and \eposlhc are unable to reproduce the
inclusive photon production in \pp collisions at \five at high multiplicities. On the other hand,
the Perugia 2011 tune of PYTHIA~6 and \phojet are found to fairly describe the data within uncertainties.

The multiplicity distributions in \pPb collisions (Fig.~\ref{MultDist_pPb}) are compared to predictions
obtained from \hijing, \dpmjet, and \ampt~(v2.25)~\cite{ampt}. None of the models could reproduce the shape of
the distributions in the full multiplicity range. \hijing is closer to the data points at intermediate
multiplicities. The tail of the distribution at high multiplicity can be described by both \hijing and
\ampt models within uncertainties. For \Pbp collisions (Fig.~\ref{MultDist_Pbp}), \dpmjet strongly
underestimates the data at high multiplicity. \hijing describes the data slightly better in the full
range of multiplicity compared to \ampt.

These experimental results are expected to provide new constraints on inclusive photon production mechanisms
implemented in theoretical models.

\begin{figure}[h!]
  \subfigure[]{
    \label{MultDistNBDfit_pp}
    \includegraphics[scale=0.4]{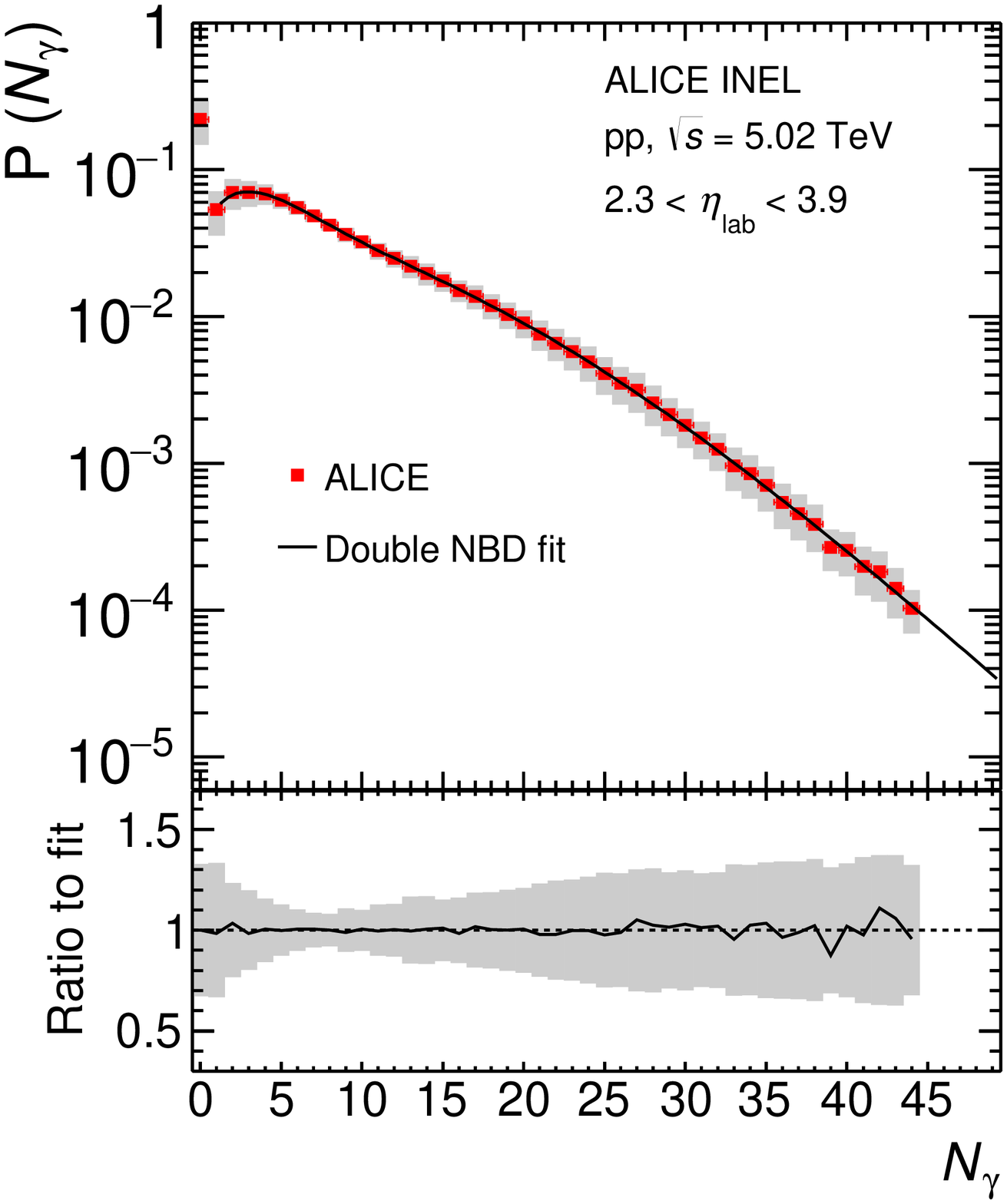}}
  \subfigure[]{
    \label{MultDistNBDfit_pPb}
    \includegraphics[scale=0.4]{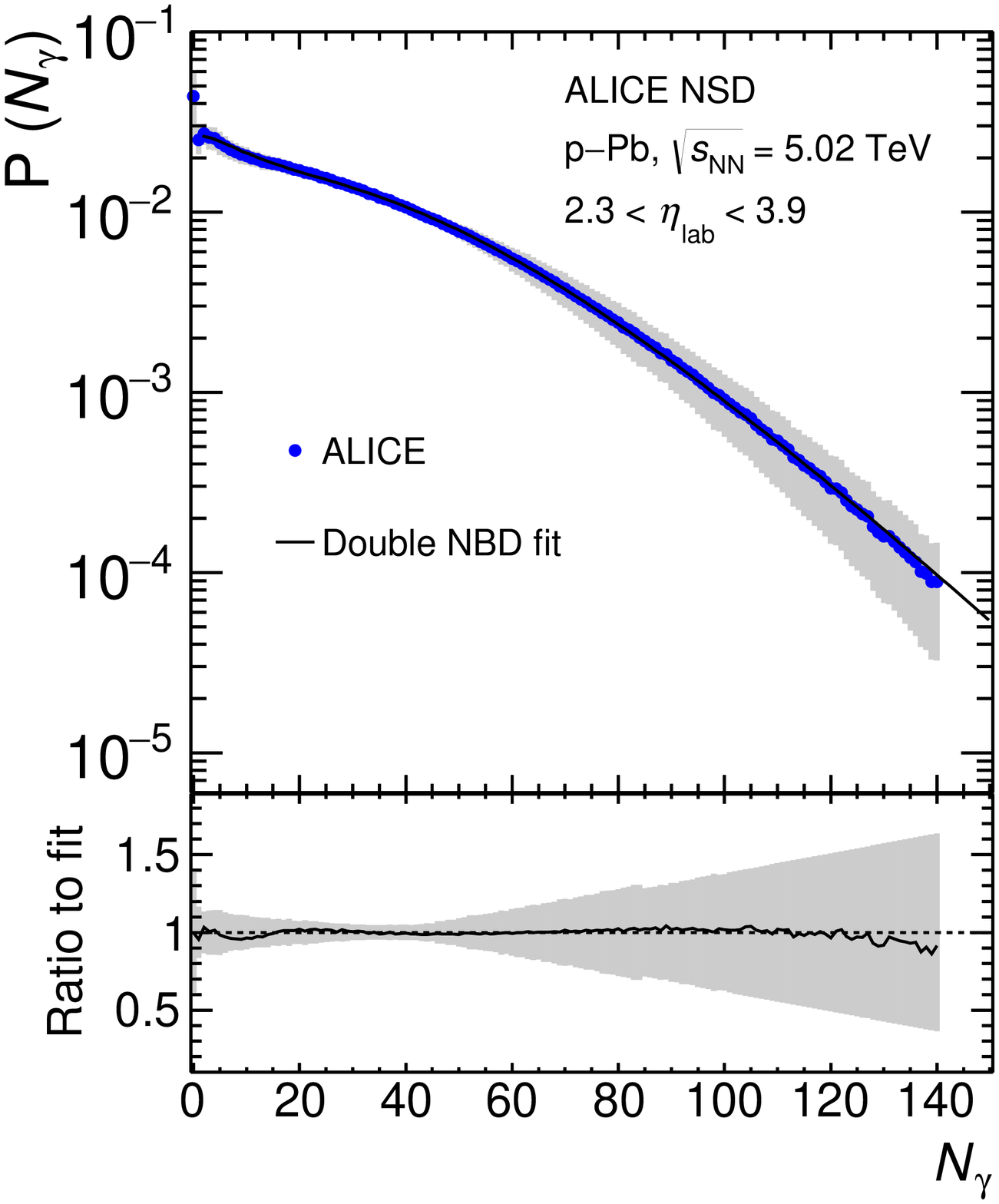}}
  \subfigure[]{
    \label{MultDistNBDfit_Pbp}
    \includegraphics[scale=0.4]{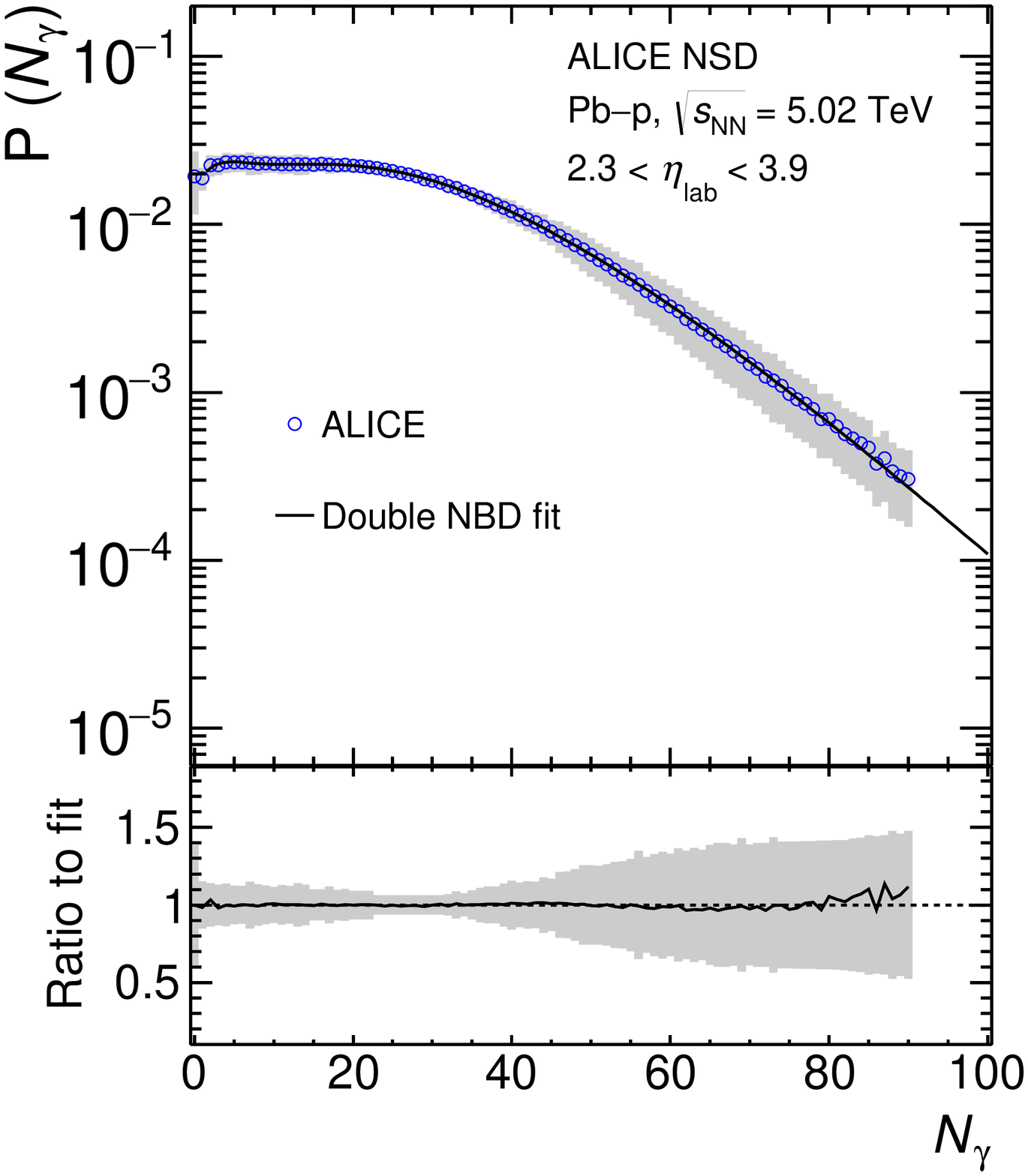}}
  \caption{Top panels: Multiplicity distribution of inclusive photons fitted to double NBDs for
    \pp (a), \pPb (b), and \Pbp (c) collisions at \fivenn. Bottom panels: The ratio of the
    data to the fit is presented. Shaded boxes represent the systematic uncertainties.}
  \label{MultDistNBDfit}
\end{figure}
In Fig.~\ref{MultDistNBDfit}, the measured multiplicity distributions are fitted by a weighted
sum of two Negative Binomial Distributions (NBDs) (Eq.~\ref{DoubleNBDEqn}) to extract the relative
contributions from soft and semihard processes in the particle production mechanisms.
In Eq.~\ref{DoubleNBDEqn}, the subscripts and superscripts in the parameters indicate the respective
components. The $\alpha_{\rm soft}$ parameter gives the fraction of soft events. The NBD distribution,
$\mathrm{P}_{\rm NBD}$, has the following parameters: the average multiplicity $\langle n \rangle$,
which is found to increase with increasing \s, and the shape parameter, $k$, which decreases with
increasing \s in hadronic collisions~\cite{ALICE:ppMidChMultpaper900to8000,ALICE:ppFrdChMultpaper,ALICE:PMDpaper}.
\begin{equation}
  \mathrm{P}(n) = \lambda[\alpha_{\rm soft} \mathrm{P}_{\rm NBD} (n,\langle n^{\rm soft} \rangle,k^{\rm soft}) +
    (1 - \alpha_{\rm soft})\mathrm{P}_{\rm NBD} (n,\langle n^{\rm semihard} \rangle,k^{\rm semihard})]
  \label{DoubleNBDEqn}
\end{equation}
where
\begin{equation}
\mathrm{P}_{\rm NBD}(n,\langle n \rangle,k) = \frac{\Gamma(n+k)}{\Gamma(k)\Gamma(n+1)}\frac{(\langle n \rangle/k)^{n}}{(1 + \langle n \rangle/k)^{n+k}} .
\end{equation}

\begin{table}[h!]
  \small
  \begin{center}
    \caption{Double NBD fit parameters for the inclusive photon multiplicity distributions.
      The obtained parameters from fits to the data by considering only uncorrelated
      uncertainties are printed in bold. The other set of parameters is obtained when
      both correlated and uncorrelated uncertainties are considered in the fitting procedure.}
    \label{DoubleNBDfitPara}
    \renewcommand{\arraystretch}{1.3}
    \begin{tabular}{c c c c c c c}
      \toprule
      Collision& $\lambda$ & $\alpha_{\rm soft}$ & $k^{\rm soft}$ & $\langle n^{\rm soft} \rangle$ & $k^{\rm semihard}$ &
      $\langle n^{\rm semihard} \rangle$ \\
      System& & & & & \\
      \midrule
      \pp& \textbf{0.79 $\pm$ 0.03} & \textbf{0.60 $\pm$ 0.19} & \textbf{2.97 $\pm$ 1.16} & \textbf{4.65 $\pm$ 1.01} & \textbf{4.56 $\pm$ 1.79} & \textbf{13.47 $\pm$ 2.60} \\
      & 0.79 $\pm$ 0.04 & 0.61 $\pm$ 0.28 & 2.97 $\pm$ 1.74 & 4.67 $\pm$ 1.60 & 4.64 $\pm$ 2.76 & 13.55 $\pm$ 4.07 \\[0.15cm]
      \hline
      \Pbp& \textbf{0.99 $\pm$ 0.01} & \textbf{0.29 $\pm$ 0.10} & \textbf{1.65 $\pm$ 0.30} & \textbf{8.92 $\pm$ 2.10} & \textbf{4.09 $\pm$ 0.39} & \textbf{31.08 $\pm$ 1.31} \\
      & 0.99 $\pm$ 0.01 & 0.31 $\pm$ 0.13 & 1.60 $\pm$ 0.33 & 9.70 $\pm$ 4.00 & 4.22 $\pm$ 0.65 & 31.56 $\pm$ 2.14 \\[0.15cm]
      \hline
      \pPb& \textbf{0.99 $\pm$ 0.01} & \textbf{0.62 $\pm$ 0.10} & \textbf{1.10 $\pm$ 0.05} & \textbf{18.76 $\pm$ 2.50} & \textbf{3.89 $\pm$ 0.30} & \textbf{47.21 $\pm$ 1.28}\\
      & 0.98 $\pm$ 0.01 & 0.60 $\pm$ 0.11 & 1.10 $\pm$ 0.10 & 18.80 $\pm$ 3.60 & 3.64 $\pm$ 0.40 & 46.18 $\pm$ 1.72 \\[0.15cm]
      \bottomrule
    \end{tabular}
  \end{center}
\end{table}

Double NBDs do not describe the value P(0), therefore, the bin $N_{\gamma}$ $=$ 0 was excluded from
the fit and a normalization factor $\lambda$ was introduced to account for this. The systematic
uncertainties associated with a) the change of photon--hadron discrimination thresholds, b) the
tuning of event generator for diffraction, c) the effect of upstream material, and d) the correction
of event selection efficiency produce a correlated shift of the multiplicity distributions. These
correlated uncertainties are therefore not considered in the fitting procedure. The other sources
of systematic uncertainties (uncorrelated) are taken into account in the fit. The obtained
parameters from these fits are printed in bold in Table~\ref{DoubleNBDfitPara}. The other set of
parameters reported in Table~\ref{DoubleNBDfitPara} is obtained from the fitting of the data by
considering both correlated and uncorrelated uncertainties to provide an estimate of how much the fit
parameters change due to the presence of correlations in the systematic uncertainties.

The double NBD fit describes the data within uncertainties. The results show an increase in the
mean multiplicity of inclusive photons with system size whereas the shape parameter decreases with system size.
Moreover, it is observed that $\langle n^{\rm semihard} \rangle$ $\approx$ 3$\langle n^{\rm soft} \rangle$
which is consistent with the results of charged-particle production at central
and forward rapidities in \pp collisions~\cite{ALICE:ppMidChMultpaper900to8000,ALICE:ppFrdChMultpaper}.

\begin{figure}[h!]
  \subfigure[]{
    \label{dndetaDistMB_pp}
    \includegraphics[scale=0.4]{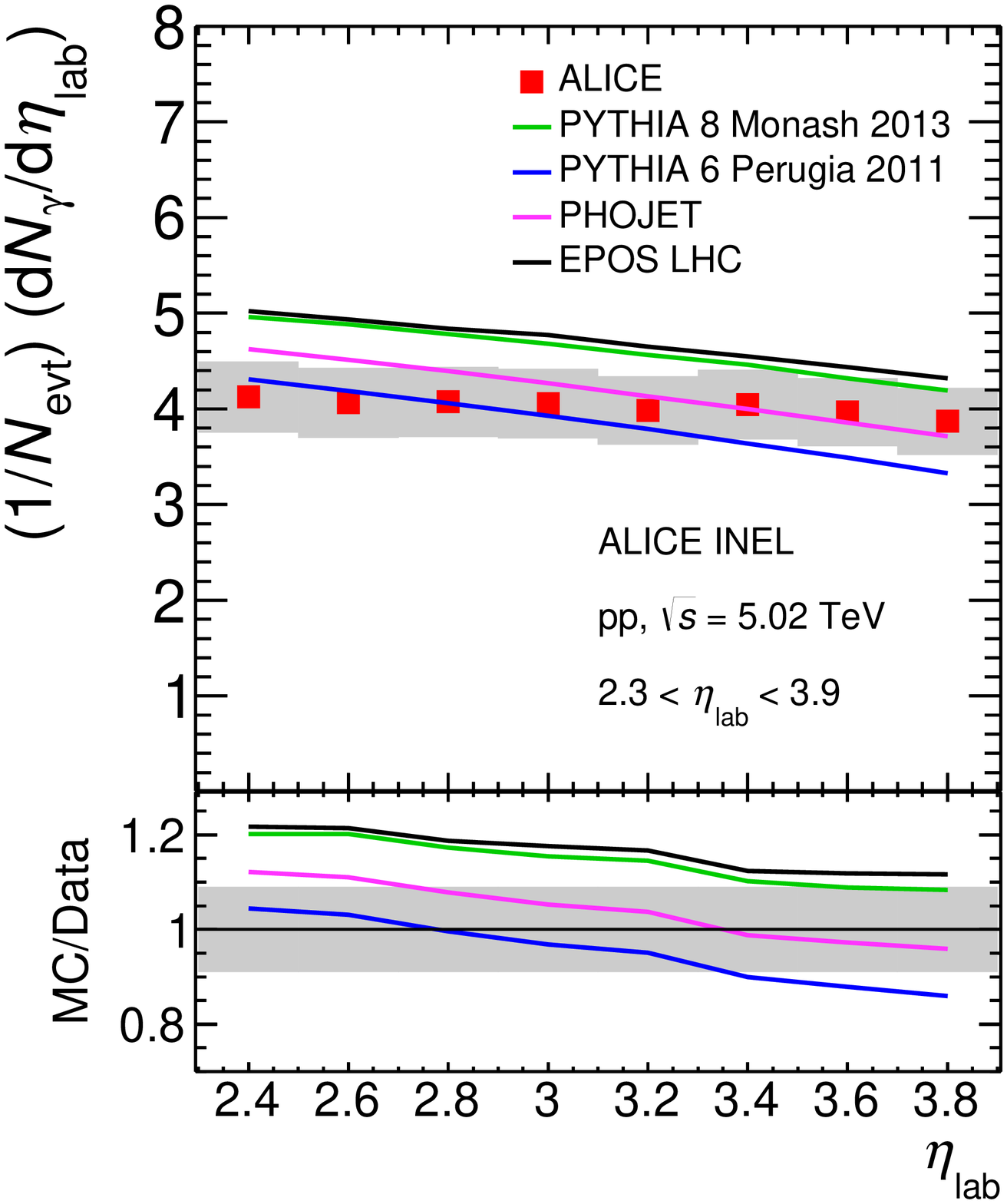}}
  \subfigure[]{
    \label{dndetaDistMB_pPb}
    \includegraphics[scale=0.4]{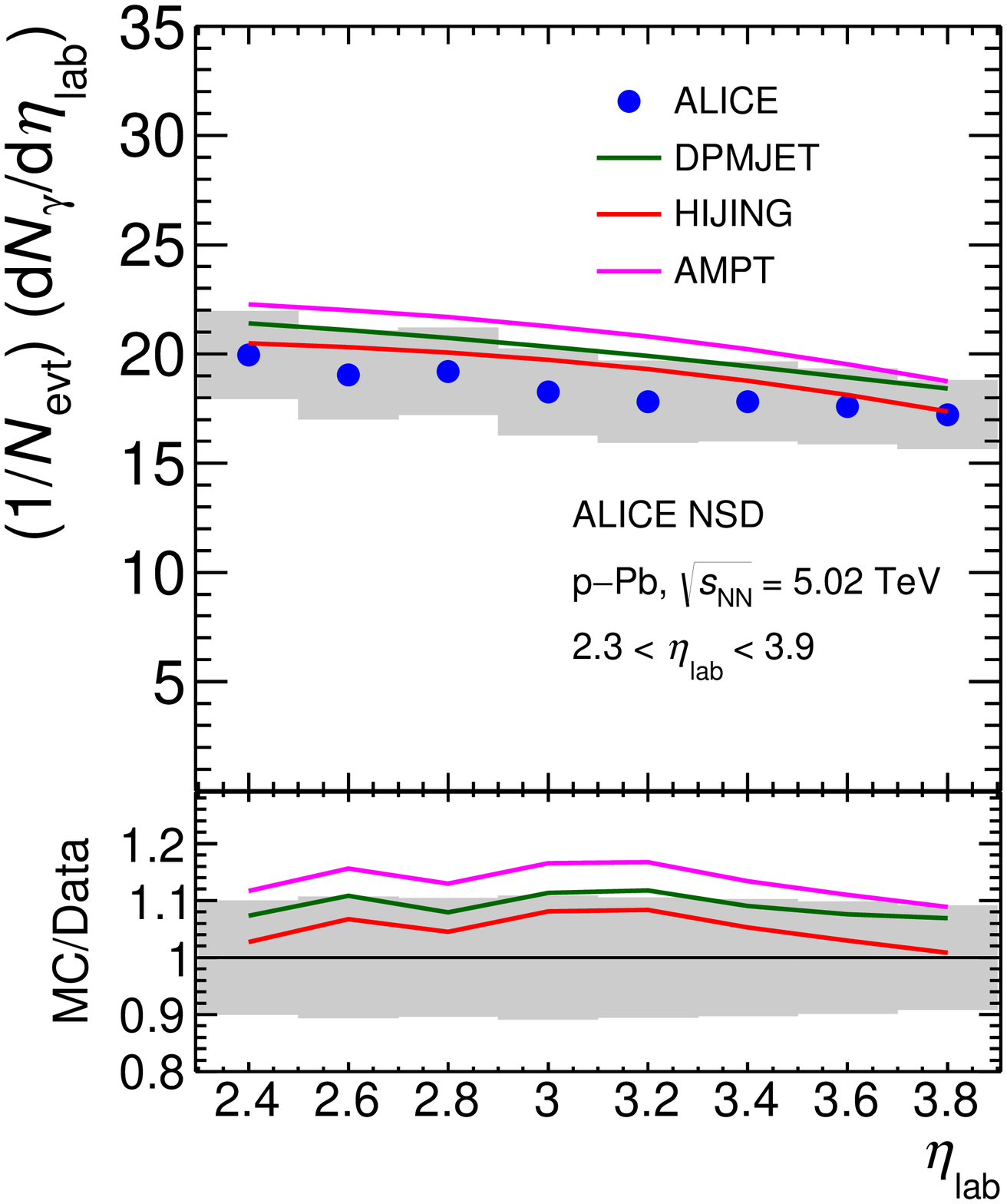}}
  \subfigure[]{
    \label{dndetaDistMB_Pbp}
    \includegraphics[scale=0.4]{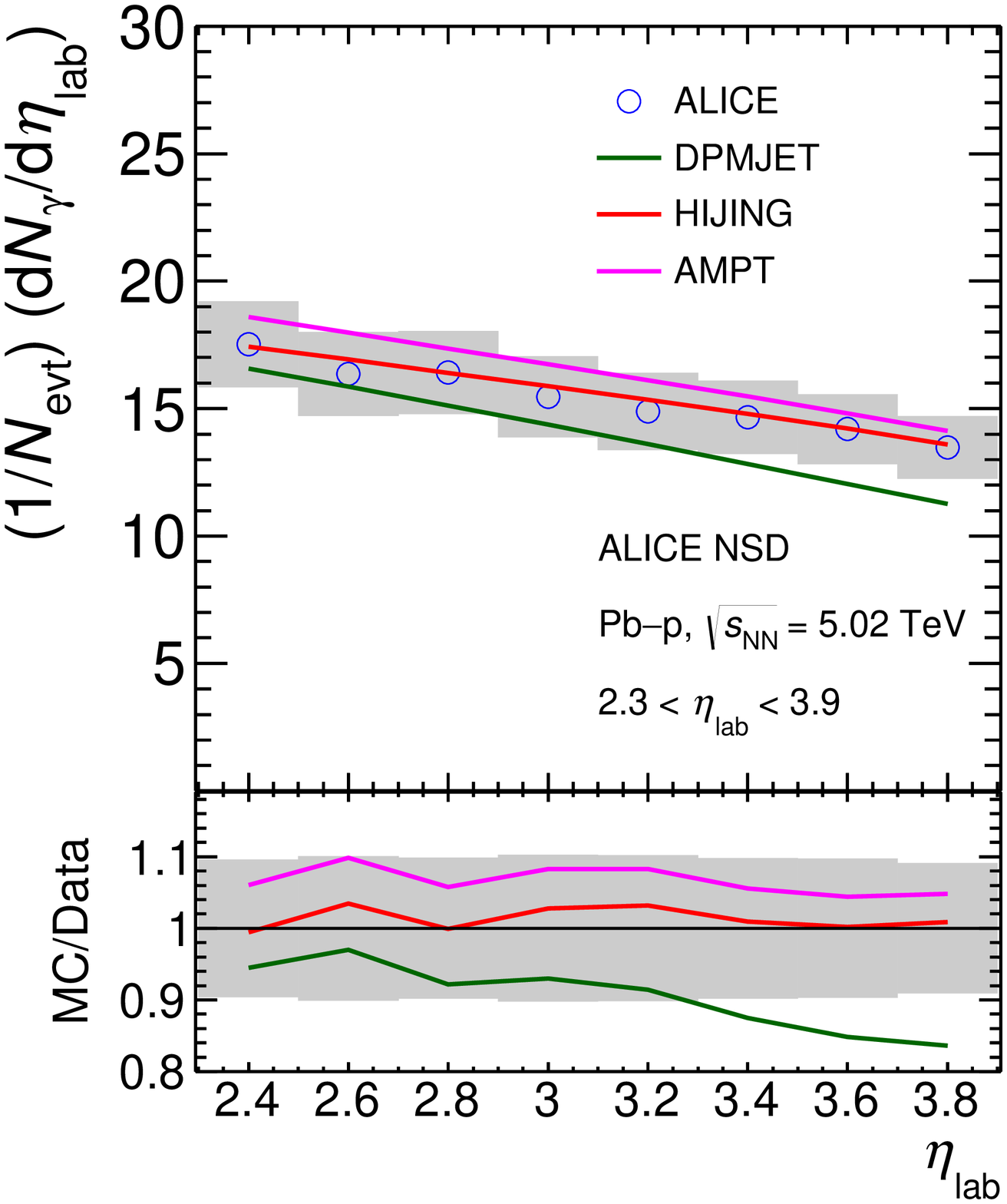}}
  \caption{Top panels: Pseudorapidity distribution of inclusive photons measured within 2.3~$<$~$\etalab$~$<$~3.9
    in \pp (a), \pPb (b), and \Pbp (c) collisions at \fivenn. Results from various MC predictions are
    superimposed. Bottom panels: The ratios between MC results and data are shown. Shaded boxes represent
    the systematic uncertainties.}
  \label{dndetaDistMB}
\end{figure}
\subsection{Pseudorapidity distributions in NSD and INEL events}
\label{results-dndeta-MB}
The pseudorapidity distributions of inclusive photons measured within $2.3~<~\etalab~<~3.9$
in \pp, \pPb, and \Pbp collisions are presented in Fig.~\ref{dndetaDistMB}. The gray bands
represent the systematic uncertainty and the statistical uncertainty is smaller than the
marker size. The measurements are compared to the predictions from MC event
generators and the ratios between MC results and data are shown in the bottom panels.

In \pp collisions (Fig.~\ref{dndetaDistMB_pp}), \phojet and PYTHIA~6 with the Perugia
2011 tune models show better agreement with the data compared to \eposlhc and PYTHIA~8
with the Monash 2013 tune. All model predictions are however found to be compatible within about
20\% from the measured values, but the $\etalab$ dependence is flatter in the data than
in the models.

In \pPb collisions the value of pseudorapidity density of inclusive photons is found to be slightly larger than
that for \Pbp collisions as expected due to the fact that the PMD measures particles produced in the Pb-going
direction in the former configuration and in the p-going direction in the latter, and also due to the different
rapidity ranges covered by the \PMD in the center-of-mass frame for \pPb and \Pbp collisions as discussed
in Sec.~\ref{results-mult}. The data are compared to several models with different descriptions of particle
production, all shifted by $\Delta y$~$=$~0.465 to take into account the shift to the laboratory system.
It is observed that \hijing describes both \pPb and \Pbp results within systematic uncertainties. \ampt
overpredicts the data in the Pb-going side and reproduces the measurements within uncertainties in the
p-going side. \dpmjet underestimates the multiplicity in the p-going side. One can observe that the \dpmjet
curve lies in between \ampt and \hijing in the Pb-going side and goes below in the p-going side. This
suggests that \dpmjet predicts a slightly narrower distribution. It is also noted that all models lie
within about 15\% from the data points.

The same models that do not reproduce the high-multiplicity part of the P($N_{\gamma}$) distribution in MB collisions,
provide a fair description of the multiplicity of inclusive photons, \dndetaphotonlab, at forward rapidities in
\pp, \pPb, and \Pbp collisions.

The simultaneous comparison of the measured P($N_{\gamma}$) and \dndetaphotonlab distributions to the predictions
of different event generators has therefore the potential of further constraining the models of inclusive photon
production in high energy hadronic collisions.

\begin{figure}[h!]
  \subfigure[]{
    \label{dndeta_pp_allcms_compare}
    \includegraphics[scale=0.41]{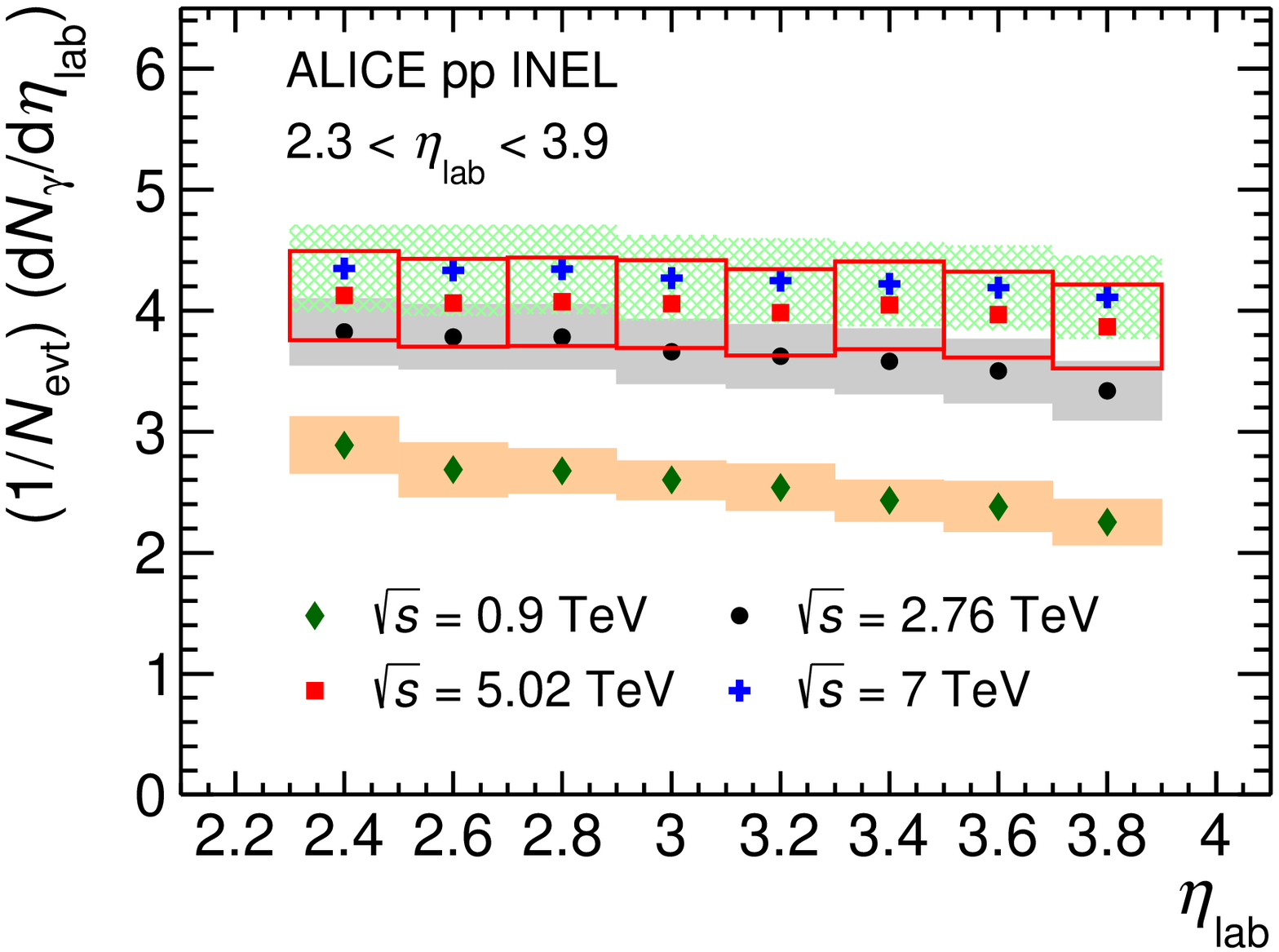}}
  \subfigure[]{
    \label{dndeta_pp_allcms_roots}
    \includegraphics[scale=0.41]{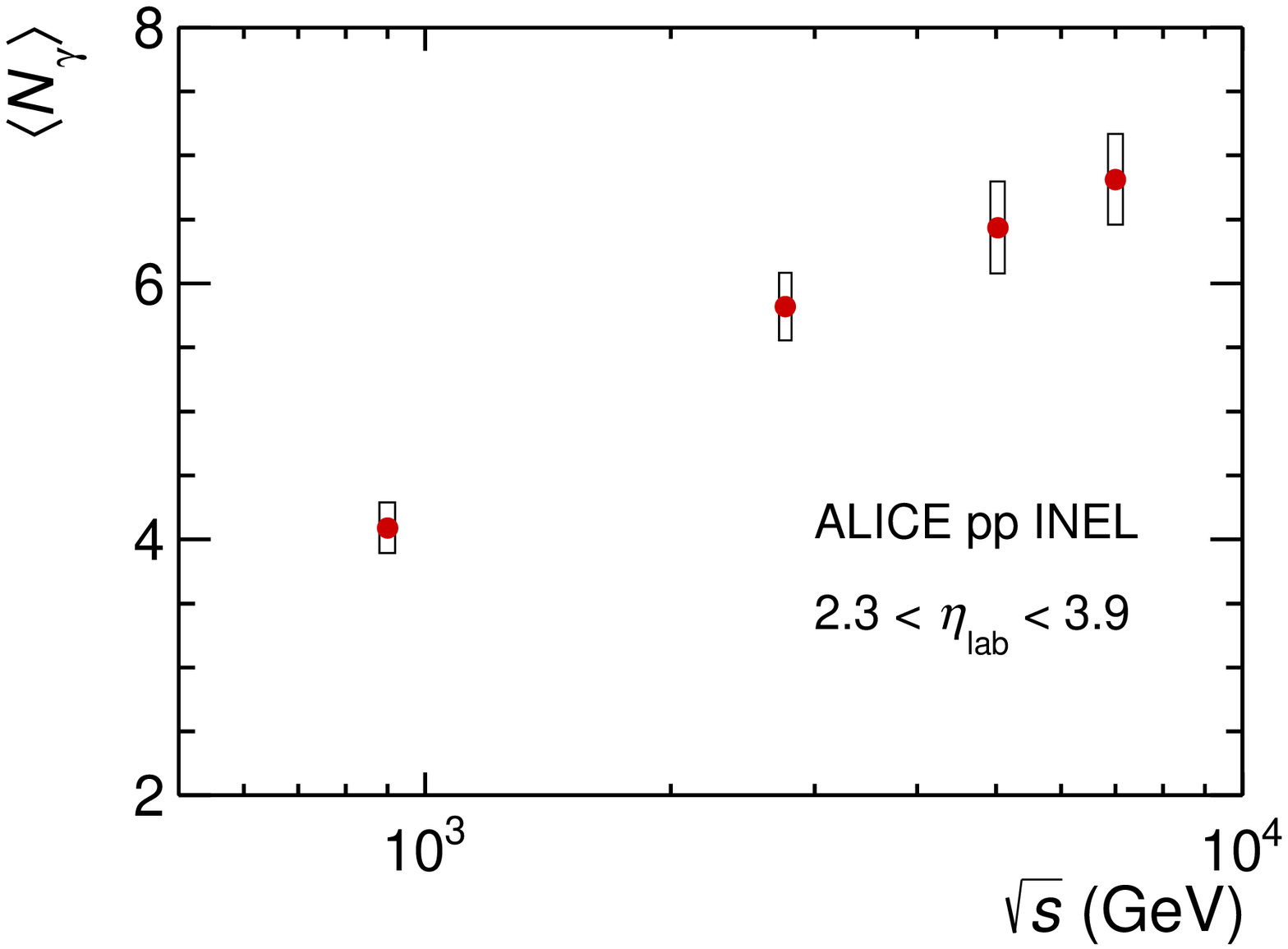}}
  \caption{(a) Comparison of \dndetaphotonlab measured at forward rapidity in INEL pp
    collisions for various center-of-mass energies, (b) average photon multiplicity within
    2.3~$<~\etalab~<$~3.9 as a function of collision energy in INEL pp collisions. The results
    at 0.9, 2.76, and 7~\TeV are taken from Ref.~\cite{ALICE:PMDpaper}.}
  \label{dndeta_pp_allcms}
\end{figure}

The \dndetaphotonlab distributions measured in the rapidity interval 2.3~$<~\etalab~<$~3.9 in INEL pp collisions at
various center-of-mass energies are compared in Fig.~\ref{dndeta_pp_allcms_compare}. Two additional systematic
uncertainties (diffraction shape and diffraction ratio) are considered in this analysis and they lead to a total
systematic uncertainty that is slightly larger at \five with respect to the other energies.
However, the total systematic uncertainty at all energies mostly comes from the 10\% uncertainty
on the upstream material in front of the PMD. The material uncertainty shows essentially no \s variation
and is therefore fully correlated over all energies. A smooth increase of the inclusive photon multiplicity
with increasing collision energy is observed. The average photon multiplicity $\langle N_{\rm \gamma} \rangle$
within 2.3~$<~\etalab~<$~3.9 as a function of $\sqrt{s}$ is shown in Fig.~\ref{dndeta_pp_allcms_roots}. The
results of the measurements of $\langle N_{\rm \gamma} \rangle$ are given
in Table~\ref{dndeta_pp_all_cms_table}.

\begin{table}[h!]
  \caption{The average photon multiplicity $\langle N_{\rm \gamma} \rangle$ within 2.3~$<~\etalab~<$~3.9
    for various center-of-mass energies in INEL pp collisions. The quoted errors are systematic uncertainties.
    Statistical uncertainties are negligible. Data points at 0.9, 2.76, and 7~\TeV are taken from Ref.~\cite{ALICE:PMDpaper}.}
  \label{dndeta_pp_all_cms_table}
  \begin{center}
    \begin{tabular}{l l}
      \toprule
      \s (\TeV) & $\langle N_{\rm \gamma} \rangle$\\
      \midrule
      0.9 & 4.09 $\pm$~0.20 \\[0.15cm]
      2.76 & 5.82 $\pm$~0.26 \\[0.15cm]
      5.02 & 6.44 $\pm$~0.36 \\[0.15cm]
      7 & 6.81 $\pm$~0.35 \\[0.15cm]
      \bottomrule
    \end{tabular}
  \end{center}	
\end{table}
\begin{figure}[h!]
  \subfigure[]{
    \label{dndetaDistMBChCompare_pp}
    \includegraphics[scale=0.41]{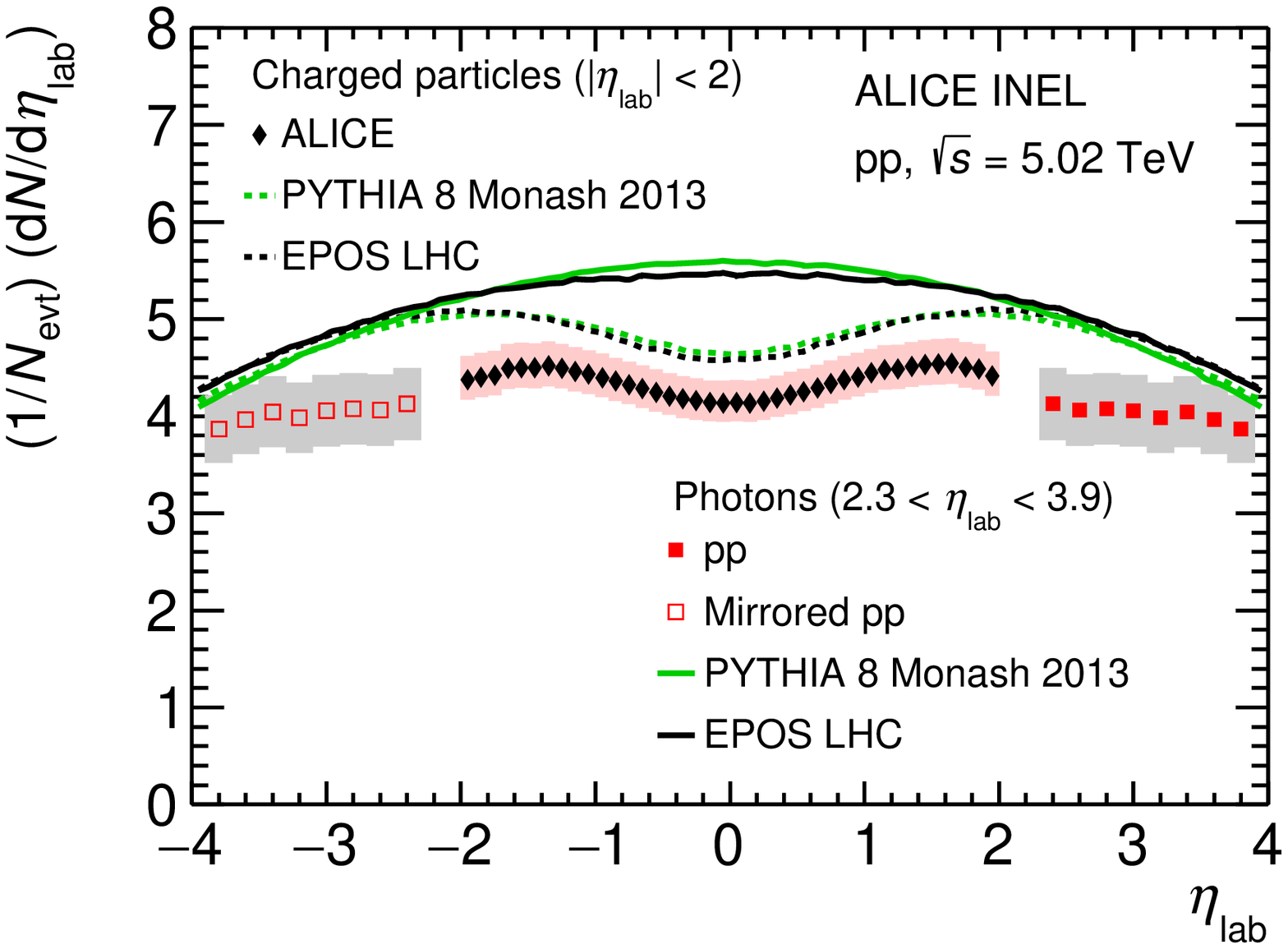}}
  \subfigure[]{
    \label{dndetaDistMBChCompare_pPb}
    \includegraphics[scale=0.41]{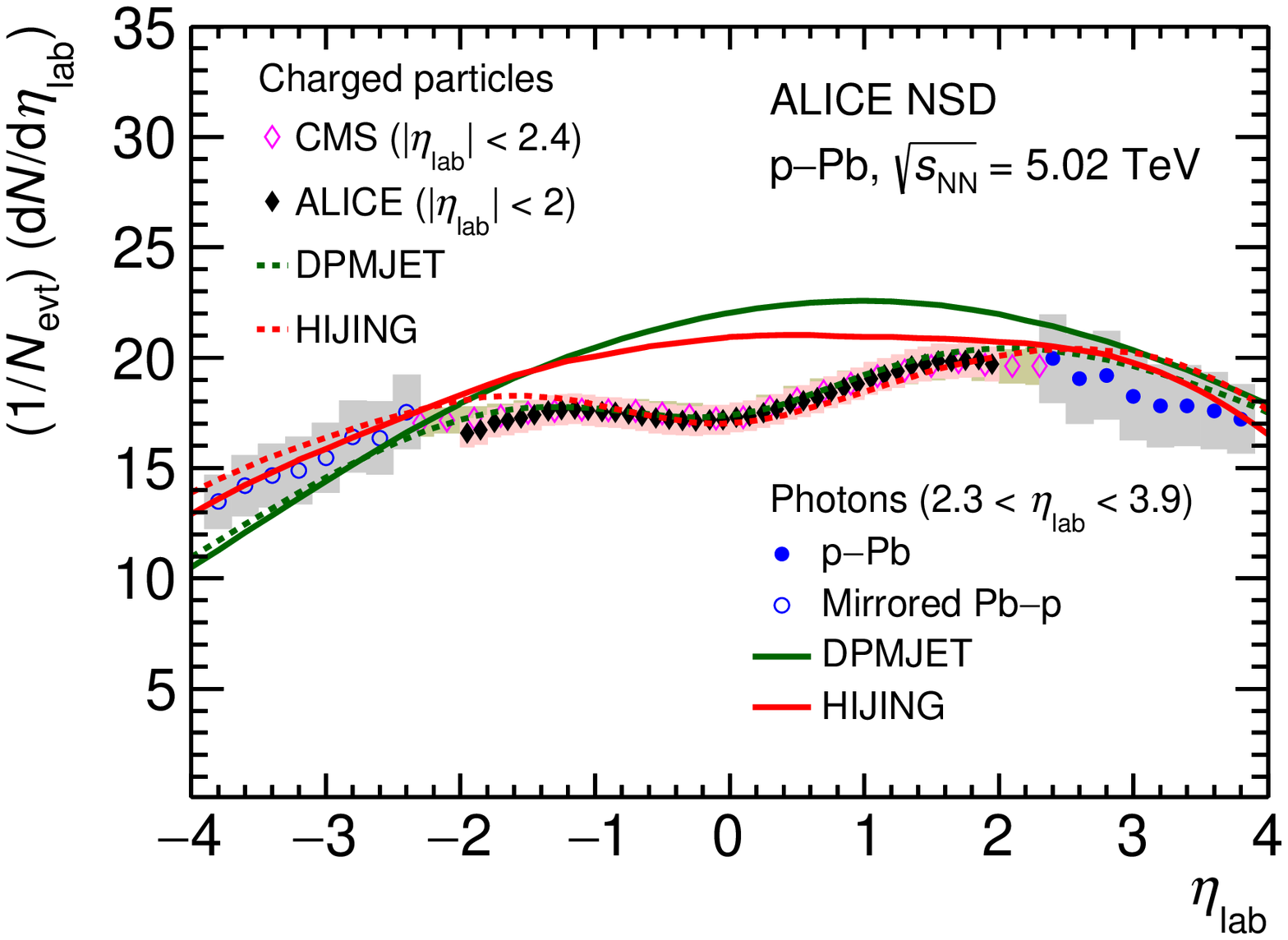}}
  \caption{The pseudorapidity distributions of inclusive photons are compared with charged-particle
    measurements at midrapidity by ALICE~\cite{ALICE:MBChPrpp5TeV} in \pp collisions (a) and both
    ALICE~\cite{ALICE:pPbMidChdNdEtapaper5020MB} and CMS~\cite{CMS:ChPrMidpPb5TeV} in \pPb collisions
    at \fivenn (b). Mirrored data points of \pp and \Pbp results and predictions from various MC models
    are superimposed.}
  \label{dndetaDistMBChCompare}
\end{figure}

Since the dominant contribution to inclusive photon production comes from \pizero decays,
the number of produced photons should be similar to the charged-particle multiplicity.
The comparison between the pseudorapidity distribution of inclusive photons in INEL pp
and NSD \pPb (combined with results of \Pbp mirrored with respect to \etalab = 0) events
measured at forward rapidity and that of charged particles at midrapidity by
ALICE~\cite{ALICE:pPbMidChdNdEtapaper5020MB,ALICE:MBChPrpp5TeV}
and CMS~\cite{CMS:ChPrMidpPb5TeV} are presented in Fig.~\ref{dndetaDistMBChCompare}.
It is observed that the inclusive photon production at forward rapidity follows the trend
of charged-particle production at midrapidity. The predictions from the different event generators
are also displayed in Fig.~\ref{dndetaDistMBChCompare} and they show similar values for charged-particle
(dashed lines) and photon (solid lines) multiplicity at forward and backward pseudorapidities, while at
midrapidity the photon and charged-particle
pseudorapidity density differ. The origin of this difference is due to a mass effect in the transformation
between $\mathrm{d}N/\mathrm{d}y$ and $\mathrm{d}N/\mathrm{d}\eta$ at $\eta \approx 0$.
The measured \dndetalab in \pPb collisions is well described by both \hijing and \dpmjet event
generators. HIJING reproduces better the \dndetaphotonlab compared to DPMJET. For \pp collisions,
both PYTHIA~8 with the Monash 2013 tune and \eposlhc overpredict the photon and charged-particle
multiplicity.

\begin{figure}[h!]
  \centering
  \includegraphics[scale=0.5]{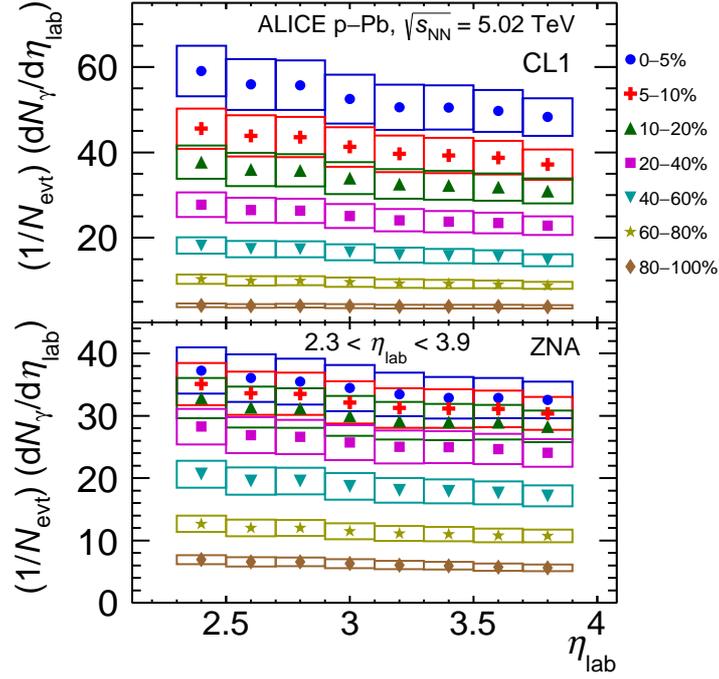}
  \caption{Pseudorapidity distributions of inclusive photons measured within 2.3~$<$~$\etalab$~$<$~3.9 in \pPb
    collisions at \fivenn for several centrality classes and for two centrality estimators:
    CL1 (top) and \ZNA (bottom).}
  \label{dndetaDistCent}
\end{figure}

\begin{figure}[h!]
  \centering
  \includegraphics[scale=0.5]{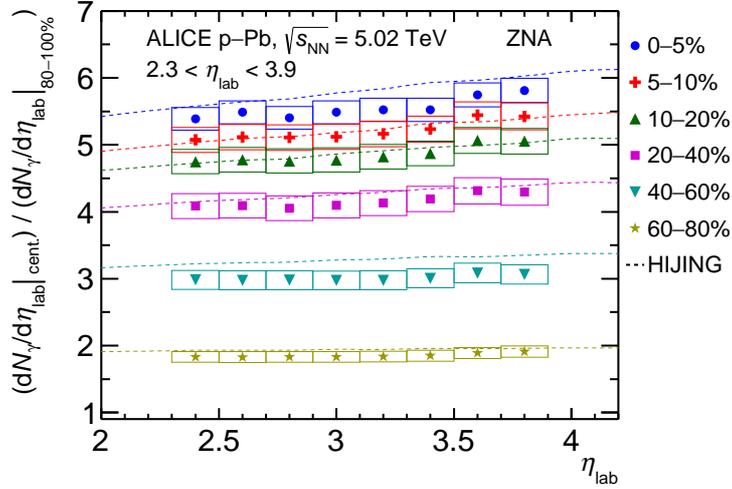}
  \caption{Ratios of \dndetaphotonlab distributions measured in different centrality
    classes to that in the most peripheral (80--100 \%) centrality class.
    The corresponding \hijing predictions are shown by dotted lines.}
  \label{dndetaRatio}
\end{figure}

\subsection{Pseudorapidity distributions in centrality classes of \pPb collisions}
\label{results-dndeta-cent}
The pseudorapidity distributions of inclusive photons as a function of pseudorapidity measured in \pPb
collisions are presented in Fig.~\ref{dndetaDistCent} for several centrality classes estimated using
two centrality estimators, CL1 (top) and \ZNA (bottom). The analysis was not performed for the \Pbp
configuration, hence only one rapidity interval is covered for the photon measurements, namely the one
in the Pb-going direction, and the ZNA measures the energy of the neutrons emitted in the direction of
the Pb beam. The multiplicity in the most central (0--5\%) collisions when considering the CL1 (\ZNA)
estimator reaches $\sim$ 3 ($\sim$ 2) times larger value compared to that measured in MB events. The lower
values of \dndetaphotonlab in case of central events selected with the \ZNA estimator is, most probably,
due to the saturation of forward neutron emission~\cite{ALICE:pPbMidChdNdEtapaper5020Cent}. The systematic
uncertainty represented by open boxes mostly comes from the uncertainty on the upstream material in front
of the PMD. To understand further the evolution of \dndetaphotonlab with centrality the distributions in each
centrality interval are divided by the distribution in the most peripheral (80--100\%) event class
and presented in Fig.~\ref{dndetaRatio}. With increasing pseudorapidity the ratios are observed to
increase linearly with a slope whose magnitude increases for central events being highest for the
most central (0--5\%) collisions. \hijing shows a similar trend and explains the data within
measured uncertainties except for the 0--5\% and 40--60\% event classes.

\begin{figure}[h!]
  \centering
  \includegraphics[scale=0.5]{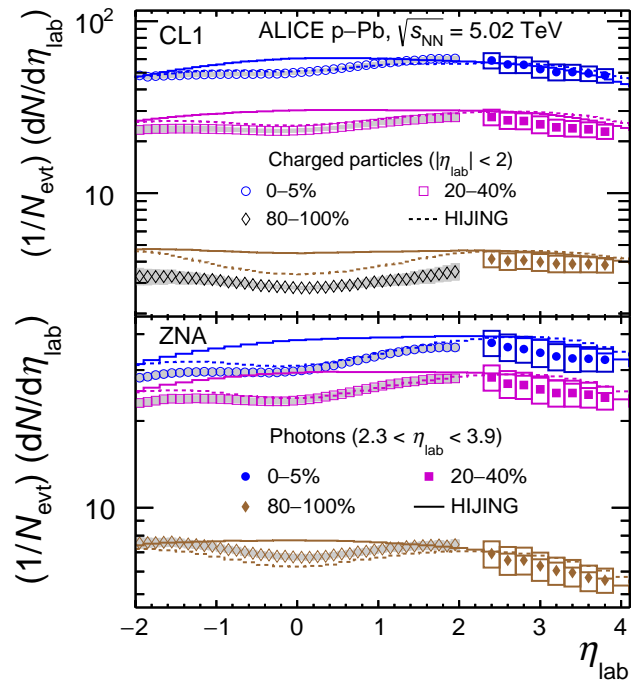}
  \caption{Pseudorapidity distribution of inclusive photons measured within 2.3~$<$~$\etalab$~$<$~3.9 in
    \pPb collisions at \fivenn compared with the charged-particle measurements~\cite{ALICE:pPbMidChdNdEtapaper5020Cent}
    and to \hijing predictions for three centrality classes and for two centrality estimators:
    CL1 (top), \ZNA (bottom).}
  \label{dndetaDistCentChcompare}
\end{figure}

The results are compared in Fig.~\ref{dndetaDistCentChcompare} to similar measurements of charged particles
at midrapidity by ALICE~\cite{ALICE:pPbMidChdNdEtapaper5020Cent} for selected centrality intervals
(0--5\%, 20--40\% and 80--100\%) determined with the CL1 (top panel) and ZNA (bottom panel) estimators.
A clear asymmetric shape of \dndetalab is observed for most central collisions and the shape becomes symmetric
(like in \pp) in the most peripheral event class. Figure~\ref{dndetaDistCentChcompare} also reports the
comparison of the measured \dndetaphotonlab with \hijing predictions for the three considered centrality classes.
The \dndetalab distributions at midrapidity are better described by the model in the case of
the \ZNA estimator. In the most peripheral (80--100\%) collisions, \hijing overpredicts (underpredicts) the
\dndetalab for the CL1 (\ZNA) estimator. The model describes the values of \dndetaphotonlab at forward rapidity to
within 15\% except for the 0--5\% CL1 class and the 80--100\% \ZNA class where the model prediction is compatible
with the data points within 5\%.

\begin{figure}[h!]
  \subfigure[]{
    \label{figNpartScale_NgammavsNpart}
    \includegraphics[scale=0.4]{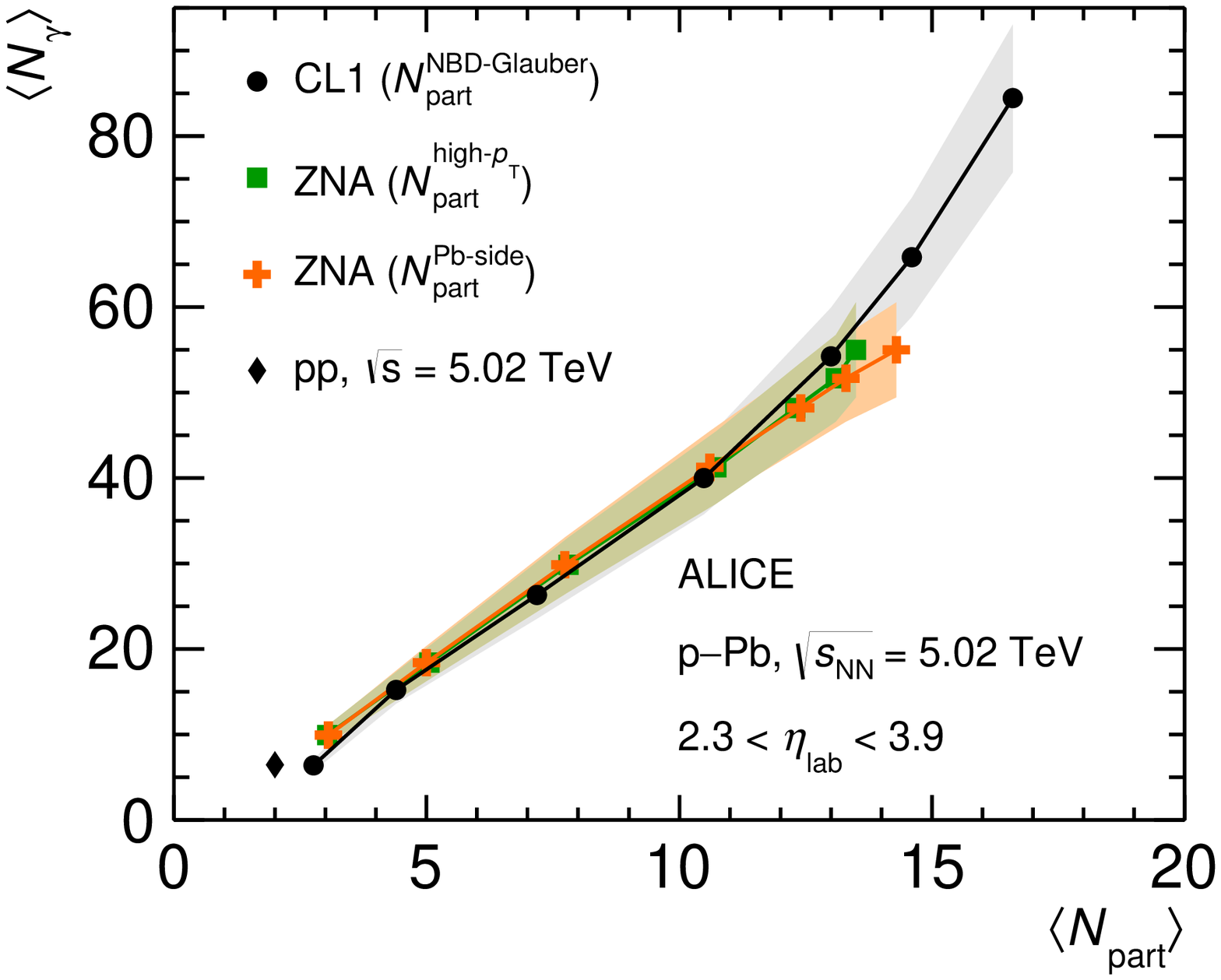}}
  \subfigure[]{
    \label{figNpartScale_NgammabyNpartvsNpart}
    \includegraphics[scale=0.4]{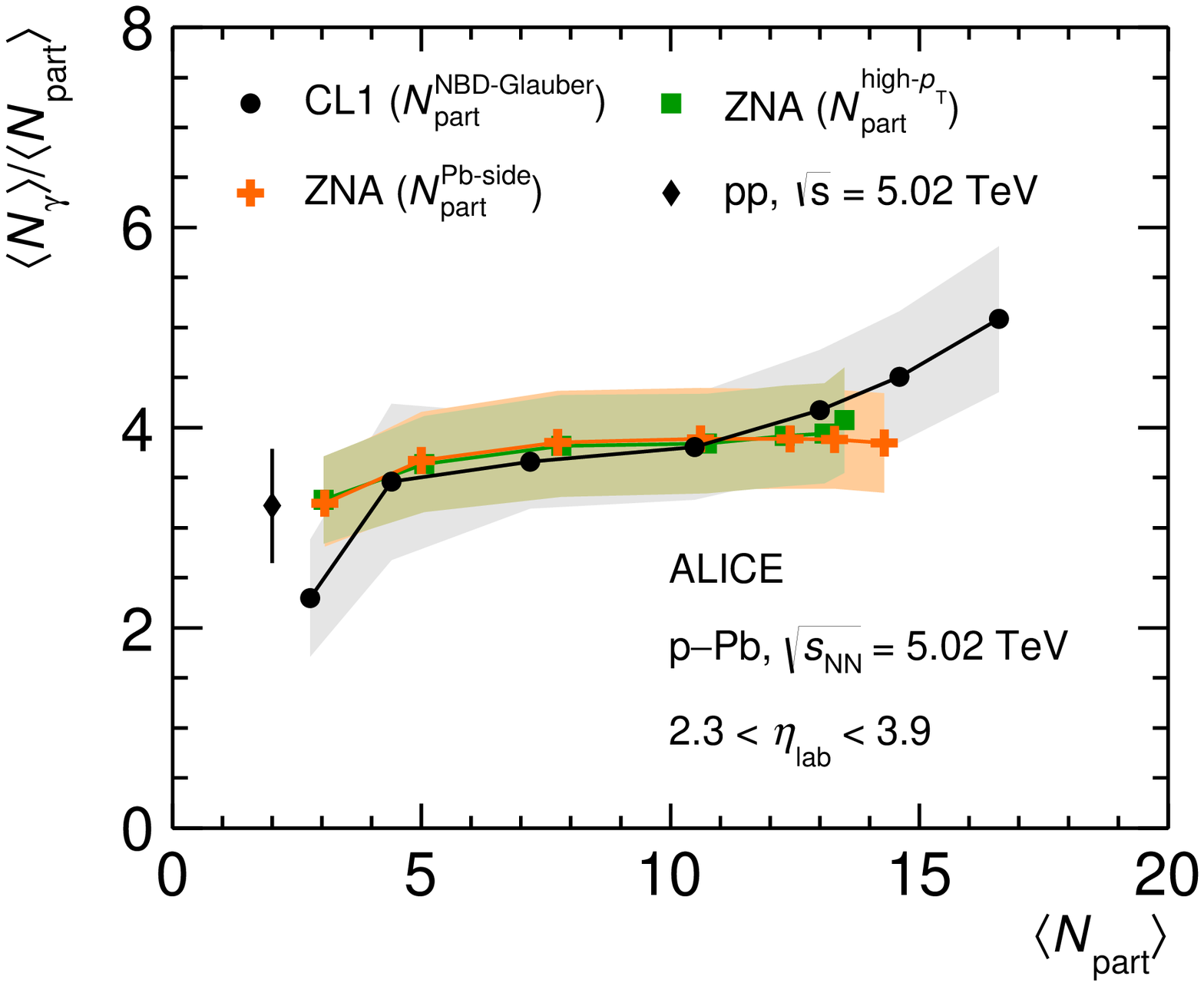}}
  \subfigure[]{
    \label{figNpartScale_NgammabyNpartvsNgamma}
    \includegraphics[scale=0.4]{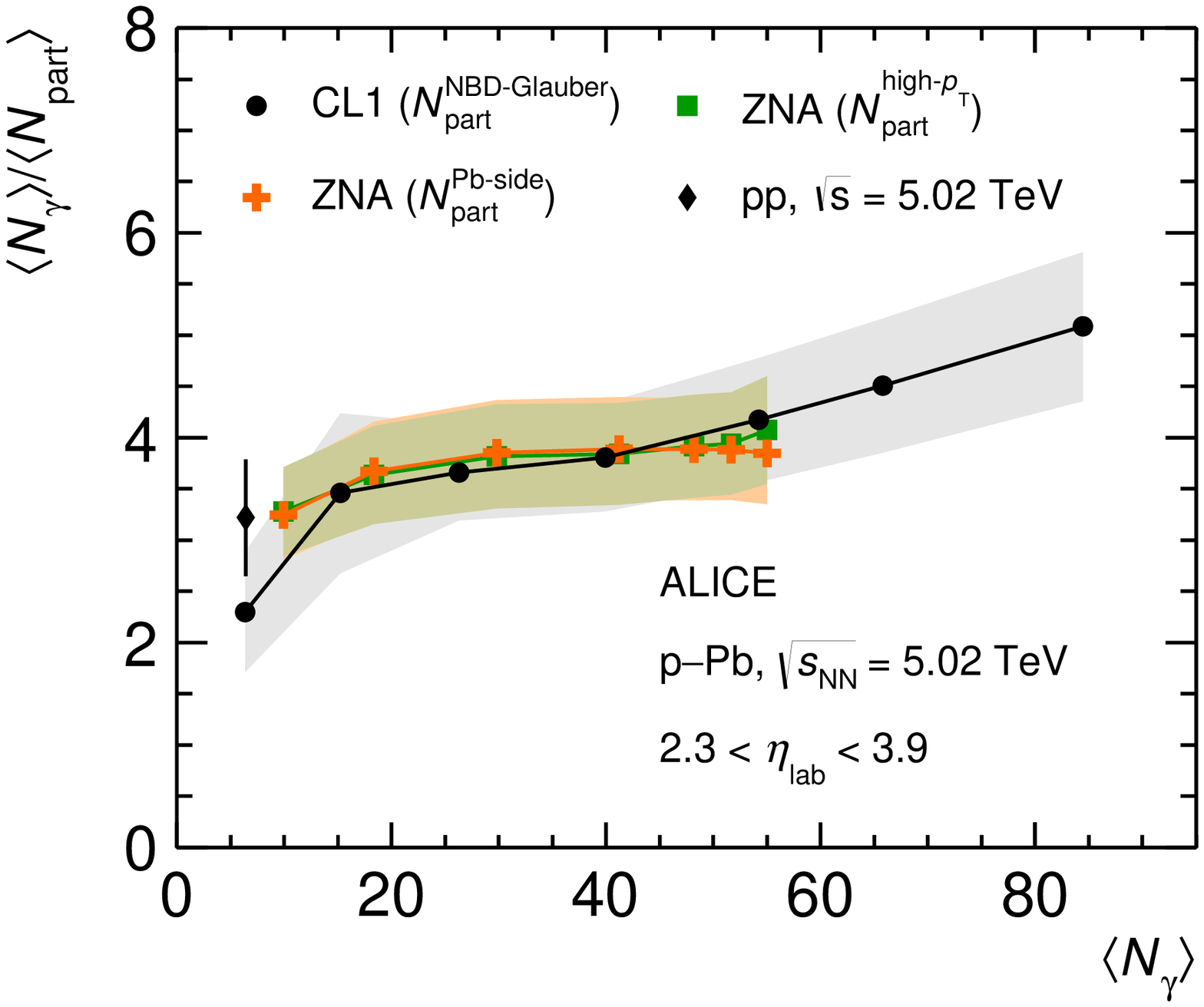}}
  \caption{(a) The average photon multiplicity $\langle N_{\rm \gamma} \rangle$ within 2.3~$<~\etalab~<$~3.9
    as a function of the average number of participants for various centrality estimators in \pPb collisions
    at \fivenn. Values of $\langle N_{\rm \gamma} \rangle$/$\langle \Npart \rangle$ are shown as a function of
    $\langle \Npart \rangle$ (b) and $\langle N_{\rm \gamma} \rangle$ (c). The lines between the data points
    are to guide the eye. The data point for \pp collisions at \five is also included for reference.}
  \label{figNpartScale}
\end{figure}

The obtained values of average photon multiplicity $\langle N_{\rm \gamma} \rangle$ at forward
rapidity (2.3~$<~\etalab~<$~3.9) as a function of $\langle N_{\rm part} \rangle$ for the two centrality estimators
in \pPb collisions at \fivenn are presented in Fig.~\ref{figNpartScale_NgammavsNpart} and listed in
Table~\ref{Npart_Ngamma_value}. The systematic uncertainties are represented by the color bands.
The values of $\langle N_{\rm part} \rangle$ for the different centrality classes for both the estimators are
taken from~\cite{ALICE:pPbMidChdNdEtapaper5020Cent}. The data point for \pp collisions at \five is also
included for reference.
The average photon multiplicity $\langle N_{\rm \gamma} \rangle$ divided by the average number of participants
is presented as a function of $\langle N_{\rm part} \rangle$ in Fig.~\ref{figNpartScale_NgammabyNpartvsNpart}.
The $\langle N_{\rm \gamma} \rangle$/$\langle \Npart \rangle$ ratio has a steeper increase with $\langle N_{\rm part} \rangle$
for the CL1 centrality estimator than for the ZNA one. This could be attributed to the strong multiplicity
bias as described in Ref.~\cite{ALICE:pPbMidChdNdEtapaper5020Cent}. It is noted that
the curve does not point towards the \pp result.
When the ZNA is used for the centrality estimation along with the hybrid method to determine the average
number of participant nucleons ($N_\mathrm{part}^{\mathrm{high}\text-\pt}$ or $N_\mathrm{part}^{\mathrm{Pb\text-side}}$),
the inclusive photon multiplicity is found to scale with $\langle N_{\rm part} \rangle$ within uncertainties
and to point towards the \pp data point at low $\langle N_{\rm part} \rangle$. Moreover, the range in
$\langle N_{\rm part} \rangle$ for the \ZNA centrality selection is more limited (that could be because of
saturation of forward neutron emission) than what is obtained by particle-multiplicity-based centrality estimators.
This effect is also emphasized in Fig.~\ref{figNpartScale_NgammabyNpartvsNgamma} where the same quantity
$\langle N_{\rm \gamma} \rangle$/$\langle N_{\rm part} \rangle$ is presented as a function of $\langle N_{\rm \gamma} \rangle$.
Similar results were observed for charged-particle multiplicity at midrapidity reported by
ALICE in Ref.~\cite{ALICE:pPbMidChdNdEtapaper5020Cent}.

\begin{table}[h!]
  \begin{center}
   \caption{The average photon multiplicity $\langle N_{\rm \gamma} \rangle$ within 2.3~$<~\etalab~<$~3.9
     for various centrality classes defined using CL1 and \ZNA centrality estimators in \pPb collisions
     at \fivenn. The corresponding values of $\langle N_{\rm part} \rangle$ are taken
     from Ref.~\cite{ALICE:pPbMidChdNdEtapaper5020Cent}. The quoted errors are systematic uncertainties.
     Statistical uncertainties are negligible.}
      \label{Npart_Ngamma_value}
      \renewcommand{\arraystretch}{1.3}
      \begin{tabular}{|c|c|c|c|c|c|}
        \hline
        \multirow{2}{*}{Centrality} & \multicolumn{2}{c|}{CL1} & \multicolumn{3}{c|}{ZNA} \\ 
        \cline{2-6}
        & $\langle N_\mathrm{part}^{\rm NBD\text-Glauber} \rangle$ & $\langle N_{\rm \gamma} \rangle$ & $\langle N_\mathrm{part}^{\mathrm{high}\text-\pt} \rangle$ & $\langle N_\mathrm{part}^{\mathrm{Pb\text-side}} \rangle$ & $\langle N_{\rm \gamma} \rangle$ \\ 
        \hline
        0--5\% & 16.60 $\pm$ 1.66 & 84.44 $\pm$ 8.65 & 13.50 $\pm$ 1.08 & 14.30 $\pm$ 1.14 & 54.98 $\pm$ 5.59 \\
        5--10\% & 14.60 $\pm$ 1.46 & 65.82 $\pm$ 6.96 & 13.10 $\pm$ 1.05 & 13.30 $\pm$ 1.06 & 51.65 $\pm$ 5.09 \\
        10--20\% & 13.00 $\pm$ 1.30 & 54.23 $\pm$ 5.67 & 12.30 $\pm$ 0.98 & 12.40 $\pm$ 0.99 & 48.18 $\pm$ 4.82 \\
        20--40\% & 10.49 $\pm$ 0.94 & 39.95 $\pm$ 4.22 & 10.73 $\pm$ 0.86 & 10.60 $\pm$ 0.85 & 41.22 $\pm$ 4.22 \\
        40--60\% & 7.18 $\pm$ 0.52 & 26.28 $\pm$ 2.79 & 7.81 $\pm$ 0.62 & 7.74 $\pm$ 0.62 & 29.82 $\pm$ 3.20 \\
        60--80\% & 4.40 $\pm$ 0.88 & 15.21 $\pm$ 1.60 & 5.05 $\pm$ 0.40 & 5.00 $\pm$ 0.40 & 18.36 $\pm$ 1.93\\
        80--100\% & 2.76 $\pm$ 0.63 & 6.34 $\pm$ 0.68 & 3.03 $\pm$ 0.24 & 3.06 $\pm$ 0.24 & 9.93 $\pm$ 1.05\\
        \hline
      \end{tabular}
  \end{center}
\end{table}

\section{Summary}
\label{summary}
In summary, we have presented results for inclusive photon
multiplicity and pseudorapidity distributions at forward rapidities
(2.3~$<~\etalab~<$~3.9) in \pp collisions at \five and \pPb and \Pbp collisions at \fivenn.
The centrality dependence of inclusive photon production is
studied in \pPb collisions for the CL1 and the \ZNA centrality estimators.
The evolution of the average photon multiplicity $\langle N_{\rm \gamma} \rangle$ with \s
in \pp collisions and with $\langle N_{\rm part} \rangle$ in \pPb collisions is presented.
A comparison of inclusive photon production at forward rapidity
with the production of charged particles at midrapidity is also discussed.
The obtained results are compared to predictions from various event generators 
(PYTHIA~6 with the Perugia 2011 tune, PYTHIA~8 with the Monash 2013 tune, \phojet, \eposlhc
for \pp collisions and \hijing, \dpmjet, \ampt for \pPb and \Pbp collisions). The multiplicity
distributions are well reproduced with double NBDs and results of the
parameterization are provided.

The PYTHIA~8 generator with the Monash 2013 tune and the \eposlhc generator do not reproduce the multiplicity
distribution in \pp collisions at high multiplicities whereas PYTHIA~6 with the Perugia 2011 tune and \phojet better
describe the data within uncertainties. For \pPb collisions, none of the considered MC models
could describe the multiplicity results in the full range.
However, \hijing and \ampt are able to 
reproduce the tail of the distribution for both \pPb and \Pbp collisions. Interestingly,
compared to multiplicity distributions, the pseudorapidity distributions of inclusive
photons are better described by all the considered MC models. The model predictions
are found to describe the data within 20\%. The pseudorapidity density of inclusive
photons at forward rapidity follows the $\etalab$ dependence of charged-particle multiplicity at midrapidity
in \pp and for various centrality classes in \pPb collisions. This can be attributed to the fact
that the major contribution to inclusive photons comes from the decay of neutral pions, and therefore
the number of produced photons should be similar to that of charged particles. The models give
similar pseudorapidity densities for photons and charged particles only at forward rapidity where
the difference between $\eta$ and $y$ is negligible.

An asymmetric \dndetalab distribution for the asymmetric \pPb collision system is evident for most
central collisions while the distributions become more symmetric for the most peripheral class.
\hijing describes the centrality evolution of both photon and charged-particle production within
about 10--12\%. The average photon multiplicity $\langle N_{\rm \gamma} \rangle$ divided by the
average number of participants scales linearly with both $N_\mathrm{part}^{\mathrm{high}\text-\pt}$ and
$N_\mathrm{part}^{\mathrm{Pb\text-side}}$ estimated by the energy deposited in the \ZDC and the trend
with $\langle \Npart \rangle$ is consistent with the data point from pp collisions at \five.

These results of inclusive photon production in \pp and \pPb collisions provide important input for the
tuning of theoretical models and MC event generators and help to establish the baseline
measurements to interpret the \PbPb data.


\newenvironment{acknowledgement}{\relax}{\relax}
\begin{acknowledgement}
\section*{Acknowledgements}

The ALICE Collaboration would like to thank all its engineers and technicians for their invaluable contributions to the construction of the experiment and the CERN accelerator teams for the outstanding performance of the LHC complex.
The ALICE Collaboration gratefully acknowledges the resources and support provided by all Grid centres and the Worldwide LHC Computing Grid (WLCG) collaboration.
The ALICE Collaboration acknowledges the following funding agencies for their support in building and running the ALICE detector:
A. I. Alikhanyan National Science Laboratory (Yerevan Physics Institute) Foundation (ANSL), State Committee of Science and World Federation of Scientists (WFS), Armenia;
Austrian Academy of Sciences, Austrian Science Fund (FWF): [M 2467-N36] and Nationalstiftung f\"{u}r Forschung, Technologie und Entwicklung, Austria;
Ministry of Communications and High Technologies, National Nuclear Research Center, Azerbaijan;
Conselho Nacional de Desenvolvimento Cient\'{\i}fico e Tecnol\'{o}gico (CNPq), Financiadora de Estudos e Projetos (Finep), Funda\c{c}\~{a}o de Amparo \`{a} Pesquisa do Estado de S\~{a}o Paulo (FAPESP) and Universidade Federal do Rio Grande do Sul (UFRGS), Brazil;
Bulgarian Ministry of Education and Science, within the National Roadmap for Research Infrastructures 2020-2027 (object CERN), Bulgaria;
Ministry of Education of China (MOEC) , Ministry of Science \& Technology of China (MSTC) and National Natural Science Foundation of China (NSFC), China;
Ministry of Science and Education and Croatian Science Foundation, Croatia;
Centro de Aplicaciones Tecnol\'{o}gicas y Desarrollo Nuclear (CEADEN), Cubaenerg\'{\i}a, Cuba;
Ministry of Education, Youth and Sports of the Czech Republic, Czech Republic;
The Danish Council for Independent Research | Natural Sciences, the VILLUM FONDEN and Danish National Research Foundation (DNRF), Denmark;
Helsinki Institute of Physics (HIP), Finland;
Commissariat \`{a} l'Energie Atomique (CEA) and Institut National de Physique Nucl\'{e}aire et de Physique des Particules (IN2P3) and Centre National de la Recherche Scientifique (CNRS), France;
Bundesministerium f\"{u}r Bildung und Forschung (BMBF) and GSI Helmholtzzentrum f\"{u}r Schwerionenforschung GmbH, Germany;
General Secretariat for Research and Technology, Ministry of Education, Research and Religions, Greece;
National Research, Development and Innovation Office, Hungary;
Department of Atomic Energy Government of India (DAE), Department of Science and Technology, Government of India (DST), University Grants Commission, Government of India (UGC) and Council of Scientific and Industrial Research (CSIR), India;
National Research and Innovation Agency - BRIN, Indonesia;
Istituto Nazionale di Fisica Nucleare (INFN), Italy;
Japanese Ministry of Education, Culture, Sports, Science and Technology (MEXT) and Japan Society for the Promotion of Science (JSPS) KAKENHI, Japan;
Consejo Nacional de Ciencia (CONACYT) y Tecnolog\'{i}a, through Fondo de Cooperaci\'{o}n Internacional en Ciencia y Tecnolog\'{i}a (FONCICYT) and Direcci\'{o}n General de Asuntos del Personal Academico (DGAPA), Mexico;
Nederlandse Organisatie voor Wetenschappelijk Onderzoek (NWO), Netherlands;
The Research Council of Norway, Norway;
Commission on Science and Technology for Sustainable Development in the South (COMSATS), Pakistan;
Pontificia Universidad Cat\'{o}lica del Per\'{u}, Peru;
Ministry of Education and Science, National Science Centre and WUT ID-UB, Poland;
Korea Institute of Science and Technology Information and National Research Foundation of Korea (NRF), Republic of Korea;
Ministry of Education and Scientific Research, Institute of Atomic Physics, Ministry of Research and Innovation and Institute of Atomic Physics and University Politehnica of Bucharest, Romania;
Ministry of Education, Science, Research and Sport of the Slovak Republic, Slovakia;
National Research Foundation of South Africa, South Africa;
Swedish Research Council (VR) and Knut \& Alice Wallenberg Foundation (KAW), Sweden;
European Organization for Nuclear Research, Switzerland;
Suranaree University of Technology (SUT), National Science and Technology Development Agency (NSTDA), Thailand Science Research and Innovation (TSRI) and National Science, Research and Innovation Fund (NSRF), Thailand;
Turkish Energy, Nuclear and Mineral Research Agency (TENMAK), Turkey;
National Academy of  Sciences of Ukraine, Ukraine;
Science and Technology Facilities Council (STFC), United Kingdom;
National Science Foundation of the United States of America (NSF) and United States Department of Energy, Office of Nuclear Physics (DOE NP), United States of America.
In addition, individual groups or members have received support from:
European Research Council, Strong 2020 - Horizon 2020, Marie Sk\l{}odowska Curie (grant nos. 950692, 824093, 896850), European Union;
Academy of Finland (Center of Excellence in Quark Matter) (grant nos. 346327, 346328), Finland;
Programa de Apoyos para la Superaci\'{o}n del Personal Acad\'{e}mico, UNAM, Mexico.

\end{acknowledgement}

\bibliographystyle{utphys}   
\bibliography{bibliography}

\providecommand{\href}[2]{#2}\begingroup\raggedright\begin{thebibliography}{10}

\bibitem{QGP-1}
P.~Braun-Munzinger, V.~Koch, T.~Sch\"afer, and J.~Stachel, ``{Properties of hot
  and dense matter from relativistic heavy-ion collisions}'',
  \href{http://dx.doi.org/10.1016/j.physrep.2015.12.003}{{\em Phys. Rept.}
  {\bfseries 621} (2016) 76--126},
  \href{http://arxiv.org/abs/1510.00442}{{\ttfamily arXiv:1510.00442
  [nucl-th]}}.

\bibitem{QGP-2}
{\bfseries ALICE} Collaboration, J.~Schukraft, ``{Heavy Ion physics with the
  ALICE experiment at the CERN LHC}'',
  \href{http://dx.doi.org/10.1098/rsta.2011.0469}{{\em Phil. Trans. Roy. Soc.
  Lond. A} {\bfseries 370} (2012) 917--932},
  \href{http://arxiv.org/abs/1109.4291}{{\ttfamily arXiv:1109.4291 [hep-ex]}}.

\bibitem{ALICE:ReviewPaper}
{\bfseries ALICE} Collaboration, ``{The ALICE experiment -- A journey through
  QCD}'', \href{http://arxiv.org/abs/2211.04384}{{\ttfamily arXiv:2211.04384
  [nucl-ex]}}.

\bibitem{CNM_effect2}
R.~Xu, W.-T. Deng, and X.-N. Wang, ``{Nuclear modification of high-$p_{\rm T}$
  hadron spectra in p--A collisions at LHC}'',
  \href{http://dx.doi.org/10.1103/PhysRevC.86.051901}{{\em Phys. Rev. C}
  {\bfseries 86} (2012) 051901},
  \href{http://arxiv.org/abs/1204.1998}{{\ttfamily arXiv:1204.1998 [nucl-th]}}.

\bibitem{CNM_effect1}
M.~Arneodo, ``{Nuclear effects in structure functions}'',
  \href{http://dx.doi.org/10.1016/0370-1573(94)90048-5}{{\em Phys. Rept.}
  {\bfseries 240} (1994) 301--393}.

\bibitem{CMS:2010ifv}
{\bfseries CMS} Collaboration, V.~Khachatryan {\em et~al.}, ``{Observation of
  long-range near-side angular correlations in proton--proton collisions at the
  LHC}'', \href{http://dx.doi.org/10.1007/JHEP09(2010)091}{{\em JHEP}
  {\bfseries 09} (2010) 091}, \href{http://arxiv.org/abs/1009.4122}{{\ttfamily
  arXiv:1009.4122 [hep-ex]}}.

\bibitem{CMS:2015fgy}
{\bfseries CMS} Collaboration, V.~Khachatryan {\em et~al.}, ``{Measurement of
  long-range near-side two-particle angular correlations in pp collisions at
  $\sqrt s =$13 TeV}'',
  \href{http://dx.doi.org/10.1103/PhysRevLett.116.172302}{{\em Phys. Rev.
  Lett.} {\bfseries 116} (2016) 172302},
  \href{http://arxiv.org/abs/1510.03068}{{\ttfamily arXiv:1510.03068
  [nucl-ex]}}.

\bibitem{ATLAS:2015hzw}
{\bfseries ATLAS} Collaboration, G.~Aad {\em et~al.}, ``{Observation of
  long-range elliptic azimuthal anisotropies in $\sqrt{s}=$13 and 2.76 TeV pp
  collisions with the ATLAS Detector}'',
  \href{http://dx.doi.org/10.1103/PhysRevLett.116.172301}{{\em Phys. Rev.
  Lett.} {\bfseries 116} (2016) 172301},
  \href{http://arxiv.org/abs/1509.04776}{{\ttfamily arXiv:1509.04776
  [hep-ex]}}.

\bibitem{ALICE:2012eyl}
{\bfseries ALICE} Collaboration, B.~Abelev {\em et~al.}, ``{Long-range angular
  correlations on the near and away side in p--Pb collisions at $\sqrt{s_{\rm
  NN}}=5.02$ TeV}'',
  \href{http://dx.doi.org/10.1016/j.physletb.2013.01.012}{{\em Phys. Lett. B}
  {\bfseries 719} (2013) 29--41},
  \href{http://arxiv.org/abs/1212.2001}{{\ttfamily arXiv:1212.2001 [nucl-ex]}}.

\bibitem{ATLAS:2012cix}
{\bfseries ATLAS} Collaboration, G.~Aad {\em et~al.}, ``{Observation of
  associated near-side and away-side long-range correlations in $\sqrt{s_{\rm
  NN}}$ = 5.02 TeV p--Pb Collisions with the ATLAS Detector}'',
  \href{http://dx.doi.org/10.1103/PhysRevLett.110.182302}{{\em Phys. Rev.
  Lett.} {\bfseries 110} (2013) 182302},
  \href{http://arxiv.org/abs/1212.5198}{{\ttfamily arXiv:1212.5198 [hep-ex]}}.

\bibitem{CMS:2012qk}
{\bfseries CMS} Collaboration, S.~Chatrchyan {\em et~al.}, ``{Observation of
  long-range near-side angular correlations in p--Pb collisions at the LHC}'',
  \href{http://dx.doi.org/10.1016/j.physletb.2012.11.025}{{\em Phys. Lett. B}
  {\bfseries 718} (2013) 795--814},
  \href{http://arxiv.org/abs/1210.5482}{{\ttfamily arXiv:1210.5482 [nucl-ex]}}.

\bibitem{ATLAS:2014qaj}
{\bfseries ATLAS} Collaboration, G.~Aad {\em et~al.}, ``{Measurement of
  long-range pseudorapidity correlations and azimuthal harmonics in
  $\sqrt{s_{\rm NN}}=5.02$ TeV p--Pb collisions with the ATLAS detector}'',
  \href{http://dx.doi.org/10.1103/PhysRevC.90.044906}{{\em Phys. Rev. C}
  {\bfseries 90} (2014) 044906},
  \href{http://arxiv.org/abs/1409.1792}{{\ttfamily arXiv:1409.1792 [hep-ex]}}.

\bibitem{ALICE:2016fzo}
{\bfseries ALICE} Collaboration, J.~Adam {\em et~al.}, ``{Enhanced production
  of multi-strange hadrons in high-multiplicity proton--proton collisions}'',
  \href{http://dx.doi.org/10.1038/nphys4111}{{\em Nature Phys.} {\bfseries 13}
  (2017) 535--539}, \href{http://arxiv.org/abs/1606.07424}{{\ttfamily
  arXiv:1606.07424 [nucl-ex]}}.

\bibitem{two_component_model1}
D.~Kharzeev, E.~Levin, and M.~Nardi, ``{Color glass condensate at the LHC:
  Hadron multiplicities in pp, p--A and A--A collisions}'',
  \href{http://dx.doi.org/10.1016/j.nuclphysa.2004.10.018}{{\em Nucl. Phys. A}
  {\bfseries 747} (2005) 609--629},
  \href{http://arxiv.org/abs/hep-ph/0408050}{{\ttfamily arXiv:hep-ph/0408050}}.

\bibitem{two_component_model2}
W.-T. Deng, X.-N. Wang, and R.~Xu, ``{Hadron production in pp, p--Pb, and
  Pb--Pb collisions with the HIJING 2.0 model at energies available at the CERN
  Large Hadron Collider}'',
  \href{http://dx.doi.org/10.1103/PhysRevC.83.014915}{{\em Phys. Rev. C}
  {\bfseries 83} (2011) 014915},
  \href{http://arxiv.org/abs/1008.1841}{{\ttfamily arXiv:1008.1841 [hep-ph]}}.

\bibitem{ALICE:PMDpaper}
{\bfseries ALICE} Collaboration, B.~B. Abelev {\em et~al.}, ``{Inclusive photon
  production at forward rapidities in proton--proton collisions at
  $\sqrt{s}~=$~0.9, 2.76 and 7~TeV}'',
  \href{http://dx.doi.org/10.1140/epjc/s10052-015-3356-2}{{\em Eur. Phys. J. C}
  {\bfseries 75} (2015) 146}, \href{http://arxiv.org/abs/1411.4981}{{\ttfamily
  arXiv:1411.4981 [nucl-ex]}}.

\bibitem{STAR:PMD_prl}
{\bfseries STAR} Collaboration, J.~Adams {\em et~al.}, ``{Multiplicity and
  pseudorapidity distributions of photons in Au + Au collisions at
  $\sqrt{s_{\mathrm{NN}}}~=~62.4$~Ge\kern-.1emV\xspace}'',
  \href{http://dx.doi.org/10.1103/PhysRevLett.95.062301}{{\em Phys. Rev. Lett.}
  {\bfseries 95} (2005) 062301},
  \href{http://arxiv.org/abs/nucl-ex/0502008}{{\ttfamily arXiv:nucl-ex/0502008
  [nucl-ex]}}.

\bibitem{STAR:PMD_prc}
{\bfseries STAR} Collaboration, J.~Adams {\em et~al.}, ``{Multiplicity and
  pseudorapidity distributions of charged particles and photons at forward
  pseudorapidity in Au + Au collisions at
  $\sqrt{s_{\mathrm{NN}}}~=~62.4$~Ge\kern-.1emV\xspace}'',
  \href{http://dx.doi.org/10.1103/PhysRevC.73.034906}{{\em Phys. Rev. C}
  {\bfseries 73} (2006) 034906},
  \href{http://arxiv.org/abs/nucl-ex/0511026}{{\ttfamily arXiv:nucl-ex/0511026
  [nucl-ex]}}.

\bibitem{ALICE:ppMidChMultpaper900}
{\bfseries ALICE} Collaboration, K.~Aamodt {\em et~al.}, ``{First
  proton--proton collisions at the LHC as observed with the ALICE detector:
  Measurement of the charged-particle pseudorapidity density at
  $\sqrt{s}~=~900$~Ge\kern-.1emV\xspace}'',
  \href{http://dx.doi.org/10.1140/epjc/s10052-009-1227-4}{{\em Eur. Phys. J. C}
  {\bfseries 65} (2010) 111--125},
  \href{http://arxiv.org/abs/0911.5430}{{\ttfamily arXiv:0911.5430 [hep-ex]}}.

\bibitem{ALICE:ppMidChMultpaper900and2360}
{\bfseries ALICE} Collaboration, K.~Aamodt {\em et~al.}, ``{Charged-particle
  multiplicity measurement in proton--proton collisions at $\sqrt{s}=0.9$ and
  2.36~TeV with ALICE at LHC}'',
  \href{http://dx.doi.org/10.1140/epjc/s10052-010-1339-x}{{\em Eur. Phys. J. C}
  {\bfseries 68} (2010) 89--108},
  \href{http://arxiv.org/abs/1004.3034}{{\ttfamily arXiv:1004.3034 [hep-ex]}}.

\bibitem{ALICE:ppMidChMultpaper7000}
{\bfseries ALICE} Collaboration, K.~Aamodt {\em et~al.}, ``{Charged-particle
  multiplicity measurement in proton--proton collisions at
  $\sqrt{s}~=~7$~Te\kern-.1emV\xspace with ALICE at LHC}'',
  \href{http://dx.doi.org/10.1140/epjc/s10052-010-1350-2}{{\em Eur. Phys. J. C}
  {\bfseries 68} (2010) 345--354},
  \href{http://arxiv.org/abs/1004.3514}{{\ttfamily arXiv:1004.3514 [hep-ex]}}.

\bibitem{ALICE:ppMidChMultpaper900to8000}
{\bfseries ALICE} Collaboration, J.~Adam {\em et~al.}, ``{Charged-particle
  multiplicities in proton\textendash{}proton collisions at $\sqrt{s} = 0.9$ to
  8~TeV}'', \href{http://dx.doi.org/10.1140/epjc/s10052-016-4571-1}{{\em Eur.
  Phys. J. C} {\bfseries 77} (2017) 33},
  \href{http://arxiv.org/abs/1509.07541}{{\ttfamily arXiv:1509.07541
  [nucl-ex]}}.

\bibitem{ALICE:ppFrdChMultpaper}
{\bfseries ALICE} Collaboration, S.~Acharya {\em et~al.}, ``{Charged-particle
  multiplicity distributions over a wide pseudorapidity range in proton--proton
  collisions at $\sqrt{s}~=$~0.9, 7, and 8~TeV}'',
  \href{http://dx.doi.org/10.1140/epjc/s10052-017-5412-6}{{\em Eur. Phys. J. C}
  {\bfseries 77} (2017) 852}, \href{http://arxiv.org/abs/1708.01435}{{\ttfamily
  arXiv:1708.01435 [hep-ex]}}.

\bibitem{ALICE:pPbMidChdNdEtapaper5020MB}
{\bfseries ALICE} Collaboration, B.~Abelev {\em et~al.}, ``{Pseudorapidity
  density of charged particles in p--Pb collisions at
  $\sqrt{s_{\mathrm{NN}}}~=~5.02$~Te\kern-.1emV\xspace}'',
  \href{http://dx.doi.org/10.1103/PhysRevLett.110.032301}{{\em Phys. Rev.
  Lett.} {\bfseries 110} (2013) 032301},
  \href{http://arxiv.org/abs/1210.3615}{{\ttfamily arXiv:1210.3615 [nucl-ex]}}.

\bibitem{ALICE:pPbMidChdNdEtapaper5020Cent}
{\bfseries ALICE} Collaboration, J.~Adam {\em et~al.}, ``{Centrality dependence
  of particle production in p--Pb collisions at
  $\sqrt{s_{\mathrm{NN}}}~=~5.02$~Te\kern-.1emV\xspace}'',
  \href{http://dx.doi.org/10.1103/PhysRevC.91.064905}{{\em Phys. Rev. C}
  {\bfseries 91} (2015) 064905},
  \href{http://arxiv.org/abs/1412.6828}{{\ttfamily arXiv:1412.6828 [nucl-ex]}}.

\bibitem{ALICE:pPbMidChdNdEtapaper8160}
{\bfseries ALICE} Collaboration, S.~Acharya {\em et~al.}, ``{Charged-particle
  pseudorapidity density at midrapidity in p--Pb collisions at
  $\sqrt{s_{\mathrm{NN}}}~=~8.16$~Te\kern-.1emV\xspace}'',
  \href{http://dx.doi.org/10.1140/epjc/s10052-019-6801-9}{{\em Eur. Phys. J. C}
  {\bfseries 79} (2019) 307}, \href{http://arxiv.org/abs/1812.01312}{{\ttfamily
  arXiv:1812.01312 [nucl-ex]}}.

\bibitem{ALICE:ppMultDependent}
{\bfseries ALICE} Collaboration, S.~Acharya {\em et~al.}, ``{Pseudorapidity
  distributions of charged particles as a function of mid- and forward rapidity
  multiplicities in pp collisions at $\sqrt{s}~=$~5.02, 7 and 13~TeV}'',
  \href{http://dx.doi.org/10.1140/epjc/s10052-021-09349-5}{{\em Eur. Phys. J.
  C} {\bfseries 81} (2021) 630},
  \href{http://arxiv.org/abs/2009.09434}{{\ttfamily arXiv:2009.09434
  [nucl-ex]}}.

\bibitem{ALICE:MBMidrapidity2760}
{\bfseries ALICE} Collaboration, K.~Aamodt {\em et~al.}, ``{Charged-particle
  multiplicity density at midrapidity in central Pb--Pb collisions at
  $\sqrt{s_{\mathrm{NN}}}~=~2.76$~Te\kern-.1emV\xspace}'',
  \href{http://dx.doi.org/10.1103/PhysRevLett.105.252301}{{\em Phys. Rev.
  Lett.} {\bfseries 105} (2010) 252301},
  \href{http://arxiv.org/abs/1011.3916}{{\ttfamily arXiv:1011.3916 [nucl-ex]}}.

\bibitem{ALICE:PbPb_CentMidrapidity2760}
{\bfseries ALICE} Collaboration, K.~Aamodt {\em et~al.}, ``{Centrality
  dependence of the charged-particle multiplicity density at midrapidity in
  Pb--Pb collisions at $\sqrt{s_{\mathrm{NN}}}~=~2.76$~Te\kern-.1emV\xspace}'',
  \href{http://dx.doi.org/10.1103/PhysRevLett.106.032301}{{\em Phys. Rev.
  Lett.} {\bfseries 106} (2011) 032301},
  \href{http://arxiv.org/abs/1012.1657}{{\ttfamily arXiv:1012.1657 [nucl-ex]}}.

\bibitem{ALICE:PbPb_CentFrdrapidity2760}
{\bfseries ALICE} Collaboration, J.~Adam {\em et~al.}, ``{Centrality evolution
  of the charged-particle pseudorapidity density over a broad pseudorapidity
  range in Pb--Pb collisions at
  $\sqrt{s_{\mathrm{NN}}}~=~2.76$~Te\kern-.1emV\xspace}'',
  \href{http://dx.doi.org/10.1016/j.physletb.2015.12.082}{{\em Phys. Lett. B}
  {\bfseries 754} (2016) 373--385},
  \href{http://arxiv.org/abs/1509.07299}{{\ttfamily arXiv:1509.07299
  [nucl-ex]}}.

\bibitem{ALICE:PbPbsatellite}
{\bfseries ALICE} Collaboration, E.~Abbas {\em et~al.}, ``{Centrality
  dependence of the pseudorapidity density distribution for charged particles
  in Pb--Pb collisions at
  $\sqrt{s_{\mathrm{NN}}}~=~2.76$~Te\kern-.1emV\xspace}'',
  \href{http://dx.doi.org/10.1016/j.physletb.2013.09.022}{{\em Phys. Lett. B}
  {\bfseries 726} (2013) 610--622},
  \href{http://arxiv.org/abs/1304.0347}{{\ttfamily arXiv:1304.0347 [nucl-ex]}}.

\bibitem{ALICE:PbPb_CentMidrapidity5020}
{\bfseries ALICE} Collaboration, J.~Adam {\em et~al.}, ``{Centrality dependence
  of the charged-particle multiplicity density at midrapidity in Pb--Pb
  collisions at $\sqrt{s_{\mathrm{NN}}}~=~5.02$~Te\kern-.1emV\xspace}'',
  \href{http://dx.doi.org/10.1103/PhysRevLett.116.222302}{{\em Phys. Rev.
  Lett.} {\bfseries 116} (2016) 222302},
  \href{http://arxiv.org/abs/1512.06104}{{\ttfamily arXiv:1512.06104
  [nucl-ex]}}.

\bibitem{ALICE:PbPb_CentFrdrapidity5020}
{\bfseries ALICE} Collaboration, J.~Adam {\em et~al.}, ``{Centrality dependence
  of the pseudorapidity density distribution for charged particles in Pb--Pb
  collisions at $\sqrt{s_{\mathrm{NN}}}~=~5.02$~Te\kern-.1emV\xspace}'',
  \href{http://dx.doi.org/10.1016/j.physletb.2017.07.017}{{\em Phys. Lett. B}
  {\bfseries 772} (2017) 567--577},
  \href{http://arxiv.org/abs/1612.08966}{{\ttfamily arXiv:1612.08966
  [nucl-ex]}}.

\bibitem{ATLAS:ChPrMidpp900}
{\bfseries ATLAS} Collaboration, G.~Aad {\em et~al.}, ``{Charged-particle
  multiplicities in pp interactions at $\sqrt{s}~=~900$~Ge\kern-.1emV\xspace
  measured with the ATLAS detector at the LHC}'',
  \href{http://dx.doi.org/10.1016/j.physletb.2010.03.064}{{\em Phys. Lett. B}
  {\bfseries 688} (2010) 21--42},
  \href{http://arxiv.org/abs/1003.3124}{{\ttfamily arXiv:1003.3124 [hep-ex]}}.

\bibitem{ATLAS:ChPrMidpp900_2760_7000}
{\bfseries ATLAS} Collaboration, G.~Aad {\em et~al.}, ``{Charged-particle
  multiplicities in pp interactions measured with the ATLAS detector at the
  LHC}'', \href{http://dx.doi.org/10.1088/1367-2630/13/5/053033}{{\em New J.
  Phys.} {\bfseries 13} (2011) 053033},
  \href{http://arxiv.org/abs/1012.5104}{{\ttfamily arXiv:1012.5104 [hep-ex]}}.

\bibitem{ATLAS:ChPrMidpp8000}
{\bfseries ATLAS} Collaboration, G.~Aad {\em et~al.}, ``{Charged-particle
  distributions in pp interactions at $\sqrt{s}~=~8$~Te\kern-.1emV\xspace
  measured with the ATLAS detector}'',
  \href{http://dx.doi.org/10.1140/epjc/s10052-016-4203-9}{{\em Eur. Phys. J. C}
  {\bfseries 76} (2016) 403}, \href{http://arxiv.org/abs/1603.02439}{{\ttfamily
  arXiv:1603.02439 [hep-ex]}}.

\bibitem{ATLAS:ChPrMidpPb5020}
{\bfseries ATLAS} Collaboration, G.~Aad {\em et~al.}, ``{Measurement of the
  centrality dependence of the charged-particle pseudorapidity distribution in
  p--Pb collisions at $\sqrt{s_{\mathrm{NN}}}~=~5.02$~Te\kern-.1emV\xspace with
  the ATLAS detector}'',
  \href{http://dx.doi.org/10.1140/epjc/s10052-016-4002-3}{{\em Eur. Phys. J. C}
  {\bfseries 76} (2016) 199}, \href{http://arxiv.org/abs/1508.00848}{{\ttfamily
  arXiv:1508.00848 [hep-ex]}}.

\bibitem{ATLAS:PbPb_CentMidrapidity2760}
{\bfseries ATLAS} Collaboration, G.~Aad {\em et~al.}, ``{Measurement of the
  centrality dependence of the charged-particle pseudorapidity distribution in
  Pb--Pb collisions at $\sqrt{s_{\mathrm{NN}}}~=~2.76$~Te\kern-.1emV\xspace
  with the ATLAS detector}'',
  \href{http://dx.doi.org/10.1016/j.physletb.2012.02.045}{{\em Phys. Lett. B}
  {\bfseries 710} (2012) 363--382},
  \href{http://arxiv.org/abs/1108.6027}{{\ttfamily arXiv:1108.6027 [hep-ex]}}.

\bibitem{CMS:ChPrMidpp900_2360_7000}
{\bfseries CMS} Collaboration, V.~Khachatryan {\em et~al.}, ``{Charged-particle
  multiplicities in pp interactions at $\sqrt{s}~=$~0.9, 2.36, and 7~TeV}'',
  \href{http://dx.doi.org/10.1007/JHEP01(2011)079}{{\em JHEP} {\bfseries 01}
  (2011) 079}, \href{http://arxiv.org/abs/1011.5531}{{\ttfamily arXiv:1011.5531
  [hep-ex]}}.

\bibitem{CMS:ChPrMidpp13}
{\bfseries CMS} Collaboration, V.~Khachatryan {\em et~al.}, ``{Pseudorapidity
  distribution of charged hadrons in proton--proton collisions at
  $\sqrt{s}~=~13$~Te\kern-.1emV\xspace}'',
  \href{http://dx.doi.org/10.1016/j.physletb.2015.10.004}{{\em Phys. Lett. B}
  {\bfseries 751} (2015) 143--163},
  \href{http://arxiv.org/abs/1507.05915}{{\ttfamily arXiv:1507.05915
  [hep-ex]}}.

\bibitem{CMS:ChPrMidpPb5TeV}
{\bfseries CMS} Collaboration, A.~M.~Sirunyan {\em et~al.}, ``{Pseudorapidity
  distributions of charged hadrons in p--Pb collisions at
  $\sqrt{s_{\mathrm{NN}}}$ = 5.02 and 8.16~TeV}'',
  \href{http://dx.doi.org/10.1007/JHEP01(2018)045}{{\em JHEP} {\bfseries 01}
  (2018) 045}, \href{http://arxiv.org/abs/1710.09355}{{\ttfamily
  arXiv:1710.09355 [hep-ex]}}.

\bibitem{CMS:ChPrMidPbPb2760}
{\bfseries CMS} Collaboration, S.~Chatrchyan {\em et~al.}, ``{Dependence on
  pseudorapidity and centrality of charged hadron production in Pb--Pb
  collisions at $\sqrt{s_{\mathrm{NN}}}$ = 2.76~TeV}'',
  \href{http://dx.doi.org/10.1007/JHEP08(2011)141}{{\em JHEP} {\bfseries 08}
  (2011) 141}, \href{http://arxiv.org/abs/1107.4800}{{\ttfamily arXiv:1107.4800
  [nucl-ex]}}.

\bibitem{LHCb:ChPrFrdpp7000}
{\bfseries LHCb} Collaboration, R.~Aaij {\em et~al.}, ``{Measurement of
  charged-particle multiplicities and densities in pp collisions at
  $\sqrt{s}~=~7$~Te\kern-.1emV\xspace in the forward region}'',
  \href{http://dx.doi.org/10.1140/epjc/s10052-014-2888-1}{{\em Eur. Phys. J. C}
  {\bfseries 74} (2014) 2888}, \href{http://arxiv.org/abs/1402.4430}{{\ttfamily
  arXiv:1402.4430 [hep-ex]}}.

\bibitem{ALICE:Exp}
{\bfseries ALICE} Collaboration, K.~Aamodt {\em et~al.}, ``{The ALICE
  experiment at the CERN LHC}'',
  \href{http://dx.doi.org/10.1088/1748-0221/3/08/S08002}{{\em JINST} {\bfseries
  3} (2008) S08002}.

\bibitem{ALICE:performance}
{\bfseries ALICE} Collaboration, B.~B. Abelev {\em et~al.}, ``{Performance of
  the ALICE experiment at the CERN LHC}'',
  \href{http://dx.doi.org/10.1142/S0217751X14300440}{{\em Int. J. Mod. Phys. A}
  {\bfseries 29} (2014) 1430044},
  \href{http://arxiv.org/abs/1402.4476}{{\ttfamily arXiv:1402.4476 [nucl-ex]}}.

\bibitem{ALICE:PMDtdr1}
{\bfseries ALICE} Collaboration, G.~Dellacasa {\em et~al.}, ``{ALICE technical
  design report: Photon multiplicity detector (PMD)}'', CERN-LHCC-99-032.
  \url{http://cds.cern.ch/record/451099}.

\bibitem{ALICE:PMDtdr2}
{\bfseries ALICE} Collaboration, P.~Cortese {\em et~al.}, ``{ALICE: addendum to
  the technical design report of the photon multiplicity detector (PMD)}'',
  CERN-LHCC-2003-038. \url{https://cds.cern.ch/record/642177}.

\bibitem{ALICE:ITStdr}
{\bfseries ALICE} Collaboration, K.~Aamodt {\em et~al.}, ``{Alignment of the
  ALICE Inner Tracking System with cosmic-ray tracks}'',
  \href{http://dx.doi.org/10.1088/1748-0221/5/03/P03003}{{\em JINST} {\bfseries
  5} (2010) P03003}, \href{http://arxiv.org/abs/1001.0502}{{\ttfamily
  arXiv:1001.0502 [physics.ins-det]}}.

\bibitem{ALICE:FrdDettdr}
{\bfseries ALICE} Collaboration, P.~Cortese {\em et~al.}, ``{ALICE forward
  detectors: FMD, TO and VO: Technical Design Report}'', CERN-LHCC-2004-025.
  \url{https://cds.cern.ch/record/781854}.

\bibitem{ALICE:V0performance}
{\bfseries ALICE} Collaboration, E.~Abbas {\em et~al.}, ``{Performance of the
  ALICE VZERO system}'',
  \href{http://dx.doi.org/10.1088/1748-0221/8/10/P10016}{{\em JINST} {\bfseries
  8} (2013) P10016}, \href{http://arxiv.org/abs/1306.3130}{{\ttfamily
  arXiv:1306.3130 [nucl-ex]}}.

\bibitem{ALICE:ZDCtdr}
{\bfseries ALICE} Collaboration, M.~Gallio, W.~Klempt, L.~Leistam, J.~De~Groot,
  and J.~Schukraft, ``{ALICE Zero Degree Calorimeter (ZDC): Technical Design
  Report}'', CERN-LHCC-99-005. \url{https://cds.cern.ch/record/381433}.

\bibitem{ALICE:CordSys}
{\bfseries ALICE} Collaboration, L.~Betev {\em et~al.}, ``{Definition of the
  ALICE coordinate system and basic rules for sub-detector components
  numbering}'',.
  \url{https://edms.cern.ch/ui/#!master/navigator/document?D:1020137949:1020137949:subDocs}.

\bibitem{GlauberModel}
M.~L. Miller, K.~Reygers, S.~J. Sanders, and P.~Steinberg, ``{Glauber modeling
  in high energy nuclear collisions}'',
  \href{http://dx.doi.org/10.1146/annurev.nucl.57.090506.123020}{{\em Ann. Rev.
  Nucl. Part. Sci.} {\bfseries 57} (2007) 205--243},
  \href{http://arxiv.org/abs/nucl-ex/0701025}{{\ttfamily arXiv:nucl-ex/0701025
  [nucl-ex]}}.

\bibitem{dEnterria:2020dwq}
D.~d'Enterria and C.~Loizides, ``{Progress in the Glauber model at collider
  energies}'', \href{http://dx.doi.org/10.1146/annurev-nucl-102419-060007}{{\em
  Ann. Rev. Nucl. Part. Sci.} {\bfseries 71} (2021) 315--344},
  \href{http://arxiv.org/abs/2011.14909}{{\ttfamily arXiv:2011.14909
  [hep-ph]}}.

\bibitem{SlowNucleons}
F.~Sikler, ``{Centrality control of hadron nucleus interactions by detection of
  slow nucleons}'', \href{http://arxiv.org/abs/hep-ph/0304065}{{\ttfamily
  arXiv:hep-ph/0304065}}.

\bibitem{ALICE:PhysicsPerfReportII}
B.~Alessandro {\em et~al.}, ``{ALICE: Physics Performance Report, Volume II}'',
  \href{http://dx.doi.org/10.1088/0954-3899/32/10/001}{{\em Journal of Physics
  G: Nuclear and Particle Physics} {\bfseries 32} (Sep, 2006) 1295--2040}.

\bibitem{hijing}
X.-N. Wang and M.~Gyulassy, ``{HIJING: A Monte Carlo model for multiple jet
  production in pp, p--A and A--A collisions}'',
  \href{http://dx.doi.org/10.1103/PhysRevD.44.3501}{{\em Phys. Rev. D}
  {\bfseries 44} (1991) 3501--3516}.

\bibitem{dpmjet}
S.~Roesler, R.~Engel, and J.~Ranft,
  \href{http://dx.doi.org/10.1007/978-3-642-18211-2_166}{``{The Monte Carlo
  event generator DPMJET-III}'',} in {\em {International Conference on Advanced
  Monte Carlo for Radiation Physics, Particle Transport Simulation and
  Applications (MC 2000)}}, pp.~1033--1038.
\newblock 12, 2000.
\newblock \href{http://arxiv.org/abs/hep-ph/0012252}{{\ttfamily
  arXiv:hep-ph/0012252 [hep-ph]}}.

\bibitem{pythia8_monash}
P.~Skands, S.~Carrazza, and J.~Rojo, ``{Tuning PYTHIA 8.1: the Monash 2013
  Tune}'', \href{http://dx.doi.org/10.1140/epjc/s10052-014-3024-y}{{\em Eur.
  Phys. J. C} {\bfseries 74} (2014) 3024},
  \href{http://arxiv.org/abs/1404.5630}{{\ttfamily arXiv:1404.5630 [hep-ph]}}.

\bibitem{EPOS_LHC}
T.~Pierog {\em et~al.}, ``{EPOS LHC: Test of collective hadronization with data
  measured at the CERN Large Hadron Collider}'',
  \href{http://dx.doi.org/10.1103/PhysRevC.92.034906}{{\em Phys. Rev. C}
  {\bfseries 92} (2015) 034906},
  \href{http://arxiv.org/abs/1306.0121}{{\ttfamily arXiv:1306.0121 [hep-ph]}}.

\bibitem{geant3}
R.~Brun {\em et~al.}, \href{http://dx.doi.org/10.17181/CERN.MUHF.DMJ1}{{\em
  {``GEANT: Detector Description and Simulation Tool"}}}.
\newblock CERN Program Library. CERN, Geneva, 1993.
\newblock \url{http://cds.cern.ch/record/1082634}.

\bibitem{aliroot}
R.~Brun {\em et~al.}, ``{Computing in ALICE}'',
  \href{http://dx.doi.org/10.1016/S0168-9002(03)00440-6}{{\em Nucl. Instrum.
  Meth. A} {\bfseries 502} (2003) 339--346}.

\bibitem{pythia8pt3}
C.~Bierlich {\em et~al.}, ``{A comprehensive guide to the physics and usage of
  PYTHIA 8.3}'', \href{http://arxiv.org/abs/2203.11601}{{\ttfamily
  arXiv:2203.11601 [hep-ph]}}.

\bibitem{MPI_pythia}
T.~Sj$\ddot{\rm o}$strand and M.~van Zijl, ``{A multiple-interaction model for
  the event structure in hadron collisions}'',
  \href{http://dx.doi.org/10.1103/PhysRevD.36.2019}{{\em Phys. Rev. D}
  {\bfseries 36} (1987) 2019--2041}.

\bibitem{pythia8pt2}
T.~Sj\"ostrand {\em et~al.}, ``{An introduction to PYTHIA 8.2}'',
  \href{http://dx.doi.org/10.1016/j.cpc.2015.01.024}{{\em Comput. Phys.
  Commun.} {\bfseries 191} (2015) 159--177},
  \href{http://arxiv.org/abs/1410.3012}{{\ttfamily arXiv:1410.3012 [hep-ph]}}.

\bibitem{Color_recon}
J.~R. Christiansen and P.~Z. Skands, ``{String Formation Beyond Leading
  Colour}'', \href{http://dx.doi.org/10.1007/JHEP08(2015)003}{{\em JHEP}
  {\bfseries 08} (2015) 003}, \href{http://arxiv.org/abs/1505.01681}{{\ttfamily
  arXiv:1505.01681 [hep-ph]}}.

\bibitem{lund_fritiof}
H.~Pi, ``{An Event generator for interactions between hadrons and nuclei:
  FRITIOF version 7.0}'',
  \href{http://dx.doi.org/10.1016/0010-4655(92)90082-A}{{\em Comput. Phys.
  Commun.} {\bfseries 71} (1992) 173--192}.

\bibitem{dpm}
A.~Capella, U.~Sukhatme, C.~I.~Tan, and J.~Tran Thanh~Van, ``{Dual parton
  model}'', \href{http://dx.doi.org/10.1016/0370-1573(94)90064-7}{{\em Phys.
  Rept.} {\bfseries 236} (1994) 225--329}.

\bibitem{regge_theory}
P.~D.~B. Collins, \href{http://dx.doi.org/10.1017/CBO9780511897603}{{\em {``An
  Introduction to Regge Theory and High-Energy Physics"}}}.
\newblock Cambridge Monographs on Mathematical Physics. Cambridge Univ. Press,
  Cambridge, UK, 5, 2009.

\bibitem{CDF_SD_1}
{\bfseries CDF} Collaboration, F.~Abe {\em et~al.}, ``{Measurement of $\mathrm
  {p\overline{p}}$ single diffraction dissociation at $\sqrt{s}~=$~546~GeV and
  1800 GeV}'', \href{http://dx.doi.org/10.1103/PhysRevD.50.5535}{{\em Phys.
  Rev. D} {\bfseries 50} (1994) 5535--5549}.

\bibitem{CDF_SD_2}
{\bfseries CDF} Collaboration, F.~Abe {\em et~al.}, ``{Observation of rapidity
  gaps in $\mathrm {p\overline{p}}$ collisions at 1.8~TeV}'',
  \href{http://dx.doi.org/10.1103/PhysRevLett.74.855}{{\em Phys. Rev. Lett.}
  {\bfseries 74} (1995) 855--859}.

\bibitem{ALICE:SD_DD_Xsections}
{\bfseries ALICE} Collaboration, B.~Abelev {\em et~al.}, ``{Measurement of
  inelastic, single- and double-diffraction cross sections in proton--proton
  collisions at the LHC with ALICE}'',
  \href{http://dx.doi.org/10.1140/epjc/s10052-013-2456-0}{{\em Eur. Phys. J. C}
  {\bfseries 73} (2013) 2456}, \href{http://arxiv.org/abs/1208.4968}{{\ttfamily
  arXiv:1208.4968 [hep-ex]}}.

\bibitem{BayesUnfolding}
G.~D'Agostini, ``{A multidimensional unfolding method based on Bayes'
  theorem}'',
  \href{http://dx.doi.org/https://doi.org/10.1016/0168-9002(95)00274-X}{{\em
  Nuclear Instruments and Methods in Physics Research Section A: Accelerators,
  Spectrometers, Detectors and Associated Equipment} {\bfseries 362} (1995)
  487--498}.

\bibitem{roounfold}
T.~Adye, ``{Unfolding algorithms and tests using RooUnfold}'',
  \href{http://arxiv.org/abs/1105.1160}{{\ttfamily arXiv:1105.1160
  [physics.data-an]}}.

\bibitem{SVDUnfolding}
A.~Hocker and V.~Kartvelishvili, ``{SVD approach to data unfolding}'',
  \href{http://dx.doi.org/10.1016/0168-9002(95)01478-0}{{\em Nucl. Instrum.
  Meth. A} {\bfseries 372} (1996) 469--481},
  \href{http://arxiv.org/abs/hep-ph/9509307}{{\ttfamily arXiv:hep-ph/9509307
  [hep-ph]}}.

\bibitem{pythia6_perugia2011}
P.~Z.~Skands, ``{Tuning Monte Carlo generators: The Perugia tunes}'',
  \href{http://dx.doi.org/10.1103/PhysRevD.82.074018}{{\em Phys. Rev. D}
  {\bfseries 82} (2010) 074018},
  \href{http://arxiv.org/abs/1005.3457}{{\ttfamily arXiv:1005.3457 [hep-ph]}}.

\bibitem{phojet}
F.~W. Bopp, R.~Engel, and J.~Ranft, ``{Rapidity gaps and the PHOJET Monte
  Carlo}'', in {\em {LAFEX International School on High-Energy Physics (LISHEP
  98) Session A: Particle Physics for High School Teachers - Session B:
  Advanced School in HEP - Session C: Workshop on Diffractive Physics}},
  pp.~729--741.
\newblock 3, 1998.
\newblock \href{http://arxiv.org/abs/hep-ph/9803437}{{\ttfamily
  arXiv:hep-ph/9803437 [hep-ph]}}.

\bibitem{ampt}
Z.-W. Lin, C.~M. Ko, B.-A. Li, B.~Zhang, and S.~Pal, ``{Multiphase transport
  model for relativistic heavy-ion collisions}'',
  \href{http://dx.doi.org/10.1103/PhysRevC.72.064901}{{\em Phys. Rev. C}
  {\bfseries 72} (2005) 064901},
  \href{http://arxiv.org/abs/nucl-th/0411110}{{\ttfamily arXiv:nucl-th/0411110
  [nucl-th]}}.

\bibitem{ALICE:MBChPrpp5TeV}
{\bfseries ALICE} Collaboration, ``{Pseudorapidity densities of charged
  particles with transverse momentum thresholds in pp collisions at $\sqrt{s} =
  5.02$ and $13$ TeV}'', \href{http://arxiv.org/abs/2211.15364}{{\ttfamily
  arXiv:2211.15364 [nucl-ex]}}.

\end{thebibliography}\endgroup

\newpage
\appendix

\section{Appendix}
\label{appendixResponse}
\begin{figure}[h!]
  \subfigure[]{
    \label{ResMatrix_AvgNgamma_pp}
    \includegraphics[scale=0.41]{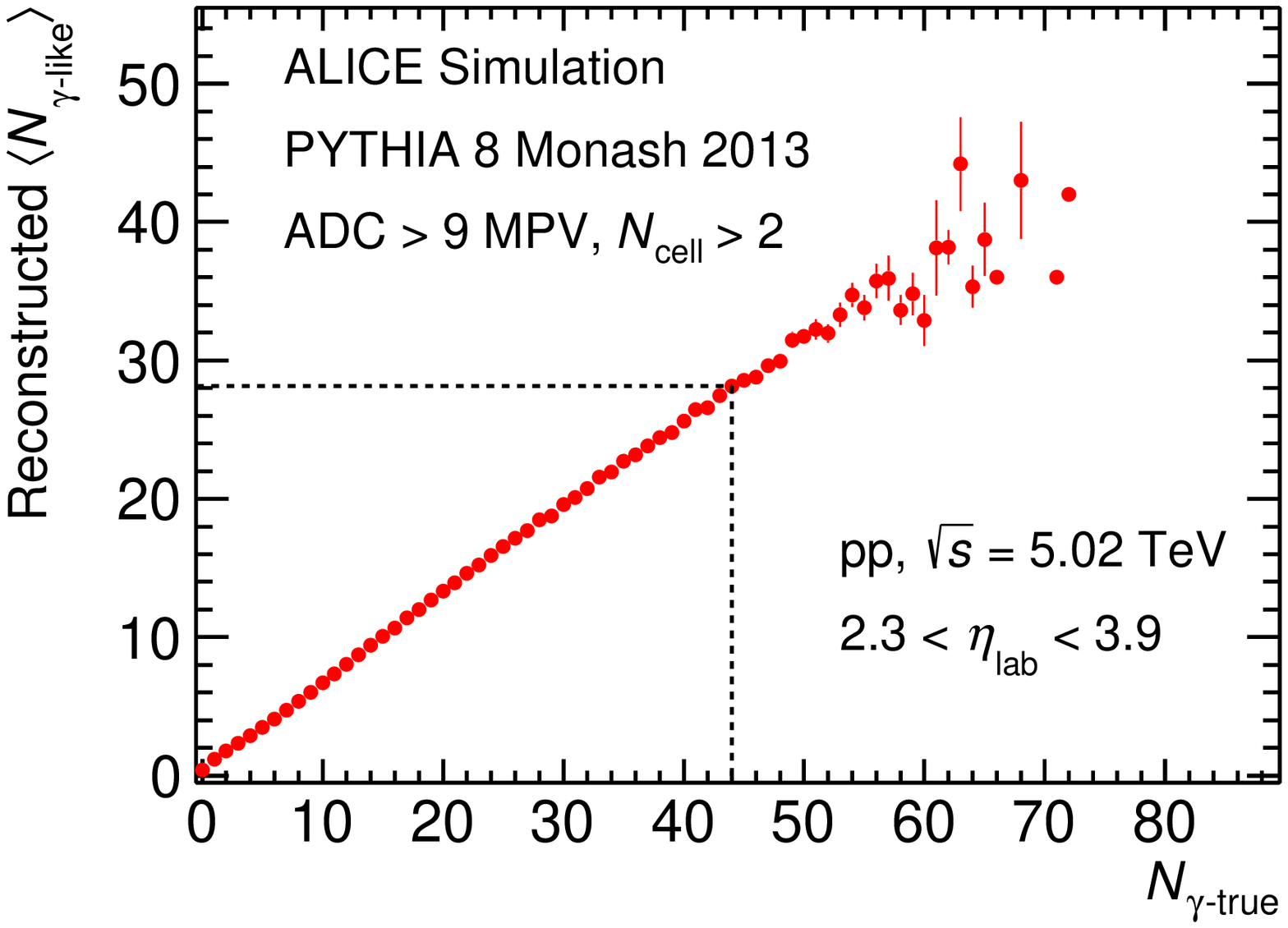}}
  \subfigure[]{
    \label{ResMatrix_SigmaNgamma_pp}
    \includegraphics[scale=0.41]{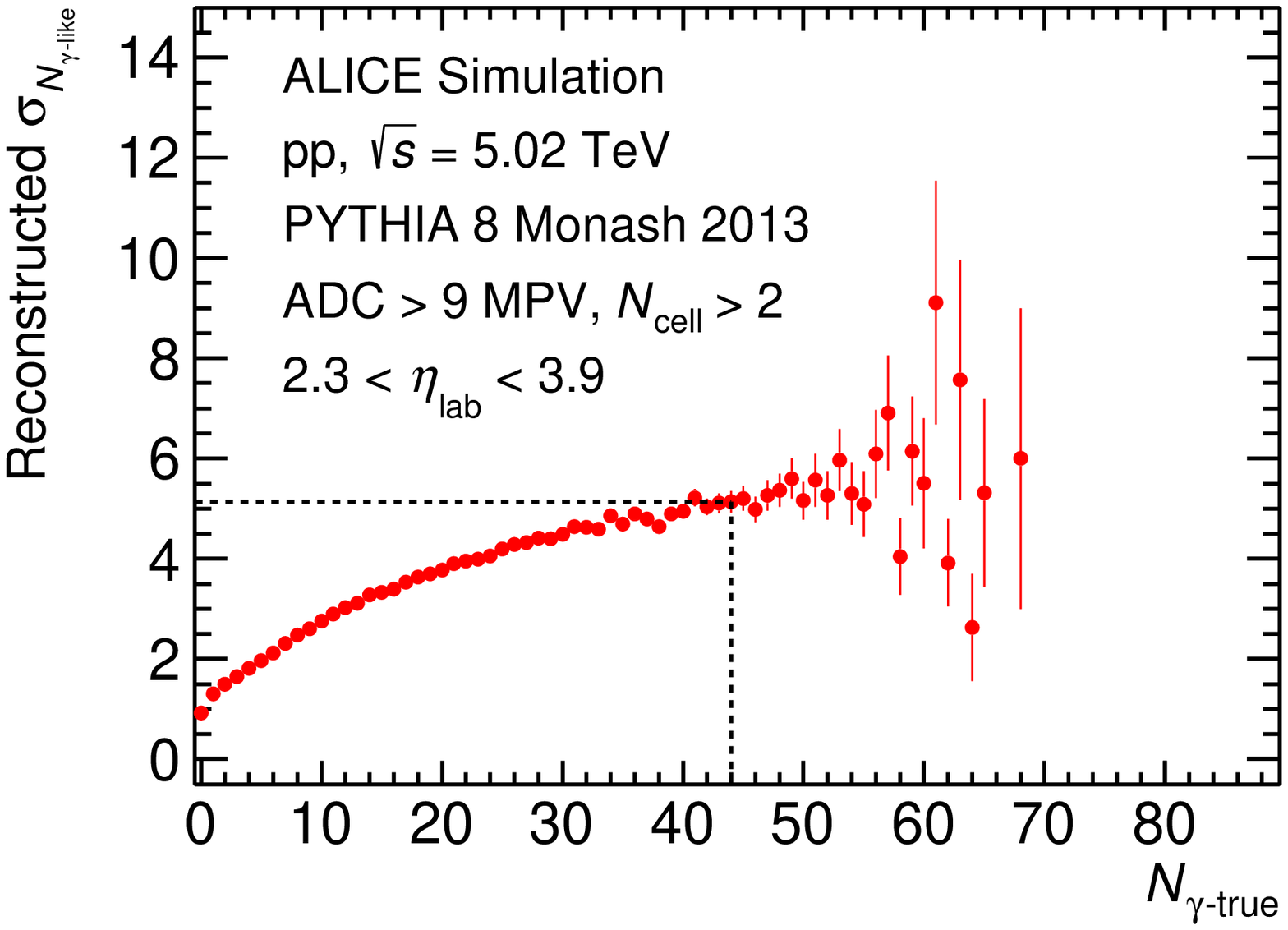}}
  \caption{The mean (a) and width (b) of the distribution of reconstructed photon
    multiplicity ($N_{\gamma\text-\rm like}$) as a function of true photon multiplicity
    ($N_{\gamma\text-\rm true}$) are presented for \pp collisions at \five. The dotted
    lines correspond to $N_{\gamma\text-\rm true}$ = 44 up to which the results are
    reported in the paper.}
  \label{ResMatrix_resolution_pp}
\end{figure}

\begin{figure}[h!]
  \subfigure[]{
    \label{ResMatrix_AvgNgamma_pPb}
    \includegraphics[scale=0.41]{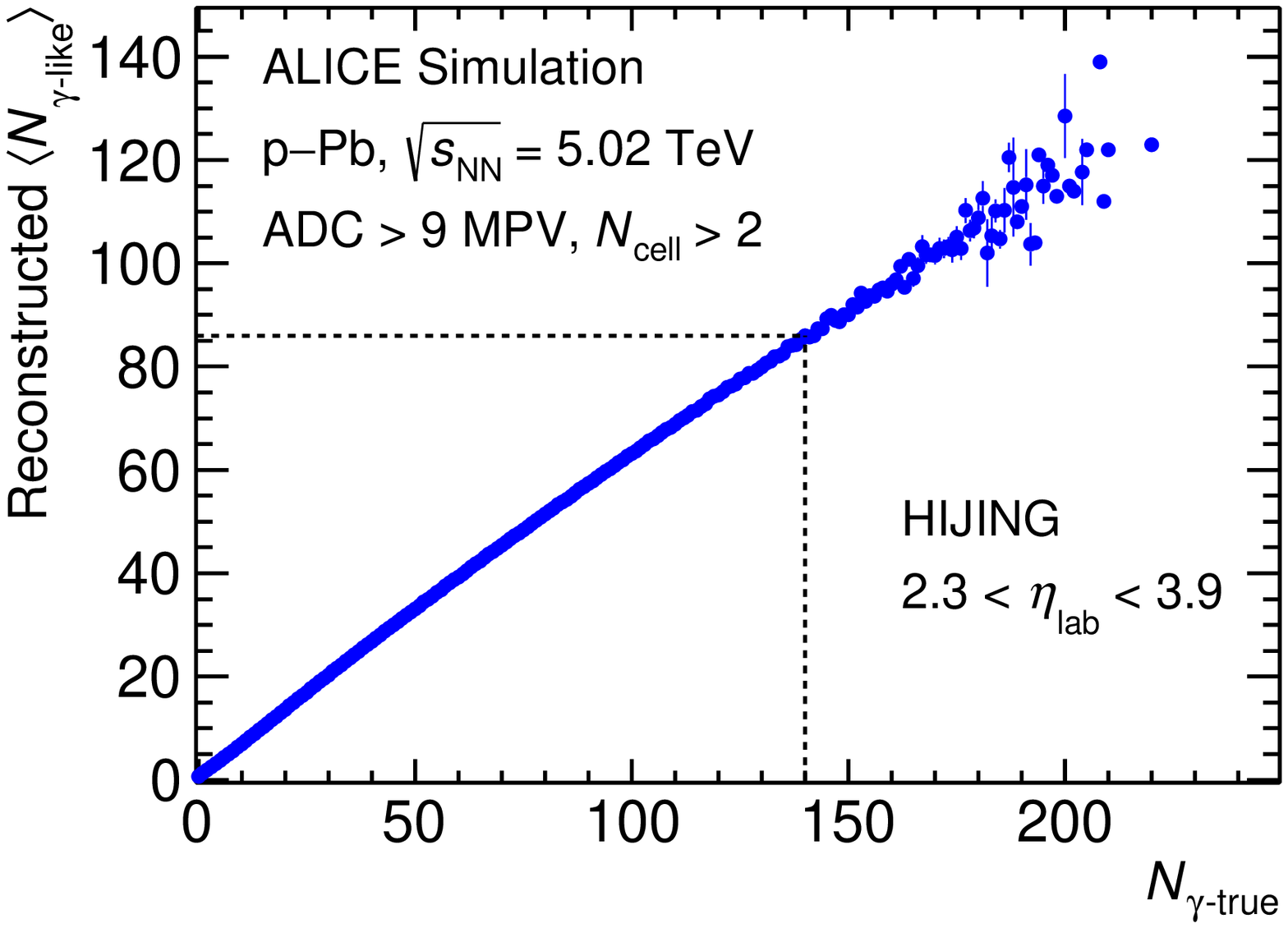}}
  \subfigure[]{
    \label{ResMatrix_SigmaNgamma_pPb}
    \includegraphics[scale=0.41]{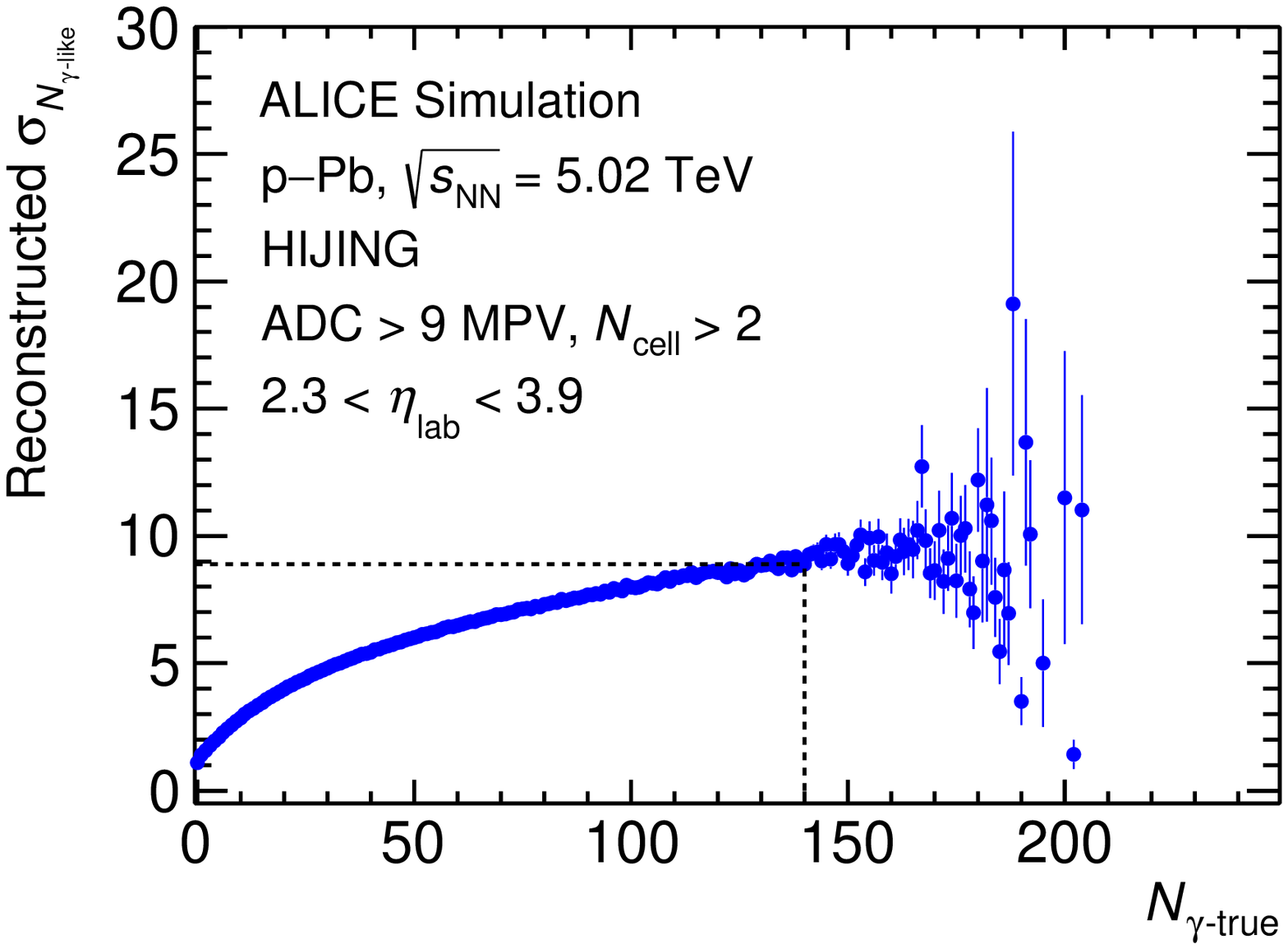}}
  \caption{The mean (a) and width (b) of the distribution of reconstructed photon
    multiplicity ($N_{\gamma\text-\rm like}$) as a function of true photon multiplicity
    ($N_{\gamma\text-\rm true}$) are presented for \pPb collisions at \fivenn. The dotted
    lines correspond to $N_{\gamma\text-\rm true}$ = 140 up to which the results are
    reported in the paper.}
  \label{ResMatrix_resolution_pPb}
\end{figure}

\clearpage

\section{The ALICE Collaboration}
\label{app:collab}
\begin{flushleft} 
\small

S.~Acharya\,\orcidlink{0000-0002-9213-5329}\,$^{\rm 125}$, 
D.~Adamov\'{a}\,\orcidlink{0000-0002-0504-7428}\,$^{\rm 86}$, 
A.~Adler$^{\rm 69}$, 
G.~Aglieri Rinella\,\orcidlink{0000-0002-9611-3696}\,$^{\rm 32}$, 
M.~Agnello\,\orcidlink{0000-0002-0760-5075}\,$^{\rm 29}$, 
N.~Agrawal\,\orcidlink{0000-0003-0348-9836}\,$^{\rm 50}$, 
Z.~Ahammed\,\orcidlink{0000-0001-5241-7412}\,$^{\rm 132}$, 
S.~Ahmad\,\orcidlink{0000-0003-0497-5705}\,$^{\rm 15}$, 
S.U.~Ahn\,\orcidlink{0000-0001-8847-489X}\,$^{\rm 70}$, 
I.~Ahuja\,\orcidlink{0000-0002-4417-1392}\,$^{\rm 37}$
% ...

\section*{Affiliation Notes}

$^{\rm I}$ Deceased\\
$^{\rm II}$ Also at: Max-Planck-Institut f\"{u}r Physik, Munich, Germany\\
$^{\rm III}$ Also at: Italian National Agency for New Technologies, Energy and Sustainable Economic Development (ENEA), Bologna, Italy\\
$^{\rm IV}$ Also at: Department of Applied Physics, Aligarh Muslim University, Aligarh, India\\
$^{\rm V}$ Also at: Institute of Theoretical Physics, University of Wroclaw, Poland\\
$^{\rm VI}$ Also at: An institution covered by a cooperation agreement with CERN\\

\section*{Collaboration Institutes}

$^{1}$ A.I. Alikhanyan National Science Laboratory (Yerevan Physics Institute) Foundation, Yerevan, Armenia\\
$^{2}$ AGH University of Science and Technology, Cracow, Poland\\
$^{3}$ Bogolyubov Institute for Theoretical Physics, National Academy of Sciences of Ukraine, Kiev, Ukraine\\
$^{4}$ Bose Institute, Department of Physics  and Centre for Astroparticle Physics and Space Science (CAPSS), Kolkata, India\\
$^{5}$ California Polytechnic State University, San Luis Obispo, California, United States\\
$^{6}$ Central China Normal University, Wuhan, China\\
$^{7}$ Centro de Aplicaciones Tecnol\'{o}gicas y Desarrollo Nuclear (CEADEN), Havana, Cuba\\
$^{8}$ Centro de Investigaci\'{o}n y de Estudios Avanzados (CINVESTAV), Mexico City and M\'{e}rida, Mexico\\
$^{9}$ Chicago State University, Chicago, Illinois, United States\\
$^{10}$ China Institute of Atomic Energy, Beijing, China\\
$^{11}$ Chungbuk National University, Cheongju, Republic of Korea\\
$^{12}$ Comenius University Bratislava, Faculty of Mathematics, Physics and Informatics, Bratislava, Slovak Republic\\
$^{13}$ COMSATS University Islamabad, Islamabad, Pakistan\\
$^{14}$ Creighton University, Omaha, Nebraska, United States\\
$^{15}$ Department of Physics, Aligarh Muslim University, Aligarh, India\\
$^{16}$ Department of Physics, Pusan National University, Pusan, Republic of Korea\\
$^{17}$ Department of Physics, Sejong University, Seoul, Republic of Korea\\
$^{18}$ Department of Physics, University of California, Berkeley, California, United States\\
$^{19}$ Department of Physics, University of Oslo, Oslo, Norway\\
$^{20}$ Department of Physics and Technology, University of Bergen, Bergen, Norway\\
$^{21}$ Dipartimento di Fisica, Universit\`{a} di Pavia, Pavia, Italy\\
$^{22}$ Dipartimento di Fisica dell'Universit\`{a} and Sezione INFN, Cagliari, Italy\\
$^{23}$ Dipartimento di Fisica dell'Universit\`{a} and Sezione INFN, Trieste, Italy\\
$^{24}$ Dipartimento di Fisica dell'Universit\`{a} and Sezione INFN, Turin, Italy\\
$^{25}$ Dipartimento di Fisica e Astronomia dell'Universit\`{a} and Sezione INFN, Bologna, Italy\\
$^{26}$ Dipartimento di Fisica e Astronomia dell'Universit\`{a} and Sezione INFN, Catania, Italy\\
$^{27}$ Dipartimento di Fisica e Astronomia dell'Universit\`{a} and Sezione INFN, Padova, Italy\\
$^{28}$ Dipartimento di Fisica `E.R.~Caianiello' dell'Universit\`{a} and Gruppo Collegato INFN, Salerno, Italy\\
$^{29}$ Dipartimento DISAT del Politecnico and Sezione INFN, Turin, Italy\\
$^{30}$ Dipartimento di Scienze MIFT, Universit\`{a} di Messina, Messina, Italy\\
$^{31}$ Dipartimento Interateneo di Fisica `M.~Merlin' and Sezione INFN, Bari, Italy\\
$^{32}$ European Organization for Nuclear Research (CERN), Geneva, Switzerland\\
$^{33}$ Faculty of Electrical Engineering, Mechanical Engineering and Naval Architecture, University of Split, Split, Croatia\\
$^{34}$ Faculty of Engineering and Science, Western Norway University of Applied Sciences, Bergen, Norway\\
$^{35}$ Faculty of Nuclear Sciences and Physical Engineering, Czech Technical University in Prague, Prague, Czech Republic\\
$^{36}$ Faculty of Physics, Sofia University, Sofia, Bulgaria\\
$^{37}$ Faculty of Science, P.J.~\v{S}af\'{a}rik University, Ko\v{s}ice, Slovak Republic\\
$^{38}$ Frankfurt Institute for Advanced Studies, Johann Wolfgang Goethe-Universit\"{a}t Frankfurt, Frankfurt, Germany\\
$^{39}$ Fudan University, Shanghai, China\\
$^{40}$ Gangneung-Wonju National University, Gangneung, Republic of Korea\\
$^{41}$ Gauhati University, Department of Physics, Guwahati, India\\
$^{42}$ Helmholtz-Institut f\"{u}r Strahlen- und Kernphysik, Rheinische Friedrich-Wilhelms-Universit\"{a}t Bonn, Bonn, Germany\\
$^{43}$ Helsinki Institute of Physics (HIP), Helsinki, Finland\\
$^{44}$ High Energy Physics Group,  Universidad Aut\'{o}noma de Puebla, Puebla, Mexico\\
$^{45}$ Horia Hulubei National Institute of Physics and Nuclear Engineering, Bucharest, Romania\\
$^{46}$ Indian Institute of Technology Bombay (IIT), Mumbai, India\\
$^{47}$ Indian Institute of Technology Indore, Indore, India\\
$^{48}$ INFN, Laboratori Nazionali di Frascati, Frascati, Italy\\
$^{49}$ INFN, Sezione di Bari, Bari, Italy\\
$^{50}$ INFN, Sezione di Bologna, Bologna, Italy\\
$^{51}$ INFN, Sezione di Cagliari, Cagliari, Italy\\
$^{52}$ INFN, Sezione di Catania, Catania, Italy\\
$^{53}$ INFN, Sezione di Padova, Padova, Italy\\
$^{54}$ INFN, Sezione di Pavia, Pavia, Italy\\
$^{55}$ INFN, Sezione di Torino, Turin, Italy\\
$^{56}$ INFN, Sezione di Trieste, Trieste, Italy\\
$^{57}$ Inha University, Incheon, Republic of Korea\\
$^{58}$ Institute for Gravitational and Subatomic Physics (GRASP), Utrecht University/Nikhef, Utrecht, Netherlands\\
$^{59}$ Institute of Experimental Physics, Slovak Academy of Sciences, Ko\v{s}ice, Slovak Republic\\
$^{60}$ Institute of Physics, Homi Bhabha National Institute, Bhubaneswar, India\\
$^{61}$ Institute of Physics of the Czech Academy of Sciences, Prague, Czech Republic\\
$^{62}$ Institute of Space Science (ISS), Bucharest, Romania\\
$^{63}$ Institut f\"{u}r Kernphysik, Johann Wolfgang Goethe-Universit\"{a}t Frankfurt, Frankfurt, Germany\\
$^{64}$ Instituto de Ciencias Nucleares, Universidad Nacional Aut\'{o}noma de M\'{e}xico, Mexico City, Mexico\\
$^{65}$ Instituto de F\'{i}sica, Universidade Federal do Rio Grande do Sul (UFRGS), Porto Alegre, Brazil\\
$^{66}$ Instituto de F\'{\i}sica, Universidad Nacional Aut\'{o}noma de M\'{e}xico, Mexico City, Mexico\\
$^{67}$ iThemba LABS, National Research Foundation, Somerset West, South Africa\\
$^{68}$ Jeonbuk National University, Jeonju, Republic of Korea\\
$^{69}$ Johann-Wolfgang-Goethe Universit\"{a}t Frankfurt Institut f\"{u}r Informatik, Fachbereich Informatik und Mathematik, Frankfurt, Germany\\
$^{70}$ Korea Institute of Science and Technology Information, Daejeon, Republic of Korea\\
$^{71}$ KTO Karatay University, Konya, Turkey\\
$^{72}$ Laboratoire de Physique des 2 Infinis, Ir\`{e}ne Joliot-Curie, Orsay, France\\
$^{73}$ Laboratoire de Physique Subatomique et de Cosmologie, Universit\'{e} Grenoble-Alpes, CNRS-IN2P3, Grenoble, France\\
$^{74}$ Lawrence Berkeley National Laboratory, Berkeley, California, United States\\
$^{75}$ Lund University Department of Physics, Division of Particle Physics, Lund, Sweden\\
$^{76}$ Nagasaki Institute of Applied Science, Nagasaki, Japan\\
$^{77}$ Nara Women{'}s University (NWU), Nara, Japan\\
$^{78}$ National and Kapodistrian University of Athens, School of Science, Department of Physics , Athens, Greece\\
$^{79}$ National Centre for Nuclear Research, Warsaw, Poland\\
$^{80}$ National Institute of Science Education and Research, Homi Bhabha National Institute, Jatni, India\\
$^{81}$ National Nuclear Research Center, Baku, Azerbaijan\\
$^{82}$ National Research and Innovation Agency - BRIN, Jakarta, Indonesia\\
$^{83}$ Niels Bohr Institute, University of Copenhagen, Copenhagen, Denmark\\
$^{84}$ Nikhef, National institute for subatomic physics, Amsterdam, Netherlands\\
$^{85}$ Nuclear Physics Group, STFC Daresbury Laboratory, Daresbury, United Kingdom\\
$^{86}$ Nuclear Physics Institute of the Czech Academy of Sciences, Husinec-\v{R}e\v{z}, Czech Republic\\
$^{87}$ Oak Ridge National Laboratory, Oak Ridge, Tennessee, United States\\
$^{88}$ Ohio State University, Columbus, Ohio, United States\\
$^{89}$ Physics department, Faculty of science, University of Zagreb, Zagreb, Croatia\\
$^{90}$ Physics Department, Panjab University, Chandigarh, India\\
$^{91}$ Physics Department, University of Jammu, Jammu, India\\
$^{92}$ Physics Program and International Institute for Sustainability with Knotted Chiral Meta Matter (SKCM2), Hiroshima University, Hiroshima, Japan\\
$^{93}$ Physikalisches Institut, Eberhard-Karls-Universit\"{a}t T\"{u}bingen, T\"{u}bingen, Germany\\
$^{94}$ Physikalisches Institut, Ruprecht-Karls-Universit\"{a}t Heidelberg, Heidelberg, Germany\\
$^{95}$ Physik Department, Technische Universit\"{a}t M\"{u}nchen, Munich, Germany\\
$^{96}$ Politecnico di Bari and Sezione INFN, Bari, Italy\\
$^{97}$ Research Division and ExtreMe Matter Institute EMMI, GSI Helmholtzzentrum f\"ur Schwerionenforschung GmbH, Darmstadt, Germany\\
$^{98}$ Saga University, Saga, Japan\\
$^{99}$ Saha Institute of Nuclear Physics, Homi Bhabha National Institute, Kolkata, India\\
$^{100}$ School of Physics and Astronomy, University of Birmingham, Birmingham, United Kingdom\\
$^{101}$ Secci\'{o}n F\'{\i}sica, Departamento de Ciencias, Pontificia Universidad Cat\'{o}lica del Per\'{u}, Lima, Peru\\
$^{102}$ Stefan Meyer Institut f\"{u}r Subatomare Physik (SMI), Vienna, Austria\\
$^{103}$ SUBATECH, IMT Atlantique, Nantes Universit\'{e}, CNRS-IN2P3, Nantes, France\\
$^{104}$ Sungkyunkwan University, Suwon City, Republic of Korea\\
$^{105}$ Suranaree University of Technology, Nakhon Ratchasima, Thailand\\
$^{106}$ Technical University of Ko\v{s}ice, Ko\v{s}ice, Slovak Republic\\
$^{107}$ The Henryk Niewodniczanski Institute of Nuclear Physics, Polish Academy of Sciences, Cracow, Poland\\
$^{108}$ The University of Texas at Austin, Austin, Texas, United States\\
$^{109}$ Universidad Aut\'{o}noma de Sinaloa, Culiac\'{a}n, Mexico\\
$^{110}$ Universidade de S\~{a}o Paulo (USP), S\~{a}o Paulo, Brazil\\
$^{111}$ Universidade Estadual de Campinas (UNICAMP), Campinas, Brazil\\
$^{112}$ Universidade Federal do ABC, Santo Andre, Brazil\\
$^{113}$ University of Cape Town, Cape Town, South Africa\\
$^{114}$ University of Houston, Houston, Texas, United States\\
$^{115}$ University of Jyv\"{a}skyl\"{a}, Jyv\"{a}skyl\"{a}, Finland\\
$^{116}$ University of Kansas, Lawrence, Kansas, United States\\
$^{117}$ University of Liverpool, Liverpool, United Kingdom\\
$^{118}$ University of Science and Technology of China, Hefei, China\\
$^{119}$ University of South-Eastern Norway, Kongsberg, Norway\\
$^{120}$ University of Tennessee, Knoxville, Tennessee, United States\\
$^{121}$ University of the Witwatersrand, Johannesburg, South Africa\\
$^{122}$ University of Tokyo, Tokyo, Japan\\
$^{123}$ University of Tsukuba, Tsukuba, Japan\\
$^{124}$ University Politehnica of Bucharest, Bucharest, Romania\\
$^{125}$ Universit\'{e} Clermont Auvergne, CNRS/IN2P3, LPC, Clermont-Ferrand, France\\
$^{126}$ Universit\'{e} de Lyon, CNRS/IN2P3, Institut de Physique des 2 Infinis de Lyon, Lyon, France\\
$^{127}$ Universit\'{e} de Strasbourg, CNRS, IPHC UMR 7178, F-67000 Strasbourg, France, Strasbourg, France\\
$^{128}$ Universit\'{e} Paris-Saclay Centre d'Etudes de Saclay (CEA), IRFU, D\'{e}partment de Physique Nucl\'{e}aire (DPhN), Saclay, France\\
$^{129}$ Universit\`{a} degli Studi di Foggia, Foggia, Italy\\
$^{130}$ Universit\`{a} del Piemonte Orientale, Vercelli, Italy\\
$^{131}$ Universit\`{a} di Brescia, Brescia, Italy\\
$^{132}$ Variable Energy Cyclotron Centre, Homi Bhabha National Institute, Kolkata, India\\
$^{133}$ Warsaw University of Technology, Warsaw, Poland\\
$^{134}$ Wayne State University, Detroit, Michigan, United States\\
$^{135}$ Westf\"{a}lische Wilhelms-Universit\"{a}t M\"{u}nster, Institut f\"{u}r Kernphysik, M\"{u}nster, Germany\\
$^{136}$ Wigner Research Centre for Physics, Budapest, Hungary\\
$^{137}$ Yale University, New Haven, Connecticut, United States\\
$^{138}$ Yonsei University, Seoul, Republic of Korea\\
$^{139}$  Zentrum  f\"{u}r Technologie und Transfer (ZTT), Worms, Germany\\
$^{140}$ Affiliated with an institute covered by a cooperation agreement with CERN\\
$^{141}$ Affiliated with an international laboratory covered by a cooperation agreement with CERN.\\

\end{flushleft} 

\end{document}